\documentclass[%
 reprint,
superscriptaddress,
 amsmath,amssymb,
 aps,
]{revtex4-2}

\usepackage{graphicx}
\usepackage{dcolumn}
\usepackage{bm}
\usepackage{multirow}

\usepackage[caption=false]{subfig}

\usepackage[hidelinks,unicode=true]{hyperref}

\usepackage{xr}

\usepackage{xcolor}

\begin{document}


\title{Habituation determines spatial position in Hyphessobrycon herbertaxelrodi shoals through altered rules of interaction}

\author{Andreu Puy}
\email{andreu.puy@upc.edu}
\affiliation{Departament de Física, Universitat Politècnica de Catalunya, Campus Nord B4, 08034 Barcelona, Spain}

\author{Elisabet Gimeno}
\affiliation{Departament de Física, Universitat Politècnica de Catalunya, Campus Nord B4, 08034 Barcelona, Spain}
\affiliation{Departament de Física de la Matèria Condensada, Universitat de Barcelona, Martí i Franquès 1, 08028 Barcelona, Spain}

\author{Francesc S. Beltran}
\affiliation{Departament de
Psicologia Social i Psicologia Quantitativa, Universitat de Barcelona, Passeig Vall d'Hebron, 171, Barcelona 08035,
Spain}

\author{Ruth Dolado}
\affiliation{Departament de
Psicologia Social i Psicologia Quantitativa, Universitat de Barcelona, Passeig Vall d'Hebron, 171, Barcelona 08035,
Spain}

\author{Christos C. Ioannou}
\affiliation{University of Bristol, School of Biological Sciences, Life Sciences Building, 24 Tyndall Avenue, Bristol BS8 1TQ, UK}

\author{M. Carmen Miguel}
\affiliation{Departament de Física de la Matèria Condensada, Universitat de Barcelona, Martí i Franquès 1, 08028 Barcelona, Spain}
\affiliation{Institute of Complex System (UBICS), Universitat de Barcelona, Barcelona 08028, Spain}

\author{Romualdo Pastor-Satorras}
\affiliation{Departament de Física, Universitat Politècnica de Catalunya, Campus Nord B4, 08034 Barcelona, Spain}


\date{14 September, 2026}

\begin{abstract} 
  Habituation plays a key role in shaping the collective behavior of moving
  animal groups, yet the effects of variation in habituation within groups are
  unknown. Here, we merge two subgroups of fish with different levels of
  habituation and quantitatively analyze their movement. Non-habituated fish
  (not familiar with their swimming environment) tend to occupy more central
  positions within the group and position themselves closer to neighbors,
  consistent with the selfish herd hypothesis. This behavior appears to be
  driven by adjusted effective local interactions with nearby
  individuals. Compared to more habituated fish, these individuals also exhibit
  greater coordination, shorter burst-and-coast dynamics, more efficient
  information transmission, and a stronger tendency to adopt follower roles. To
  validate the findings, we develop a machine learning tool that successfully
  classifies the subgroup identity of individual fish. Our study demonstrates
  how differences in local-scale inter-individual interactions can drive spatial
  assortment within groups, based on heterogeneity in their manipulated exposure
  to the experimental setup.
\end{abstract}

\maketitle


\section*{Introduction}

Flocks of birds and schools of fish can move with such cohesion and coordination
that they appear to act with a single `collective
mind'~\cite{couzinCollectiveCognitionAnimal2009}. This can give the impression
that individuals within the group are homogeneous, including  the behaviors
that they use to interact with others in the group. However, variation between
individuals within animal groups is the rule rather than the
exception~\cite{jollesRoleIndividualHeterogeneity2020}, and variation forms the
bedrock upon which evolutionary selection can act to shape the traits that
determine collective behavior~\cite{ioannouMultiscaleReviewDynamics2023}. The
variation between individuals within groups that can impact collective behavior
varies widely from traits that are fixed within the lifetime of an individual
such as sex, to those that vary slowly over time such as age, and those that can
change within seconds, such as the perception of predation
risk~\cite{jollesRoleIndividualHeterogeneity2020}.

Collective behavior is impacted by the level of habituation of the
group. Habituation is a simple form of learning in which an animal’s behavioral
response to a repeatedly presented, benign stimulus decreases over time. Indeed,
the extent to which individuals perceive their environment to be risky depends
on their evolutionary history and early-life
experience~\cite{herbert-readHowPredationShapes2017}, recent or current cues
that indicate risk~\cite{sosnaIndividualCollectiveEncoding2019}, and familiarity
with their environment~\cite{macgregorCollectiveMotionDiminishes2021}.
Generally, an increased perception of predation risk in natural
environments increases group size and cohesion, which is consistent with
multiple mechanisms that reduce the rate and success of attacks from
predators~\cite{krauseLivingGroups2002, ioannouGroupingPredation2017}, and is
also expected from Hamilton's 1971 selfish herd
hypothesis~\cite{hamiltonGeometrySelfishHerd1971}, where individuals reduce the
distance to their near neighbors and move to central locations of the group to
minimize their exposure to unknown factors in a new environment. Individuals can
also rely more on social information~\cite{websterSocialLearningStrategies2008}
and become more responsive to one another when the perception of risk is
greater~\cite{schaerfEffectsExternalCues2017}.

Heterogeneity between individuals in their perception of different stimuli can
generate non-random spatial structure within groups, known as assortment. For
instance, some spatial positions are associated with a greater risk of
predation, usually the front and periphery of the group, but these positions
also allow greater access to food
resources~\cite{krauseDifferentialFitnessReturns1994}. This is supported by
bolder individuals, i.e. those with a consistently lower perception of risk,
tending to be at the front of moving
groups~\cite{balaban-feldInfluencePredationRisk2018}, as well as larger
individuals that are less at risk from
predators~\cite{roseCodSpawningMigration1993}. Non-random assortment within
groups has also been demonstrated with informational state, e.g. the extent to
which individuals are experienced with their
environment~\cite{reebsInfluenceBodySize2001,
  brentEcologicalKnowledgeLeadership2015}. However, in all of these studies,
variation between individuals was unmanipulated and often correlated to other
traits including body size, age and reproductive status, thus a direct effect of
differences in informational state on spatial position has not been
demonstrated.

Research suggests that in most species of fish and birds that form highly
coordinated, cohesive and dynamic groups, this behavior is achieved through
rapid responses to passive cues of neighbors changing their speed and direction,
rather than active signals such as alarm
calls~\cite{ioannouMultiscaleReviewDynamics2023}. These are commonly explained
by simple effective interactions arising from decision making processes of
individuals. A standard framework to study these interactions is based on the
concept of effective forces acting on individuals. Social forces arise from
interactions among group members and are typically described by a combination of
attraction, repulsion, and alignment~\cite{sumpterCollectiveAnimalBehavior2010,
  vicsekCollectiveMotion2012,
  herbert-readUnderstandingHowAnimal2016}. Attraction brings individuals
together into groups, repulsion prevents collisions, and alignment ensures
individuals move in the same direction as their neighbors. Furthermore,
interactions between pairs can be asymmetric and
non-reciprocal~\cite{puySelectiveSocialInteractions2024}, leading to a net
information flow with leader-follower
relationships~\cite{zafeirisWhyWeLive2018}.

Previous studies have explored how risk perception affects collective behavior,
how pre-existing differences between individuals in risk perception result in
spatial assortment within groups, and the fine-scale ‘rules of interaction’ that
underpin collective behavior~\cite{limaBehavioralDecisionsMade1990}. Here, we
instead manipulated the informational state of individuals within fish shoals,
testing whether fish habituated to the test environment occupied different
spatial positions when shoaling with non-habituated individuals. Based on the selfish herd hypothesis, we predicted that non-habituated individuals, expected to perceive the novel environment as more uncertain or risky, would exhibit more risk-averse spatial behavior than habituated individuals, including occupying more central positions within the group and maintaining shorter distances to neighbors. Through
detailed analysis of the fish’s responses to other fish, we then demonstrate the
differences in interaction rules, coordination, leader-follower behavior, and
individual self-propulsion forces between habituated and non-habituated fish,
and how local-scale individual behaviors can result in global-scale patterns of
spatial assortment as predicted from an influential model of collective
motion~\cite{couzinCollectiveMemorySpatial2002}.  We also investigate the
temporal evolution of interactions throughout the two-hour duration of the
experiment. Finally, to conclusively demonstrate behavioral differences, we
develop a machine learning tool to classify individuals into their respective
subgroups based on their observed behaviors.

We used black neon tetras (\emph{Hyphessobrycon herbertaxelrodi}), a highly
social fish species that forms polarized, compact, and planar schools. Sixteen
individuals, already familiar with each other, were divided into two subgroups
of eight individuals, designated as \emph{habituated} (Hab) and
\emph{non-habituated} (Non-hab). The habituated subgroup underwent a habituation
process in an experimental tank with an approximately two-dimensional setting,
where individuals swam freely for two hours each day over five days. In
contrast, the non-habituated subgroup was not exposed to the experimental
tank. On the sixth day, both subgroups were merged in the experimental tank, and
their individual trajectories were recorded as they swam freely for two
hours. The experiment was replicated three times with different sets of
individuals, labeled series A, B, and C. Supplementary Video S1 provides a
sample of the school's behavior, with digitized trajectories overlaid onto the
video. See Methods for the full experimental details.

\section*{Results}

\subsection*{Group-level spatial interactions}

\begin{figure*}[t!]
\centering
\subfloat[]{%
  \includegraphics[width=0.25\textwidth]{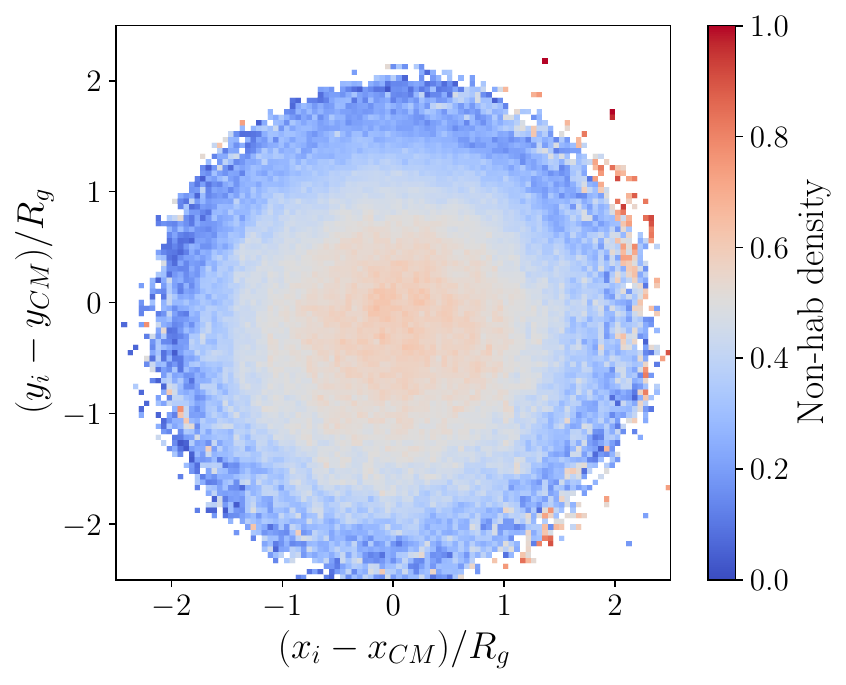}%

}
\subfloat[]{%
  \includegraphics[width=0.25\textwidth]{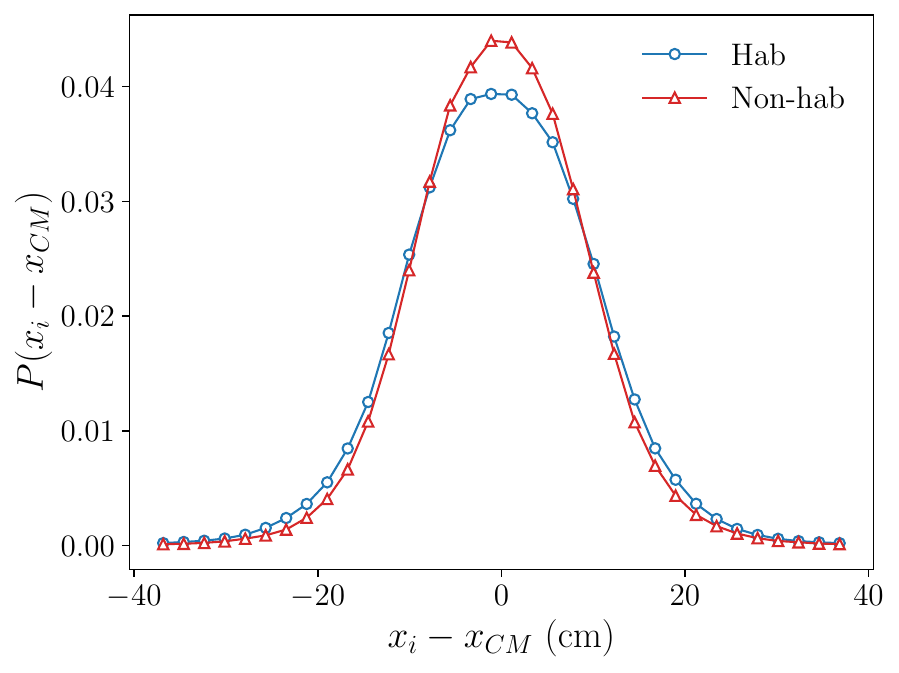}
}
\subfloat[]{%
  \includegraphics[width=0.25\textwidth]{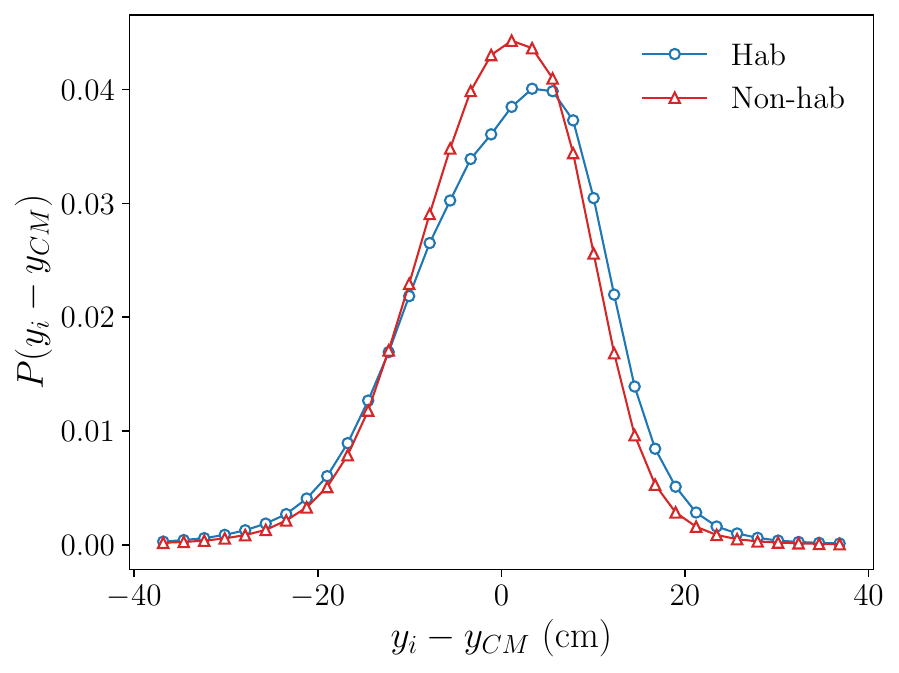}%
}

\subfloat[]{%
  \includegraphics[width=0.25\textwidth]{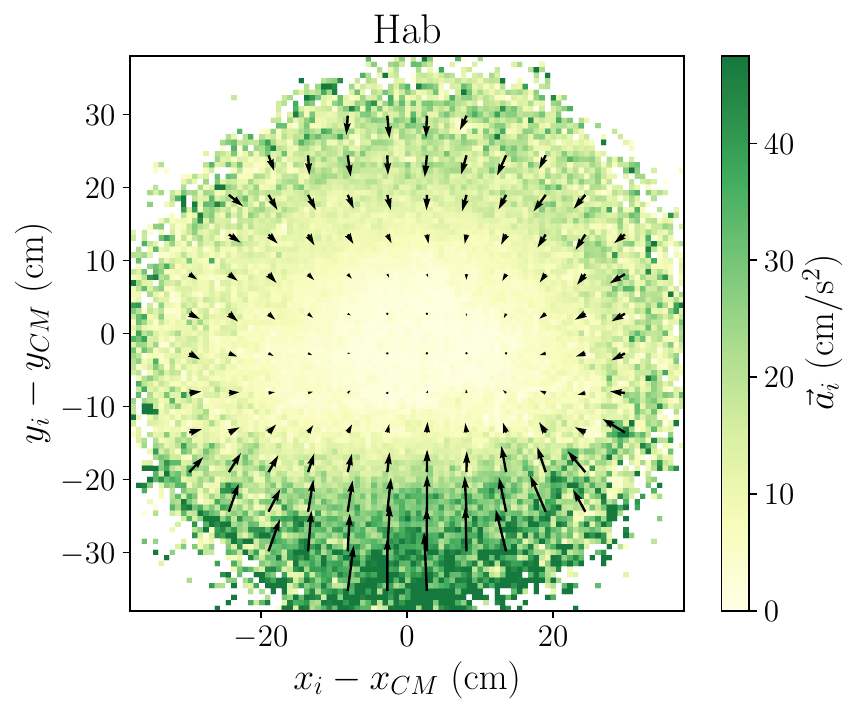}%
}
\subfloat[]{%
  \includegraphics[width=0.25\textwidth]{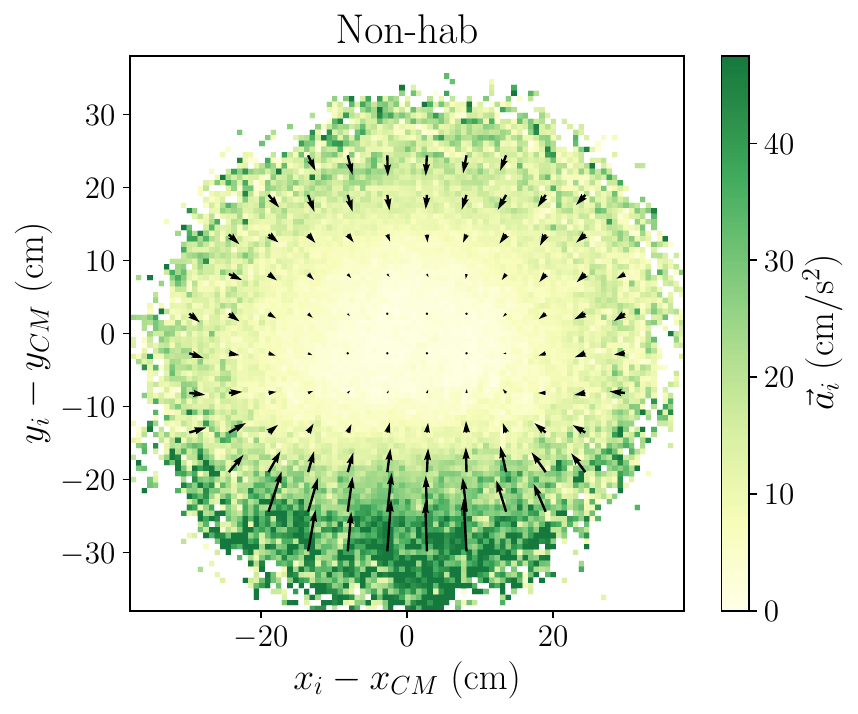}%
}
  \caption{\textbf{Group-level spatial interactions.} (a) Spatial density
    distribution of non-habituated individuals $i$ relative to the center of
    mass CM, $(\vec{r}_i - \vec{r}_{CM})/R_g$. For each bin, the value is computed as the ratio between the counts of non-habituated individuals and the total counts of all individuals, aggregated across all time frames and experimental series. The $y$-axis is
    oriented along the direction of motion of the group given by the
    center-of-mass velocity $\vec{v}_{CM} \equiv \frac{1}{N} \sum_i
    \vec{v}_i$. (b) and (c) Probability density functions (PDFs) of the relative
    position of individuals with respect to the center of mass,
    $\vec{r}_{i}-\vec{r}_{CM}$, along the (b) $x$-coordinate and (c)
    $y$-coordinate. 
    (d) and (e) Average acceleration
    (force map), $\vec{a}_i$, depending on the relative position with respect to
    the center of mass, $\vec{r}_i - \vec{r}_{CM}$, for (d) habituated and (e)
    non-habituated individuals. The force components are displayed with arrows
    and the modulus with the colormap. 
    In all 2-dimensional plots, bins with fewer than 15 counts are excluded to
    reduce statistical noise. For all PDFs, we show error bands calculated from
    the standard deviation of a Bernoulli distribution, with the probability
    given by the fraction of counts in each bin. 
    }\label{fig:group_spatial}
\end{figure*}

We initially tested whether non-habituated individuals occupied different
spatial positions within the group compared to habituated individuals. We thus
examined the spatial density distribution of individuals based on their
normalized relative positions within the group,
$(\vec{r}_i - \vec{r}_{CM})/R_g$. Here, $\vec{r}_i \equiv (x_i, y_i)$ denotes
the positions of the individuals,
$\vec{r}_{CM} \equiv \frac{1}{N} \sum_i \vec{r}_i$ represents the position of
the center of mass CM of a group with $N$ individuals, and
$R_g\equiv \sqrt{ \frac{1}{N} \sum_i \left| \vec{r}_{i} - \vec{r}_{CM}
  \right|^2}$ is the radius of gyration, indicating the typical length scale of
the group. The $y$-coordinate is oriented along the direction of motion of the group given by the center-of-mass velocity $\vec{v}_{CM} \equiv \frac{1}{N} \sum_i \vec{v}_i$. Consistent with the selfish herd hypothesis, non-habituated
individuals occupy peripheral positions less frequently than habituated
individuals (Fig.~\ref{fig:group_spatial}a). This is robust across the different
series (Supplementary
figs.~\ref{supp:fig:group_spatial_series}a-c) and remains
consistent throughout the two-hour duration of the experiment (Supplementary
figs.~\ref{supp:fig:group_spatial_tEvo}a and b).

This analysis can be further refined by separating it into the $x$ and
$y$-coordinates, revealing significant differences between habituated and
non-habituated groups PDFs, as measured by the Kolmogorov–Smirnov (KS)
test~\cite{zar2014biostatistical}. We obtain the values $D=0.03$ for $x$,
$D=0.06$ for $y$; both $p<0.001$, indicating that the distributions are
statistically different. In both axes, the center of the group is dominated by
non-habituated individuals (Figs.~\ref{fig:group_spatial}b and c; Supplementary
figs.~\ref{supp:fig:group_spatial_series}d-i for different series). However,
while in the $x$-coordinate distributions are symmetric, we find that in the
$y$-coordinate, habituated individuals tend to
be positioned more towards the front of the group (positive $y_i - y_{CM}$
values).

The selfish herd hypothesis predicts that individuals will dynamically change
their spatial positions to avoid being on the periphery (i.e. border) of the
group. We define an individual is on the border at a given frame if it belongs
to the convex hull of the group, which is the smallest convex polygon that
contains all individuals. With this definition, we have, on average, $7.2$
individuals on the border and $8.8$ individuals in the inner part of the group
at any given frame (Supplementary
fig.~\ref{supp:fig:border_definitions}a). Consistent with our previous results,
there were more habituated individuals on the border ($4.0$ on average) than
non-habituated individuals ($3.2$ on average) (Supplementary
fig.~\ref{supp:fig:border_definitions}b). This difference was found to be significant in a permutation test in which subgroup labels were randomly reassigned within each experimental series while preserving subgroup sizes ($p=0.019$, one-sided test).

 To infer effective forces acting on individuals based on their spatial position
 within the group, we use the force map
 method~\cite{katzInferringStructureDynamics2011,
   herbert-readInferringRulesInteraction2011,
   puySelectiveSocialInteractions2024}. This infers forces from the dependency
 of the acceleration on some relevant variables, while averaging over other
 variables (see Methods). Despite the differences in spatial
 position between habituated and non-habituated fish, we find the force maps for
 habituated (Fig.~\ref{fig:group_spatial}d) and non-habituated
 (Fig.~\ref{fig:group_spatial}e) individuals are very similar, with no net force
 for central regions of the group and arrows pointing inwards at larger
 distances, signaling an attractive force towards the center. This effective
 force is responsible for individuals forming cohesive groups. Additionally, we
 find forces are stronger for individuals at the rear of the group (negative
 $y$-values). These results are consistent across experimental series
 (Supplementary figs.~\ref{supp:fig:group_spatial_series}j-l).
To take a more detailed view of the trends in the force maps, we consider the average modulus of the acceleration in the radial direction of the force map, $a_i^r$, as a function of the distance to the center-of-mass, $\left| \vec{r}_i - \vec{r}_{CM} \right|$. We do not find significant differences between non-habituated and habituated individuals (Supplementary figs.~\ref{supp:fig:group_spatial_series}m-o).
Nonetheless, attractive forces decrease over time (Supplementary fig.~\ref{supp:fig:group_spatial_tEvo}e). This is consistent with the temporal trend that the size occupied by the group expands (Supplementary fig.~\ref{supp:fig:group_spatial_tEvo}c and d), i.e. the group becomes less cohesive as time progresses, consistent with previous work~\cite{macgregorCollectiveMotionDiminishes2021}.

\subsection*{Local-level spatial interactions}

Group-level behaviors, such as spatial positioning, are typically understood to
arise from the superposition of local interactions between individuals. In this
section, we study local-level spatial interactions between individuals and their
nearest neighbors. We analyze behavioral differences depending on both the
individual's subgroup identity and the nearest-neighbor's subgroup identity, to
determine whether individuals can recognize and respond differently to these
identities.

\begin{figure*}[t!] 
  \centering
\subfloat[]{%
  \includegraphics[width=0.25\textwidth]{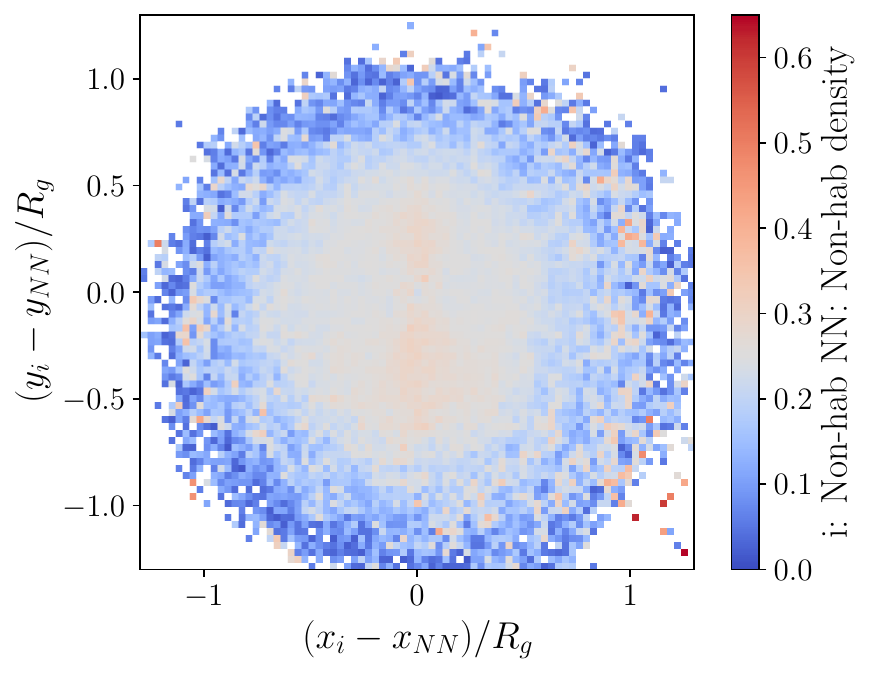}%
}
\subfloat[]{%
  \includegraphics[width=0.25\textwidth]{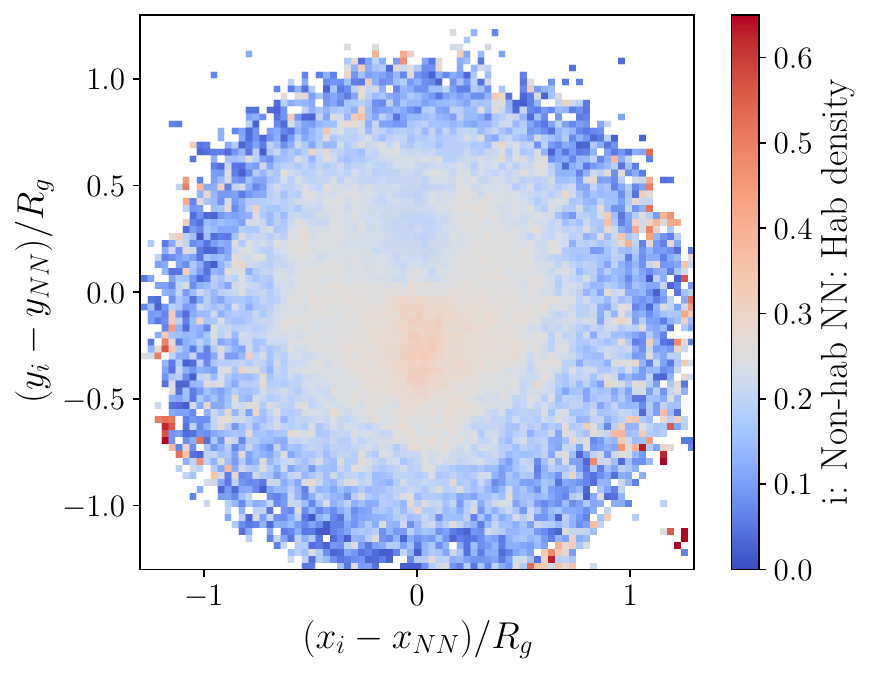}%
}
\subfloat[]{%
  \includegraphics[width=0.25\textwidth]{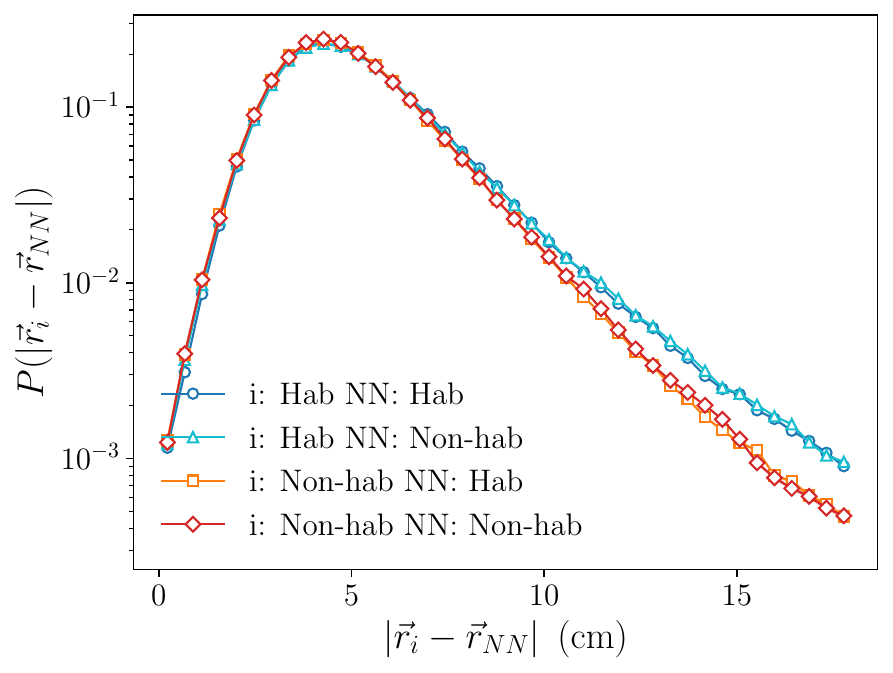}%
}

\subfloat[]{%
  \includegraphics[width=0.25\textwidth]{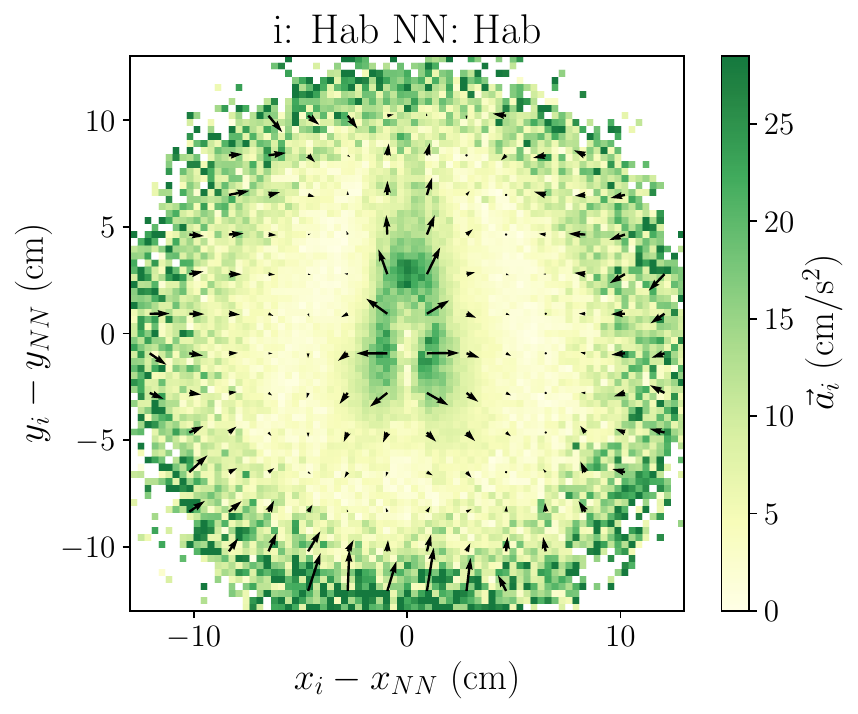}%
}
\subfloat[]{%
  \includegraphics[width=0.25\textwidth]{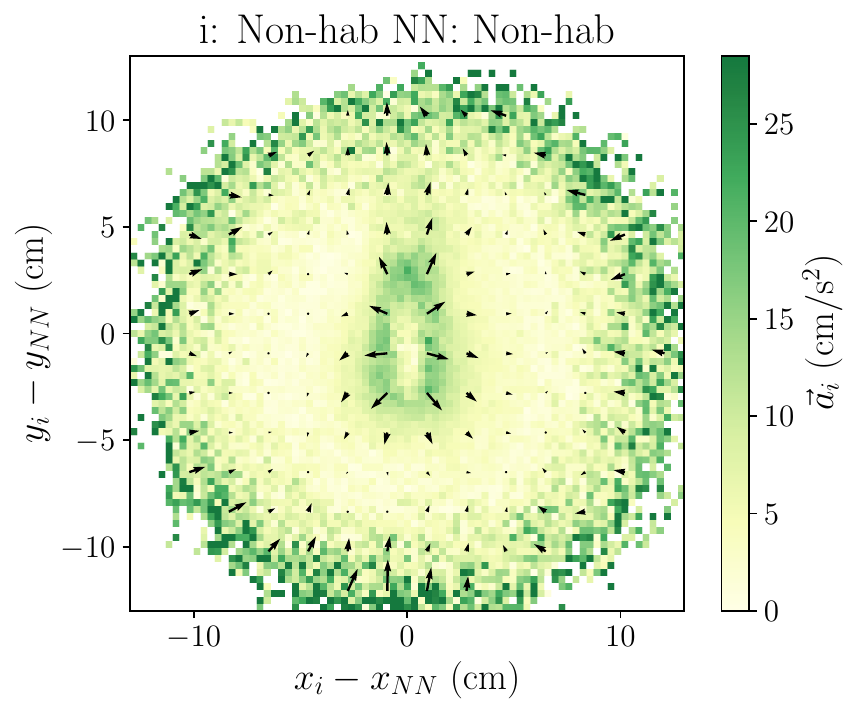}%
}
\subfloat[]{%
  \includegraphics[width=0.25\textwidth]{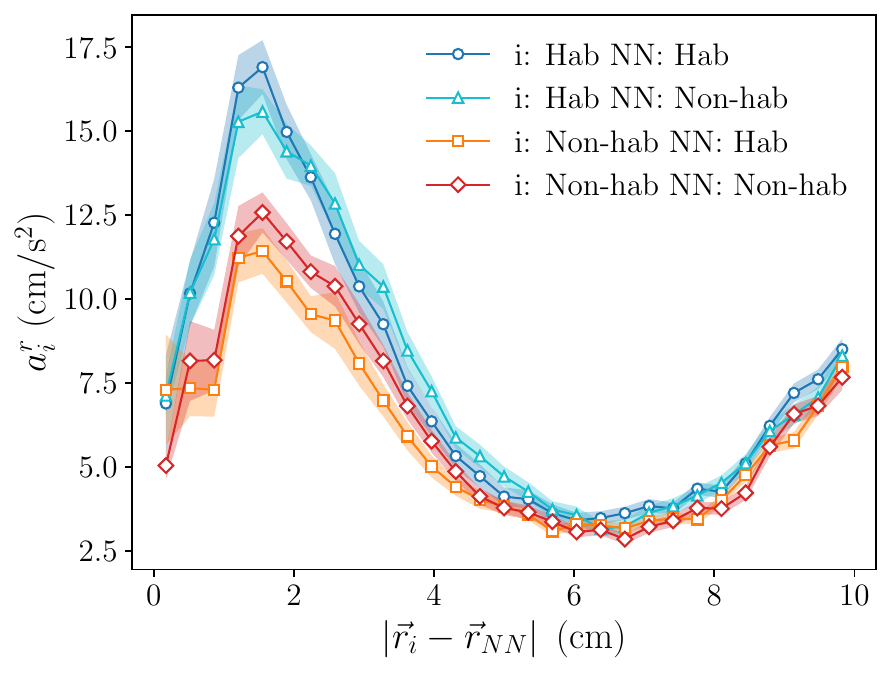}%
}
  \caption{\textbf{Local-level spatial interactions.} Spatial density
    distribution of non-habituated individuals $i$ with either (a)
    non-habituated or (b) habituated nearest neighbors $NN$ as a function of
    their normalized relative positions $(\vec{r}_i - \vec{r}_{NN})/R_g$. For each bin, the value is computed as the ratio between the counts of the corresponding focal individual--nearest neighbor pair type and the total counts of all pairs, aggregated across all time frames and experimental series.  The
    $y$-axis is oriented along the direction of motion of individual
    $i$. (c) PDFs of the distance between the individual $i$ and its nearest
    neighbor $NN$, $\left| \vec{r}_i - \vec{r}_{NN}\right|$. Average
    acceleration (attraction-repulsion force map) of individual $i$ depending on
    its relative position to its nearest neighbor $NN$,
    $\vec{r}_i - \vec{r}_{NN}$, for both (d) habituated and (e) non-habituated
    pairs. (f) Average modulus of the acceleration in the radial direction of
    the attraction-repulsion force map, $a_i^r$, depending on the distance to
    the nearest neighbor,
    $\left| \vec{r}_i - \vec{r}_{NN} \right|$. Error bands represent the standard error of the mean.}\label{fig:local_spatial}
\end{figure*}

We first analyze the spatial density distribution of individuals depending on
their normalized relative positions with respect to their nearest neighbor $NN$,
$(\vec{r}_i - \vec{r}_{NN})/R_g$. The $y$-axis is aligned with the direction of motion of the focal individual. Figs.~\ref{fig:local_spatial}a and b show this distribution for non-habituated focal individuals with non-habituated and habituated nearest neighbors, respectively; the corresponding panels for habituated focal individuals are shown in Supplementary Fig.~\ref{supp:fig:density_habituated_positions}. By contrasting these distributions, we find that non-habituated individuals position themselves closer to their nearest neighbors than habituated individuals, exhibiting a higher density of neighbors at short distances and a lower density at larger distances, regardless of the neighbor’s subgroup identity. This behavior is consistent with selfish-herd-like spatial organization at the local level. Furthermore, we observe a subtle but systematic difference in the
central region of the density distributions:
non-habituated individuals tend to be located behind habituated nearest
neighbors. These findings are robust across series (Supplementary figs.~\ref{supp:fig:local_spatial_series}a-f) and
remain consistent throughout the 2-hour duration of the experiment (Supplementary
figs.~\ref{supp:fig:local_spatial_tEvo}a-d).

To better quantify local-level spatial interactions, Fig.~\ref{fig:local_spatial}c (Supplementary figs.~\ref{supp:fig:local_spatial_series}g-i for different series) shows the PDF of the distances between individuals and their nearest neighbors. The peak of the PDF, indicating the preferred nearest-neighbor distance, is identical for all individuals and neighbors. However, there are fewer non-habituated individuals at the longer distances, regardless of the nearest neighbor identity (KS test: $D=0.003$ between habituated individuals with habituated vs non-habituated neighbors, $D=0.04$ between habituated–habituated and non-habituated–non-habituated pairs; both $p<0.001$). Over time, nearest-neighbor distances tend to increase (Supplementary figs.~\ref{supp:fig:local_spatial_tEvo}e).

We infer effective spatial forces at the local level with the
attraction-repulsion force map~\cite{katzInferringStructureDynamics2011,
  herbert-readInferringRulesInteraction2011,
  puySelectiveSocialInteractions2024}, which extracts attraction-repulsion
forces from the average individual acceleration $\vec{a}_i$ based on the
relative position of nearest neighbors, $\vec{r}_i - \vec{r}_{NN}$. The force
maps (Figs.~\ref{fig:local_spatial}d and e) for interactions between habituated
and non-habituated individuals are qualitatively similar: for near distances
arrows point outward, indicating the focal individual experiences a repulsive
force away from the neighbor; for intermediate distances, there is an area with
no net force (the equilibrium distance); and for longer distances, arrows point
inward, signaling an attractive force toward the neighbor. However, the
repulsive force between non-habituated individuals appears weaker than for
habituated individuals, which helps explain the observed patterns in the spatial
density distributions. To quantify this difference, we also analyze the average
modulus of the acceleration in the radial direction of the force map, $a_i^r$ as
a function of the distance to the nearest neighbor,
$\left| \vec{r}_i - \vec{r}_{NN} \right|$ (Fig.~\ref{fig:local_spatial}f). We
find that the peak for small nearest-neighbor distances, associated with the
repulsive forces, is considerably lower for non-habituated individuals compared
to habituated individuals, regardless of the subgroup identity of the nearest
neighbor ($\chi^2=24$ and $p=0.75$ between habituated individuals with
habituated vs non-habituated neighbors, $\chi^2=120$ and $p<0.001$ between
habituated–habituated and non-habituated–non-habituated pairs; both $df =
29$). In contrast, their behaviors are similar within the equilibrium distance
and attractive regions. These results are robust across different series
(Supplementary fig.~\ref{supp:fig:local_spatial_series}j-l). Across time,
repulsive forces increase, the equilibrium distance expands, and attractive
forces decrease (Supplementary fig.~\ref{supp:fig:local_spatial_tEvo}f).


\subsection*{Group-level coordination}

\begin{figure}[t!]
  \centering
  \subfloat[]{%
  \includegraphics[width=0.25\textwidth]{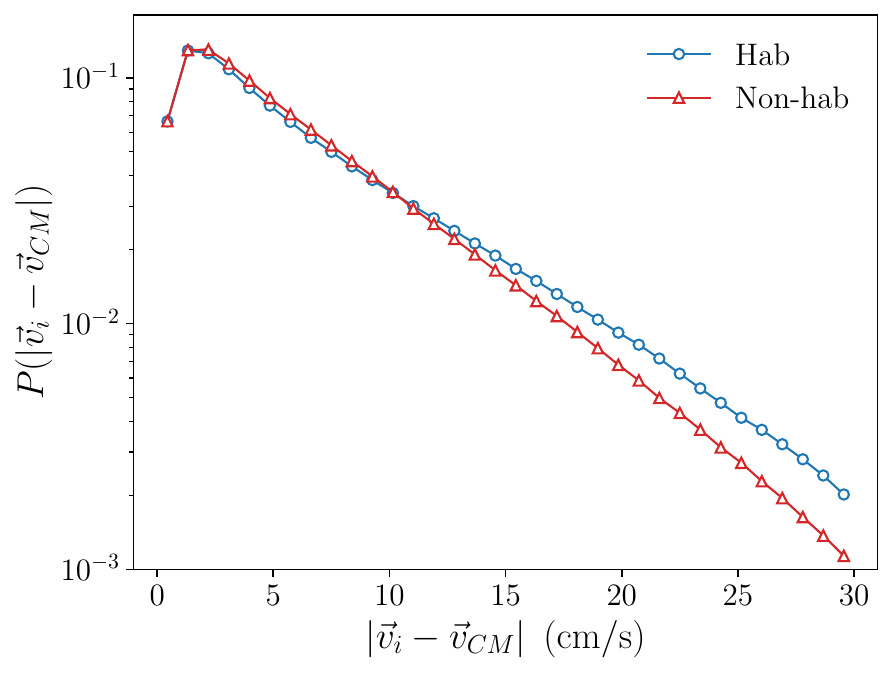}%
}
\subfloat[]{%
  \includegraphics[width=0.25\textwidth]{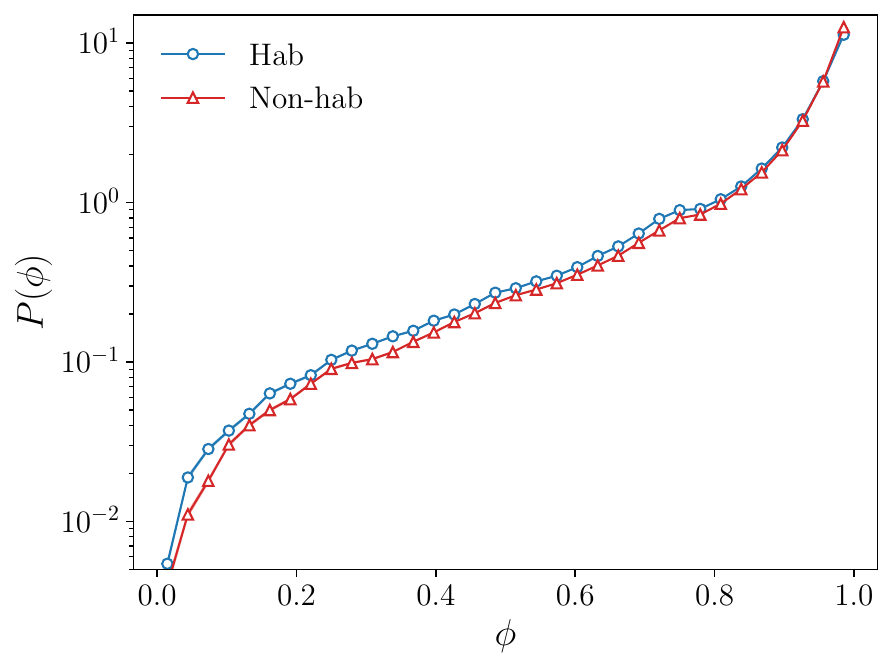}%
}

\subfloat[]{%
  \includegraphics[width=0.25\textwidth]{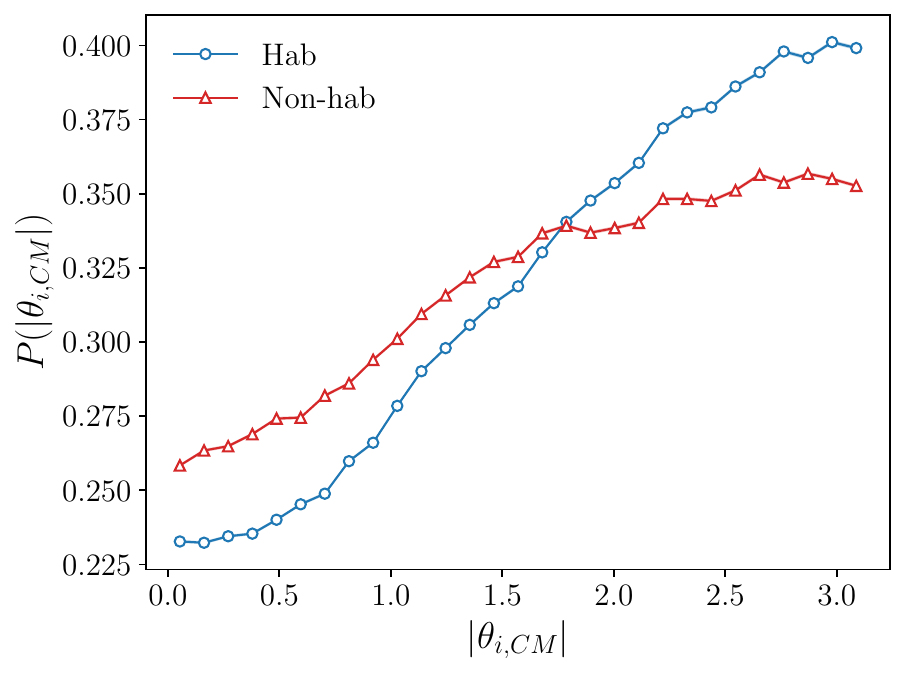}%
}
  \caption{\textbf{Group-level coordination.} PDFs of (a) the modulus of the
    velocity differences between individuals and the center of mass,
    $\left|\vec{v}_i - \vec{v}_{CM}\right|$, (b) the polarization $\phi$ and (c)
    the absolute value of the individuals' orientation relative to the CM,
    $\theta_{i, CM}$.}\label{fig:group_velocity}
\end{figure}

In this section, we aim to study the speed and orientation of individuals
relative to the school. We begin by examining the modulus of the velocity
differences between the individual and the group,
$\left| \vec{v}_i - \vec{v}_{CM} \right|$. We find velocities of non-habituated
fish are more aligned with the average group motion
(Fig.~\ref{fig:group_velocity}a) (KS test: $D = 0.04$, $p<0.001$). Similarly, we
analyze the polarization $\phi$~\cite{vicsekNovelTypePhase1995} of the subgroup,
\begin{equation*}
  \phi \equiv \left | \frac{1}{N_s} \sum_i \frac{ \vec{v}_i  }{  v_i } \right|,
  \label{eq:4}
\end{equation*}
where we sum over the $N_s$ individuals of the subgroup. Polarization approaches
1 when the subgroup is ordered and aligned, with all individuals moving in the
same direction, and approaches $0$ when it is disordered, with
individuals moving in random and independent directions. Non-habituated individuals exhibit a reduced frequency of low-polarization states, leading to a slightly higher overall polarization
(Fig.~\ref{fig:group_velocity}b), indicating they are more aligned with each
other than habituated individuals (KS test: $D = 0.04$, $p<0.001$). Both
measures are consistent across experimental series (Supplementary
figs.~\ref{supp:fig:group_velocity_series}a-f). Additionally, over time, the
modulus of the velocity differences increases (Supplementary
fig.~\ref{supp:fig:group_velocity_tEvo}a), and polarization decreases
(Supplementary fig.~\ref{supp:fig:group_velocity_tEvo}b), indicating that
alignment with the group decreases over time, again consistent with previous
work on fish shoals~\cite{macgregorCollectiveMotionDiminishes2021}.

According to the selfish herd hypothesis, non-habituated individuals may orient
themselves towards the center of the group to reach the safer central
positions. This can be quantified as the orientation of individuals with respect
to the group,
\begin{equation*}
    \theta_{i, CM} = \arctan\left(\frac{(\vec{r}_{CM} - \vec{r}_i) \times \vec{v}_i}{(\vec{r}_{CM} - \vec{r}_i)\cdot \vec{v}_i}  \right),
\end{equation*}
where an individual pointing directly towards the center of mass has
$\theta_{i,CM} = 0$, while an individual pointing directly away from the center
of mass has $\left| \theta_{i,CM} \right| = \pi$ (Supplementary
fig.~\ref{supp:fig:theta_i_CM}). Despite only a few individuals tending to point
towards the center, these are more likely to be non-habituated rather than
habituated individuals (Fig.~\ref{fig:group_velocity}c, Supplementary
figs.~\ref{supp:fig:group_velocity_series}g-i) (KS test: $D = 0.04$,
$p<0.001$). Differences between the subgroups tend to decrease over time, but
they still remain noticeable at the end of our two-hour recordings
(Supplementary fig.~\ref{supp:fig:group_velocity_tEvo}c).

\subsection*{Local-level coordination}

\begin{figure}[t!]
  \centering
  \subfloat[]{%
  \includegraphics[width=0.25\textwidth]{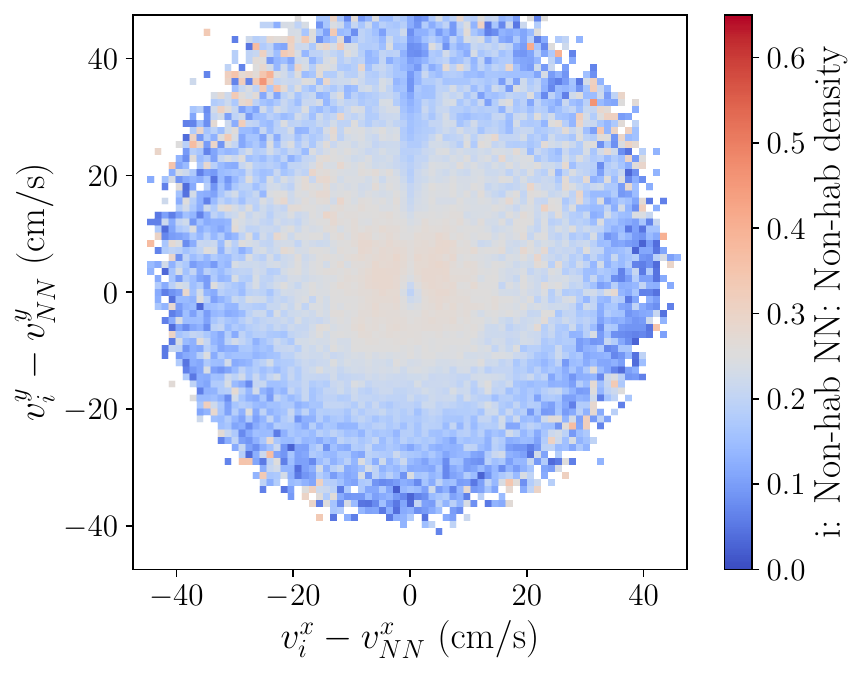}%
}
\subfloat[]{%
  \includegraphics[width=0.25\textwidth]{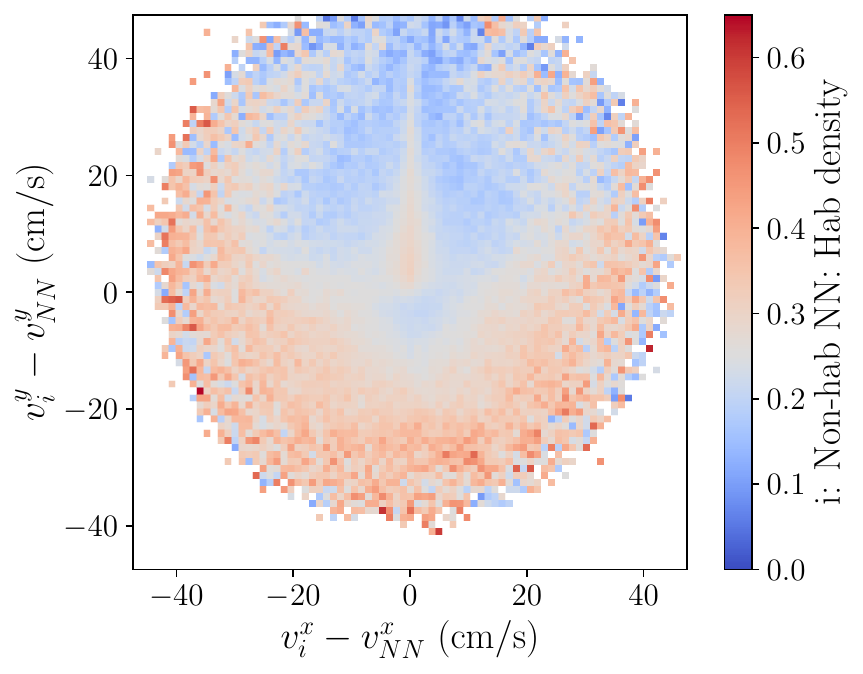}%
}

\subfloat[]{%
  \includegraphics[width=0.25\textwidth]{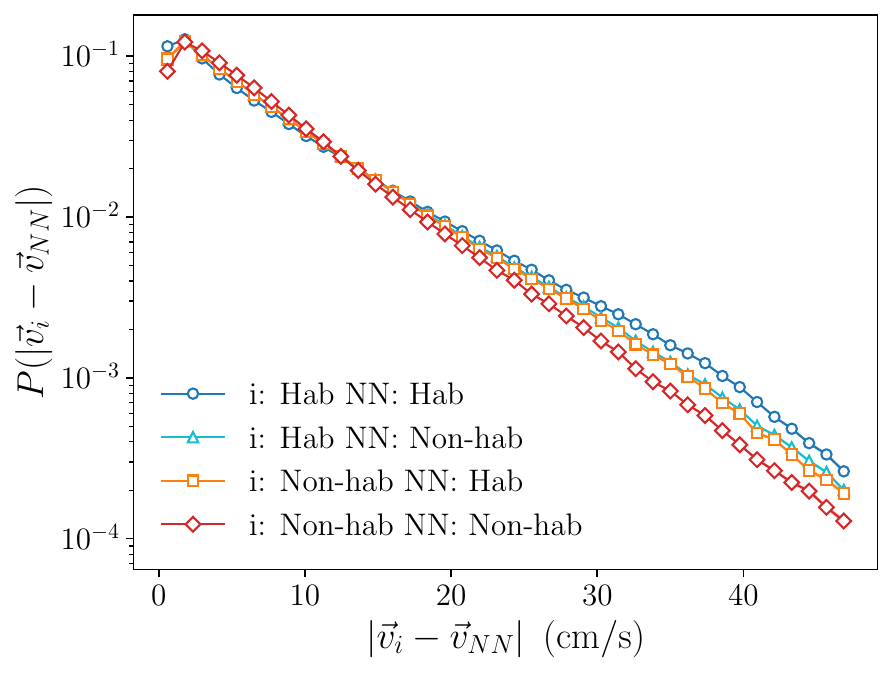}%
}
  \caption{\textbf{Local-level coordination.} Velocity density distribution of
    non-habituated individuals $i$ with (a) non-habituated and (b) habituated
    nearest neighbors $NN$ depending on their relative velocities
    $\vec{v}_i - \vec{v}_{NN}$. The $y$-coordinate is oriented along the
    direction of motion of individual $i$. (c) PDFs of the modulus of the
    relative velocity
    $\left|\vec{v}_i - \vec{v}_{NN}\right|$.}\label{fig:local_velocity}
\end{figure}

In addition to coordination with the group and subgroup, coordination can also
be quantified at a local level using relative velocities. The probability
distribution of relative velocities between non-habituated individuals and their
non-habituated nearest neighbors (Fig.~\ref{fig:local_velocity}a) tends to
cluster in the central region of the plot, reflecting lower velocity differences
compared to other types of neighbor pairs. In contrast, non-habituated
individuals with habituated nearest neighbors (Fig.~\ref{fig:local_velocity}b)
are more frequently located in the bottom-half region of the plot, i.e. they
tend to be slower than habituated fish. The findings are consistent for
complementary plots with habituated individuals (Supplementary
fig.~\ref{supp:fig:density_habituated_velocities}), across series (Supplementary
figs.~\ref{supp:fig:local_velocity_series}a-f) and over time (Supplementary
figs.~\ref{supp:fig:local_velocity_tEvo}a-d).

Fig.~\ref{fig:local_velocity}c shows the probability density of velocity
differences for the four subgroup combinations of nearest-neighbor pairs (also
see Supplementary figs.~\ref{supp:fig:local_velocity_series}g-i). Velocity
differences are lower among non-habituated pairs, indicating higher alignment at
the local level, compared to habituated pairs. Mixed pairings, involving
habituated individuals and non-habituated nearest neighbors, fall in between (KS
test: $D=0.04$ between habituated-habituated and non-habituated-non-habituated
pairs, $D=0.011$ between habituated–non-habituated and non-habituated-habituated
pairs; both $p<0.001$). Additionally, we observe that over time, velocity
differences increase, corresponding to a decrease in local alignment
(Supplementary fig.~\ref{supp:fig:local_velocity_tEvo}e).

To interpret these patterns, we infer the effective local alignment forces using
the alignment force map~\cite{puySelectiveSocialInteractions2024}. This map
shows the average acceleration of individuals, $\vec{a}_i$, as a function of
their relative velocities with respect to their nearest neighbors,
$\vec{v}_i - \vec{v}_{NN}$ (Supplementary
fig.~\ref{supp:fig:local_aligForceMap}). There is qualitatively similar behavior
for both subgroups, i.e. in both cases we observe a selective alignment
mechanism with faster nearest neighbors, as described in
Ref.~\cite{puySelectiveSocialInteractions2024}. Notably, non-habituated
individuals with habituated nearest neighbors occupy with higher probability the
region in the velocity density distribution (Fig.~\ref{fig:local_velocity}b)
that corresponds to effective alignment interactions, characterized by
inward-pointing arrows. In contrast, habituated individuals with non-habituated
nearest neighbors are more likely to occupy areas of the plot characteristic of
apparent anti-alignment interactions (Supplementary
fig.~\ref{supp:fig:density_habituated_velocities}b), with outward-pointing
arrows. These regions and effective interactions can be rationalized in terms of
leader-follower relationships~\cite{puySelectiveSocialInteractions2024}.

\subsection*{Leader-follower relationships}

Information flow is naturally observed from individuals at the front of a group
to those behind~\cite{nagyHierarchicalGroupDynamics2010,
  romero-ferreroStudyTransferInformation2023}, and from individuals moving at
higher speeds to those with lower
velocities~\cite{pettitSpeedDeterminesLeadership2015,
  puySelectiveSocialInteractions2024}. This implies that non-habituated
individuals, tending to position themselves in the rear of the group and
exhibiting lower velocities compared to their habituated nearest neighbors, may
be following habituated individuals. To explicitly test this, we analyze the
transmission of information between pairs of individuals and determine
leader-follower relationships.

\begin{figure}[t!]
  \centering
  \subfloat[]{%
  \includegraphics[width=0.25\textwidth]{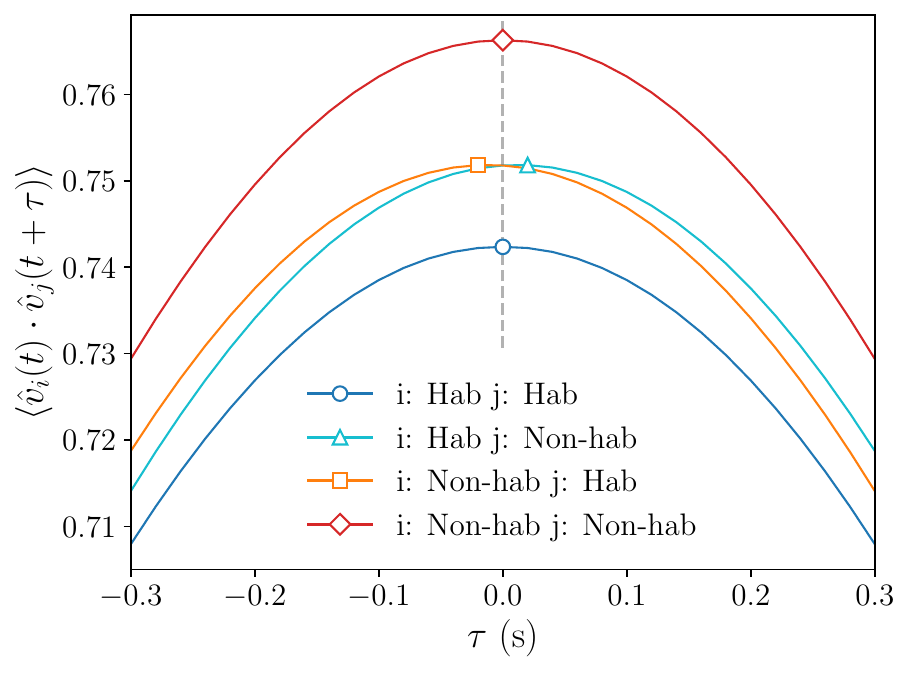}%
}
\subfloat[]{%
  \includegraphics[width=0.25\textwidth]{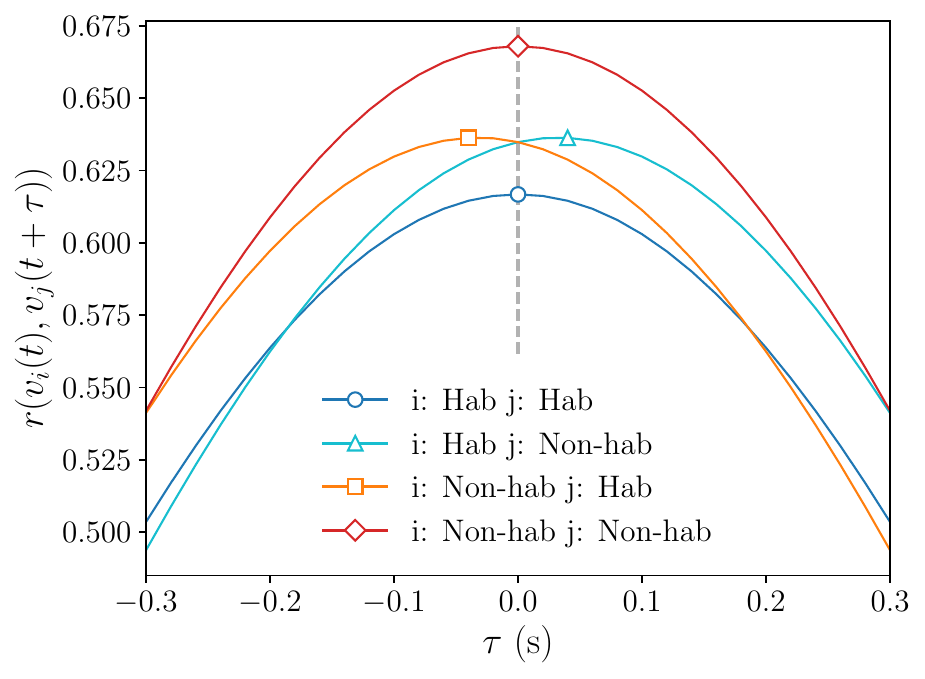}%
}
  \caption{\textbf{Leader-follower relationships.} Correlations in (a) the
    orientation and (b) the speed (Pearson correlation coefficient $r$) between
    an individual $i$ and an individual $j$ at a delayed time $\tau$. The
    markers denote the maximum of each line, while the vertical dashed line
    represents $\tau=0$. Correlation values range from $[-1,1]$. If there are
    leader-follower interactions, a follower requires some time to react and
    copy the leader. This results in correlations having a higher value at that
    delayed time, provided the sampling interval (here, the frame rate of the
    video) is higher than the reaction time of the follower. If the maximum
    occurs for a positive delay $\tau$, this implies individual $i$ is the
    leader with respect to individual $j$. While for a negative delay,
    individual $i$ is the follower. Additionally, the value of the maximum
    reflects the amount of information transmitted by the leadership
    flow.}\label{fig:leader-follower_relationships}
\end{figure}

We employ correlations between an individual $i$ and an individual $j$ at a
delayed time $\tau$ and average them across all time frames $t$ maintaining the
subgroup identities of individuals. We consider two types of correlations, in
the orientation given by
$\left< \hat{v}_i (t) \cdot \hat{v}_{j} (t+\tau)
\right>$~\cite{nagyHierarchicalGroupDynamics2010}
(Fig.~\ref{fig:leader-follower_relationships}a) and in the speed given by the
Pearson correlation coefficient
$r(v_i(t), v_{j}(t+\tau))$~\cite{katzInferringStructureDynamics2011,
  puySelectiveSocialInteractions2024}
(Fig.~\ref{fig:leader-follower_relationships}b). Habituated individuals exhibit
maximum correlations at positive time delays with respect to non-habituated
individuals ($\tau^\theta=0.02 \pm 0.01~\textrm{s}$ and $\tau^v=0.04 \pm 0.01~\textrm{s}$ for orientation and speed correlations, respectively), indicating they tend to act as leaders. Conversely, when considering
non-habituated individuals with respect to habituated individuals, both
correlations have their maxima at negative delays $\tau$
(Fig.~\ref{fig:leader-follower_relationships}), confirming that non-habituated
individuals behave as followers. Additionally, the maximum correlation between
pairs of non-habituated individuals ($c^\theta = 0.766 \pm 0.005$ and $c^v = 0.668 \pm 0.008$ for orientation and speed correlations, respectively) is higher than between pairs of habituated
individuals ($c^\theta = 0.742 \pm 0.005$ and $c^v = 0.617 \pm 0.007$), implying that non-habituated individuals transmit information on
orientations and speeds more effectively among themselves. Mixed pairs of
habituated and non-habituated individuals show intermediate levels of
information transmission ($c^\theta = 0.752 \pm 0.005$ and $c^v = 0.636 \pm 0.007$). These trends tend to be robust across different series
(Supplementary fig.~\ref{supp:fig:leader_follower_series}). Over time,
correlations decrease (Supplementary fig.~\ref{supp:fig:leader_follower_tEvo}), indicating a reduction in the transmission of
information, again consistent with previous
work~\cite{macgregorCollectiveMotionDiminishes2021}.

\subsection*{Burst-and-coast dynamics}

\begin{figure}[t!]
  \centering
  \subfloat[]{ \includegraphics[width=0.25\textwidth]{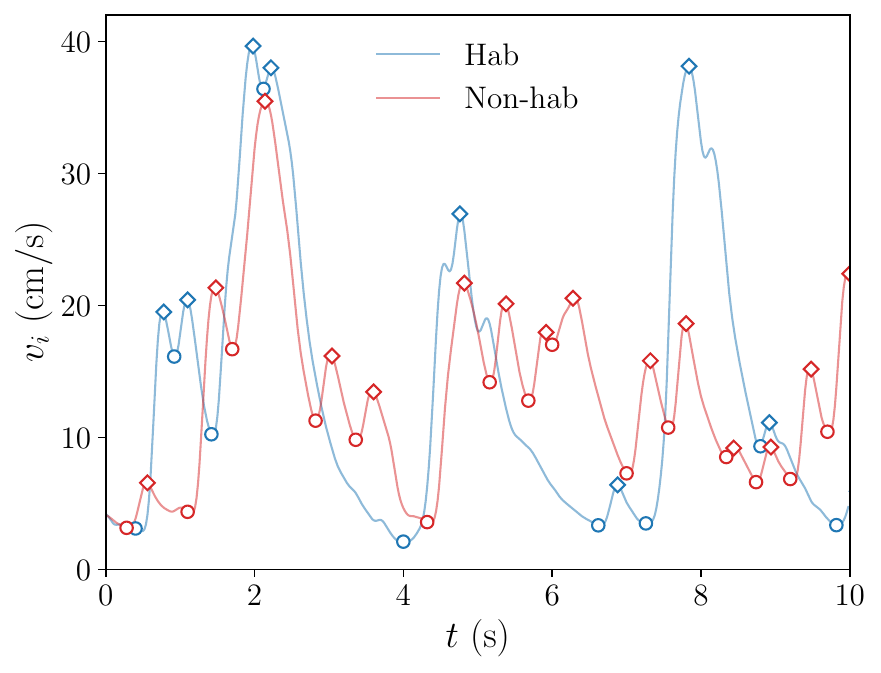}%
}
\subfloat[]{%
  \includegraphics[width=0.25\textwidth]{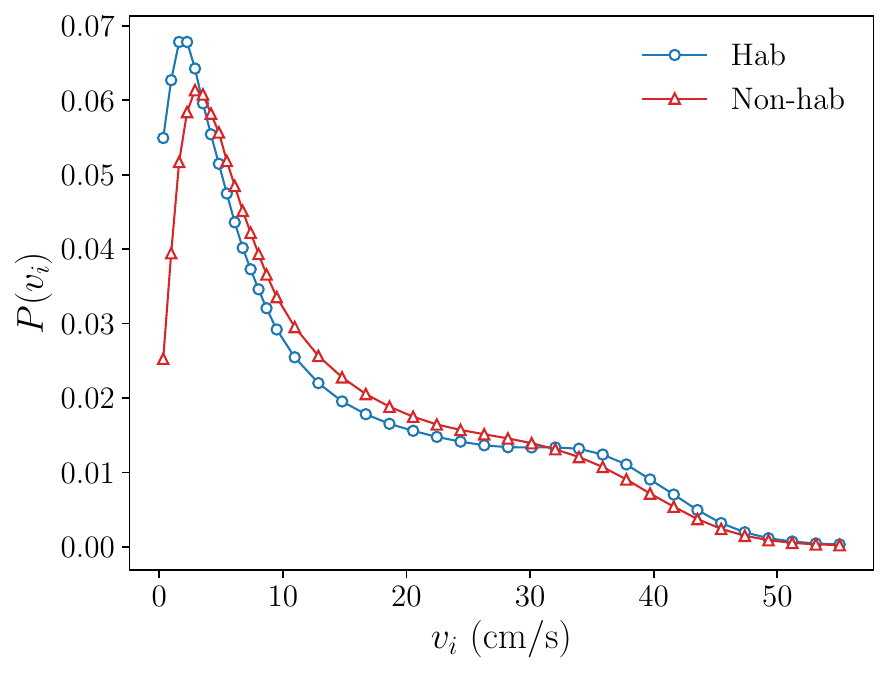}%
}

\subfloat[]{%
  \includegraphics[width=0.25\textwidth]{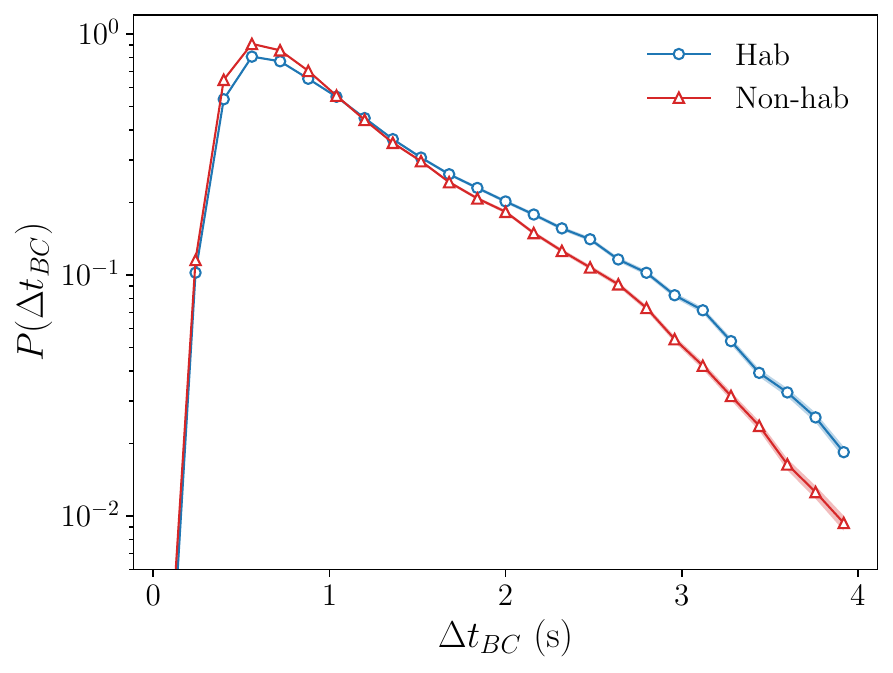}%
}
\subfloat[]{%
  \includegraphics[width=0.25\textwidth]{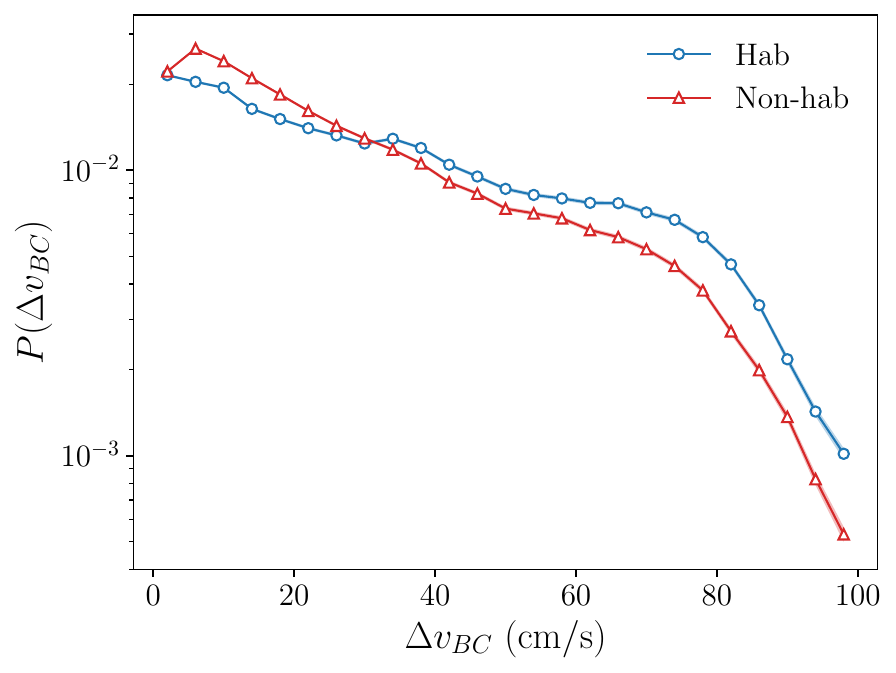}%
}
  \caption{\textbf{Burst-and-coast dynamics.} (a) Temporal evolution of the
    speed for a single habituated and non-habituated individual for a time
    interval of 10~s in series A. Local maxima and minima are marked with
    diamonds and circles, respectively. PDFs of (b) the individuals speeds
    $v_i$, (c) the time interval $\Delta t_{BC}$ and (d) the speed variability
    $\Delta v_{BC}$ between consecutive minima in the burst-and-coast
    oscillations of speeds.}\label{fig:self-propulsion}
\end{figure}

Apart from social interactions, we can examine individual behavior by studying
differences in self-propulsion forces between the two subgroups. Many fish
species exhibit oscillations in speed due to a burst-and-coast
mechanism~\cite{weihsEnergeticAdvantagesBurst1974,
  harpazDiscreteModesSocial2017, liBurstandcoastSwimmersOptimize2021}
(Fig.~\ref{fig:self-propulsion}a), where increased speed is associated with an
active phase powered by the fish muscles and speed decreases with a passive
gliding phase. Fish with different levels of experience may adapt their
burst-and-coast mechanism accordingly, as has been shown when comparing fish
from high versus low predation environments when tested in a collective decision
making assay~\cite{herbert-readCollectiveDecisionmakingAppears2019}.

We observe that both subgroups exhibit the same qualitative bimodal behavior in the PDF of individual speeds $v_i$ (Fig.~\ref{fig:self-propulsion}b), with a large narrow peak at small speeds ($v\sim3$~cm/s) corresponding to the minimum of the burst-and-coast dynamics, and a smaller, broader peak or shoulder at large speeds ($v\sim35$~cm/s) corresponding to the maximum of the burst-and-coast dynamics (Fig.~\ref{fig:self-propulsion}a). The large speed shoulder is broader because the maximum of the burst-and-coast dynamics exhibits more variability. However, non-habituated fish exhibit their two peaks at intermediate speeds compared to habituated individuals, indicating a constrained burst-and-coast movement, with a higher minimum speed and a lower maximum speed (KS test: $D=0.05$, $p<0.001$). This is consistent across different series (Supplementary figs.~\ref{supp:fig:self_propulsion_series}a-c). Additionally, we observe that speeds decrease over time (Supplementary fig.~\ref{supp:fig:self_propulsion_tEvo}a).

To quantify the burst-and-coast dynamics in more detail, we consider the time interval $\Delta t_{BC}$ and the speed variability $\Delta v_{BC}$ between consecutive minima. We define speed variability as  $\Delta v_{BC} \equiv \Delta v_\textrm{up} + \Delta v_\textrm{down}$, where $\Delta v_\textrm{up}$ and $\Delta v_\textrm{down}$ are the absolute value of the speed increments during the increase and decrease phases of the burst-and-coast dynamics, respectively. We find that non-habituated individuals, compared to habituated individuals, exhibit shorter burst-and-coast dynamics, characterized by shorter time intervals $\Delta t_{BC}$ and lower speed variabilities $\Delta v_{BC}$ (Figs.~\ref{fig:self-propulsion}c and d and Supplementary figs.~\ref{supp:fig:self_propulsion_series}d-i) (KS test: $D=0.06$ for $\Delta t_{BC}$, $D=0.10$ for $\Delta v_{BC}$; both $p<0.001$).
Over time, individuals tend to stretch their burst-and-coast motion, resulting in longer time intervals and greater speed variabilities (Supplementary figs.~\ref{supp:fig:self_propulsion_tEvo}b and c).

\subsection*{Machine learning identification tool}

In the previous sections, we identified distinct aggregated behaviors and interactions among individuals from subgroups that vary in their habituation to the experimental setup. To further confirm these differences and quantify the most prominent ones, we develop a machine learning tool to classify individuals by their subgroup identity. Our goal is to maximize the accuracy of predicting the correct subgroup identity of individuals based on observables recorded in two scenarios: at a given instant of time and over a temporal window.

In machine learning, the task of classifying data into categories belongs to the class of supervised learning methods. It consists of building a model from a dataset to predict an output label based on input variables or features. The model is initially fitted on a training dataset, and its performance is then evaluated using a separate test dataset. To ensure robustness and prevent information leakage arising from temporally correlated trajectories, we employ a leave-one-series-out cross-validation scheme. In each fold, the model is trained on two independent experimental series and evaluated on the remaining unseen series. Performance metrics are then averaged across folds. Among the many classification algorithms available, we choose the regularized gradient boosting implemented by the XGBoost library~\cite{chenXGBoostScalableTree2016} due to its strong predictive performance, as evidenced by its success in numerous machine learning competitions.
We also tested a logistic regression algorithm, but XGBoost yielded superior results.

For the first scenario, we take a snapshot of the individuals at a given instant and attempt to predict their subgroup identities. We include variables analyzed in previous sections as features (see Methods). Additionally, we reduce temporal correlations in the dataset by using samples taken every 100 frames (corresponding to 2-second intervals; see~\cite{macgregorShoalingBehaviourResponse2023}). We have checked that this sampling time is sufficient to ensure the vanishing of temporal self-correlations in  most variables considered. The performance of the instantaneous machine learning tool is shown in the confusion matrix in Table~\ref{table:machine_learning}a. The confusion matrix displays the probability of correctly predicting the subgroup identity of an individual based on its true identity. We find that the instantaneous machine learning tool achieves only modest predictive performance, indicating that subgroup identity is difficult to infer from instantaneous behavioral observables alone. Additionally, XGBoost quantifies the importance of the features used in the fitted model. In particular, we analyzed the total number of times a feature is used to split the data across all trees normalized against all features. We find the different features used share comparable weights (Supplementary fig.~\ref{supp:fig:machineLearning_weights}a).

\begin{table}[b!]
\centering
\caption{\textbf{Machine Learning classification performance.}
Confusion matrices showing the classification accuracy for predicting fish subgroup identity (habituated vs. non-habituated).
Results are averaged across leave-one-series-out cross-validation folds, where each fold trains on two experimental series and tests on the remaining independent series.}
\label{table:machine_learning}

\noindent\textbf{a) Instantaneous}

\vspace{0.3em}
\begin{tabular}{|c|c|c|c|}
\hline
\multicolumn{2}{|c|}{} & \multicolumn{2}{c|}{Predicted} \\
\cline{3-4}
\multicolumn{2}{|c|}{} & Hab & Non-hab \\
\hline
\multirow{2}{*}{True} & Hab     & $0.523(7)$ & $0.477(7)$ \\
\cline{2-4}
                      & Non-hab & $0.475(8)$ & $0.525(8)$ \\
\hline
\end{tabular}

\vspace{2em}

\textbf{b) Temporal window}

\vspace{0.3em}

\begin{tabular}{|c|c|c|c|}
\hline
\multicolumn{2}{|c|}{} & \multicolumn{2}{c|}{Predicted} \\
\cline{3-4}
\multicolumn{2}{|c|}{} & Hab & Non-hab \\
\hline
\multirow{2}{*}{True} & Hab     & $0.71(3)$ & $0.29(3)$ \\
\cline{2-4}
                      & Non-hab & $0.23(6)$ & $0.77(6)$ \\
\hline
\end{tabular}
\end{table}

In our second scenario, we observe individuals over a 10-minute time window before predicting their subgroup identities. To aggregate results across the window, we bin variables into histograms. Because this approach aggregates behavior over non-overlapping 10-minute windows, it reduces the influence of short-timescale temporal correlations. We use the bin values as features for the machine learning model, as well as measures of maximum leader-follower correlations (see Methods). This machine learning identification tool achieves a balanced classification accuracy of approximately 75\%, demonstrating that behavioral differences between habituated and non-habituated individuals are sufficiently consistent to generalize across independent experimental series (Table~\ref{table:machine_learning}b). We also rank features by their weight as provided by XGBoost (Supplementary figs.~\ref{supp:fig:machineLearning_weights}b and \ref{supp:fig:machineLearning_plots_tWindow}), which indicate that features associated with histogram maxima, as well as measures of maximum leader-follower correlations, tend to have greater importance.

\section*{Discussion}

Our comprehensive, large-scale dataset, combined with techniques to extract effective interactions, enabled a quantitative analysis of how differences in habituation to the experimental setup within groups affect behavior. We provide evidence that fish adapt their behavior based on their level of habituation, with adaptations that also evolve temporally within experimental trials and show trends consistent with previous work \cite{macgregorCollectiveMotionDiminishes2021}. These adaptations manifest at both group and local scales in spatial position and velocity coordination, leader–follower relationships, and individual self‑propulsion forces. Given that our analyses rely on large trajectory datasets with temporally correlated observations, small $p$-values can arise even for modest distributional differences. We therefore base our conclusions primarily on the magnitude and consistency of the observed effects across multiple analyses and independent experimental replicates, rather than on statistical significance alone. Although interactions among individuals evolve over time, the behavioral differences between habituated and non‑habituated subgroups remained consistent throughout the two‑hour experiment. Finally, we validated and quantified these interaction patterns by developing a machine‑learning classifier that reliably identifies individuals’ habituation subgroup.

By combining analyses of the spatial positioning of habituated and non‑habituated fish with interaction quantification, we find that individuals with lower habituation to the setup tend to occupy more central positions within the group and remain closer to their neighbors. Despite the enduring debate over the movement rules underlying the selfish herd hypothesis~\cite{hamiltonGeometrySelfishHerd1971,viscidoDilemmaSelfishHerd2002, morrellSpatialPositioningSelfish2011}, and the historical separation between selfish‑herd research and studies of interaction rules in collective behavior~\cite{herbert-readUnderstandingHowAnimal2016}, evidence for the selfish herd has accumulated in several taxa~\cite{krauseEffectSchreckstoffShoaling1993, devosSharksShapeGeometry2010, kingSelfishherdBehaviourSheep2012a}, even though recent work has questioned its generality~\cite{sankeyAbsenceSelfishHerd2021, hammerDisentanglingManyeyesDilution2023}. 
Our findings are consistent with selfish-herd-like spatial organization in schooling fish and, more importantly, identify candidate interaction rules through which such organization may emerge. Compared with habituated individuals, non-habituated fish exhibit reduced short-range repulsion and correspondingly tighter local spacing, while their effective attraction toward the group center remains comparable. This contrast suggests that central positioning arises primarily as an emergent consequence of local-scale interactions among neighbors, rather than from a group-level behavior in which individuals explicitly move toward the center, although we cannot fully exclude a contribution from such effects. Ultimately, selfish-herd-like behavior is likely modulated by system-specific factors, including environmental context and the nature of predation risk~\cite{parrishReexaminingSelfishHerd1989,jollesBothPreyPredator2022}.

The maintenance of central positions is likely facilitated by non‑habituated individuals being more ordered and aligned with both the group and their neighbors, and by exhibiting reduced speed variability with smaller, shorter‑period oscillations compared to habituated individuals. These traits likely make non‑habituated fish more prone to follow others: they occupy more rearward positions within the group and move at lower relative speeds with respect to neighbors. Using time delays in peak velocity correlations with neighbors~\cite{nagyHierarchicalGroupDynamics2010, katzInferringStructureDynamics2011}, we confirmed that habituated individuals tend to lead non‑habituated ones, while non‑habituated individuals transmit orientation and speed information more effectively. This pattern indicates that less‑habituated individuals rely more on social information, consistent with studies of social learning~\cite{websterSocialLearningStrategies2008}. We note that this rearward positioning also contributes, at least in part, to the tendency of non-habituated individuals to orient toward the group center; as such, this observation does not necessarily imply an explicit behavioral tendency to move toward central positions.

Differences in self-propulsion further suggest alternative or complementary interpretations of the observed spatial patterns. In particular, variations in individual speed and speed variability are known to modulate effective alignment and attraction and to drive spatial sorting in collective systems~\cite{kentSpeedmediatedPropertiesSchooling2019,jollesGrouplevelPatternsEmerge2020,klamserImpactVariableSpeed2021}. More generally, lower-amplitude, higher-frequency oscillatory dynamics can generate segregation in active and inertial particle systems~\cite{rosato-brazilnuts1987}, suggesting that differences in swimming amplitude and burst-and-coast frequency may also contribute to assortment within fish shoals. However, such mechanisms alone would be expected to promote primarily front-to-rear segregation rather than the robust central positioning observed here, and are therefore unlikely to fully account for our results. Instead, self-propulsion dynamics, coordination, and effective interactions are likely intrinsically coupled, with the observed spatial organization emerging from their combined influence.

An equivalent description of the observed collective patterns is that habituated individuals are comparatively more exploratory, less prone to align, less likely to follow, and less tolerant of close neighbors. We do not view these interpretations as competing explanations, but rather as complementary descriptions of the same underlying interaction-level differences, since our analysis characterizes how individuals respond to one another rather than the cognitive strategies driving those responses. More broadly, similar spatial organization can arise from different combinations of interaction rules, kinematics, and coordination dynamics, making it difficult to uniquely infer underlying mechanisms from collective patterns alone.

Our study was designed to manipulate the level of habituation to the experimental setup by repeatedly exposing one subgroup to the test conditions while keeping those conditions novel for the other subgroup. This novelty meant that non‑habituated fish had greater uncertainty about the environment than habituated fish, whose repeated testing provided more certain information that the setup was low risk. Habituation is believed to reduce fear‑related behavior and physiological stress responses to novel environments~\cite{millerSchoolingShoalingPatterns2012} and to repeated stressors~\cite{houslayHabituationIndividualVariation2019}. To rule out alternative explanations such as food expectation, fish were fed to satiation~2 h before trials and no food was introduced into the arena. By contrast, previous studies manipulated perceived risk with methods more likely to induce acute stress, including predator exposure~\cite{steinBehavioralResponseCrayfish1976, papadopoulouEmergenceSplitsCollective2022} or alarm substances~\cite{krauseEffectSchreckstoffShoaling1993, sosnaIndividualCollectiveEncoding2019}. Our habituation‑based method is less invasive, minimizes ethical concerns associated with inducing acute stress, and is easier to implement and thus more accessible to researchers.

At the same time, some limitations should be acknowledged. Because our habituation protocol necessarily combined multiple factors --- repeated exposure to the test tank, repeated handling and transfer, and separate housing of the two subgroups --- we cannot fully disentangle their individual contributions. We therefore interpret our results as reflecting differences in behavioral state induced by the treatment as a whole, rather than by tank habituation in isolation. In particular, although all individuals within a replicate were already familiar with one another prior to the habituation phase, their separation during the treatment means that residual differences in within-subgroup familiarity cannot be entirely excluded. Such familiarity effects are known to influence aggregation and social behavior in fish~\cite{jacobyEffectFamiliarityAggregation2012}. However, we consider this factor unlikely to explain our results: nearest-neighbor distances depend primarily on the focal individual’s subgroup identity, and not on that of the neighbor, which contrasts with expectations from preferential association among familiar individuals. Future experiments could help disentangle these mechanisms more conclusively. In particular, this could involve mixing subgroups with identical habituation levels, independently manipulating swimming dynamics, perceived risk, and familiarity, or introducing  a third subgroup explicitly conditioned to expect high risk (e.g., via predator cues) to compare the effects of increased certainty of danger with those of low-certainty novelty.

In summary, we show that differences in habituation to the experimental setup drive spatial positioning and social interactions, providing mechanistic insight into how behavioral heterogeneity can generate selfish-herd-like spatial organization. More broadly, quantifying effective interactions and framing results with simple, testable mechanistic arguments can resolve long‑standing debates about the movement rules underlying collective structure and offer unifying explanations across taxa. These approaches also create clear design principles for synthetic systems (e.g., robotic swarms and bioinspired collectives), enabling controlled tests of interaction rules and practical applications in engineered collective behavior.

\section*{Methods}

\subsection*{Subjects}\label{sec:methods:subjects}

We employ schooling fish of the species black neon tetra (\textit{Hyphessobrycon
  herbertaxelrodi}) wild-type, a small freshwater fish with a strong tendency to
form cohesive, highly polarized, and planar
schools~\cite{gimenoDifferencesShoalingBehavior2016}. Young adult fish (8-12
months old) were obtained from a local vendor (Bichosfera, Olvan, Barcelona,
Spain). Initially, 75 fish were kept in three separate main aquaria measuring
$40 \times 43 \times 30$ cm, with a sex ratio close to 1:1. From each main
aquarium, 16 fish were randomly selected, with 8 assigned to a \emph{habituated
  group} and the remaining 8 to a \emph{non-habituated group}. This resulted in
3 habituated groups and 3 non-habituated groups, comprising a total of 48
individuals (mean body length: $2.9$ cm). This procedure was explicitly followed to avoid possible confounding effects from familiarity between habituated and non-habituated individuals. Groups were housed in separate aquaria
of identical dimensions to the main aquaria. Water conditions were maintained at
$25\pm2 ^\circ$C, with a pH of 7.5–7.7, total hardness of 7$^\circ$-10$^\circ$
dH, and carbonate hardness of 10$^\circ$-12$^\circ$ dH, under a regular
light/dark cycle. Nitrate levels were kept below 50 mg/L, while ammonia and
nitrite concentrations were maintained at zero. Optimal water quality was
ensured through weekly 15\% water changes. Fish were fed to satiation with Ocean
Nutrition Community Formula Flakes (Ocean Nutrition Europe,
Belgium). Institutional guidelines for the care and use of animals were
followed, and all procedures were conducted in accordance with the standards
approved by the Ethics Committee of the University of Barcelona (Project
No. 343/23).

\subsection*{Apparatus}

A square experimental tank measuring $100 \times 100 \times 40$~cm and a water
column with a height of 7 cm were used, allowing for approximately
two-dimensional movement. Water in the experimental tank was maintained at
24–26$^\circ$C, with the same pH and dH levels as in the aquaria. The tank was
surrounded by translucent curtains made of white polyester that acted as a light
diffuser and was directly lit by six lamps, providing a homogeneous light
intensity. Videos were recorded at 50 fps and a resolution of $5312 \times 2988$
pixels with a digital camera (GoPro Hero 11 Black) mounted 1 m above the
tank. In the recordings, the side of the tank measured $2634$ pixels.

\subsection*{Procedure}

\subsubsection*{Habituation}

Only habituated groups participated in this phase. Each group received one trial
per day over 5 days. Habituation trials consisted of fish swimming freely in the
experimental tank for 120 minutes. To ensure fish became acclimatized to the
water in the experimental tank, prior to the habituation trial, each group was
transferred from its aquarium into a bucket containing a mixture of water from
the tank and the fishes’ aquarium and remained there for 5 min. After the trial
was completed, fish were gently captured with a net and transferred back to
their aquarium, to which water from the experimental tank was added to
facilitate reacclimatization. Water was then added to the experimental tank to
compensate for the water transferred into the aquarium, and heaters and water
filters were reintroduced for one hour to control for possible cumulative
effects of the decrease in water temperature and any pheromones or chemical
signals left in the water by the preceding group.

\subsubsection*{Test trial}

On the sixth day, habituated groups were merged with the non-habituated groups
of fish that came from the same main aquarium. Before starting each trial, the
habituated group and the non-habituated group were transferred into different
plastic containers of $20 \times 14 \times 14$~cm each. One plastic container
was placed in the upper right corner of the experimental tank and the other
container in the lower right corner, the positioning of the habituated and
non-habituated containers was counterbalanced between the trials. The test trial
began by releasing the groups from their corresponding container and allowing
the fish to swim freely in the experimental tank for 2 hours while being
recorded. 
The same procedure indicated above was repeated to acclimatize the
fish before and after introducing them into the experimental tank, as well as
filtering and heating the water in the experimental tank after each trial. We
performed three independent replicates, designated series A, B, and C.

\subsection*{Data extraction and preprocessing}

Digitized individual trajectories were obtained from the video recordings using
the software \textsc{idtracker.ai}~\cite{romero-ferreroIdtrackerAiTracking2019,
  torrentsNewIdtrackeraiRethinking2025}, which can reliably track up to 100
individuals with 99.9\% accuracy. 
Unlike tracking methods that link detections frame-to-frame, idtracker.ai identifies each individual based on its visual appearance across the entire video, preventing the accumulation of identity swaps due to transient crossings and occlusions. Video recording began immediately upon release, enabling continuous tracking from an initial configuration in which subgroup membership was known. Subgroup labels were assigned based on these initial positions and then propagated throughout the trajectories. To avoid potential bias arising from the initial spatial separation, the frames used for subgroup identification were excluded from the analysis.
We analyzed three datasets (series A, B, and
C), each lasting 120 minutes and comprising 357,001 frames. The software
reported high identity–assignment probabilities of $99.39\%$, $100\%$, and
$100\%$ for series A, B, and C, respectively, covering $98.5\%$, $98.8\%$, and
$99.0\%$ of frames. Frames with lower confidence, typically due to occlusions,
were first processed through idtracker.ai’s automatic interpolation and then
manually verified and corrected using its Validator tool. This workflow ensured that no  identity swaps remained in the final trajectories analyzed.

For better accuracy, we projected the trajectories in the plane of the fish
movement, warping the tank walls of the image into a proper square (for more
details see~\cite{puySelectiveSocialInteractions2024}). Moreover, we smoothed the
trajectories with a Gaussian filter with $\sigma = 2$, truncating the filter at
$5\sigma$. Individual velocities and accelerations were obtained from the
Gaussian filter using first and second derivatives of the Gaussian kernel,
respectively.

To mitigate spurious effects in averaging results, we discarded frames where any
fish in the group remained stationary for prolonged periods. Specifically, we
excluded frames where any individual moved at a speed lower than 0.95 cm/s for
more than 5 seconds. This led to the removal of 0.6\%, 1.0\%, and 36.8\% of the
frames in series A, B, and C, respectively. The substantially larger fraction of excluded frames in series C resulted from repeated episodes in which one individual remained nearly immobile for extended periods, occasionally disrupting collective motion. Analyses restricted to the remaining frames reveal qualitative patterns consistent with those observed in the other replicates (see Supplementary Results).

\subsection*{Force map method}\label{sec:methods:force_map_method}

The force map method assumes a system with several force terms depending on sets
of independent variables, and that the average in time of each force term is
zero. Consider we want to infer a force term $\vec{f}(x)$ as a function of only
the variable $x$. We can achieve this from Newton's second law by studying the
acceleration depending on $x$ while averaging over all other variables in the
system, $\vec{a}(x)$~\cite{katzInferringStructureDynamics2011,
  herbert-readInferringRulesInteraction2011,
  puySelectiveSocialInteractions2024}. Working in units of mass $m = 1$,
Newton's second law will become $\vec{f} (x) \simeq \vec{a} (x)$, as all other
forces will cancel out in the average. In practice, this approach is implemented
by binning over the whole dataset the relevant variable $x$ as
$\{x_k\}_{k=1, \ldots, M}$, where $k$ is the index of the bin, and averaging the
values of the acceleration on that bin,
$\vec{f}(x_k) \simeq \left< \vec{a} (x)\right>_{x \in x_k}$. See
Ref.~\cite{puySelectiveSocialInteractions2024} for more information.

Force-map approaches provide coarse-grained, effective descriptions of behavioral responses, but they also have known limitations~\cite{escobedoDatadrivenMethodReconstructing2020,herasDeepAttentionNetworks2019,mudaliarExaminationAveragingMethod2020,puySelectiveSocialInteractions2024}. Because such maps are typically visualized in at most two dimensions, they require examining lower-dimensional projections that may be poorly sampled or averaging over additional variables; they do not generally yield simple analytical expressions reproducing the observed patterns;  distinct interactions can appear mixed within a single map, making them difficult to identify and disentangle; and different underlying interaction rules can produce similar effective force patterns. We therefore interpret the inferred force maps as effective interaction responses, rather than as a unique or direct decomposition into independent physical forces.

\subsection*{Machine learning tool
  specifics}\label{sec:methods:machineLearning_details}

We employ the function \texttt{XGBClassifier} from the \textsc{XGBoost} python
library~\cite{chenXGBoostScalableTree2016}, with $1000$ number of boosting
rounds (\texttt{n\_estimators}), a $10$ maximum tree depth for base learners
(\texttt{max\_depth}), a $0.5$ boosting learning rate (\texttt{learning\_rate})
and a $0.25$ subsample ratio of the training instance (\texttt{subsample}). All
other parameters are left on their default values.

To mitigate overfitting, we verified that each feature used independently
improves the prediction accuracy and discarded other features that did not
contribute to the improvement.

For the instantaneous machine learning scenario, we employ as features: the
normalized relative position within the group $(\vec{r}_i - \vec{r}_{CM})/R_g$,
the average normalized distance to other individuals
$\left< \left| \vec{r}_i - \vec{r}_{j}\right|/R_g \right>$, the relative
velocity with the group $\vec{v}_i - \vec{v}_{CM}$, the acceleration
$\vec{a}_i$, the speed $v_i$, the group velocity $\vec{v}_{CM}$, and the rate of
change of the acceleration $\delta \vec{a}_i \equiv \frac{d\vec{a}_i}{dt}$ (to
test for possible dependencies in higher derivatives). Each component of these
vectors is treated as a separate feature. The $y$-coordinate points in the
direction of motion of the group and the individual for group and local
measures, respectively.

For the 10-minute time window machine learning scenario, we use histograms for
the normalized relative distance to the center of mass
$\left| \vec{r}_i - \vec{r}_{CM} \right|/R_g$, the average normalized distance
to other individuals $\left< \left| \vec{r}_i - \vec{r}_{j}\right|/R_g \right>$,
and the speed $v_i$. We use 20 bins in each histogram. Additionally, we include
leader-follower correlations in the orientation
$\left< \hat{v}_i (t) \cdot \hat{v}_{j} (t+\tau) \right>$ and in the speed
$r(v_i(t), v_{j}(t+\tau))$, averaged across other individuals $j$. We use as
features the values of these correlations for each discrete delay $\tau$, the
delay at which the maximum correlation occurs $\tau^\theta_i$ and $\tau^v_i$,
and the relative difference of the maximum correlation value with respect to the
group $\Delta c^\theta_i$ and $\Delta c^v_i$. This relative difference is
defined as $\Delta c_i^x \equiv c_i^x - \left< c_i^x \right>_i$, where $c_i^x$
is the value of the maximum, for $x=\theta$, $v$. For a visual representation of
the features, see Supplementary
fig.~\ref{supp:fig:machineLearning_plots_tWindow}.

\section*{Data availability}
The experimental datasets and figure data files can be accessed on Zenodo~\cite{puyDataHabituationDetermines2026}: \url{https://zenodo.org/records/17141691}.

\bibliography{all}

\section*{Acknowledgements}

We thank F. Mazzanti, E. Romero and D. Bierbach for helpful discussions.

\section*{Funding}

This work was supported by projects PID2022-137505NB-C21 (AEI/10.13039/501100011033/FEDER, UE) and PID2022-137505NB-C22 (MICIU/AEI/10.13039/501100011033), as well as by the European Regional Development Fund (ERDF), "A way of making Europe". R.P.-S. acknowledges additional support from the Acadèmia d'Excel·lència Award funded by the Generalitat de Catalunya.

\section*{Author contributions}

A.P., F.S.B., R.D., M.C.M., and R.P.-S. designed the study. E.G. acquired and
processed the empirical data. A.P. analyzed the empirical data. A.P. and
R.P.-S. developed the machine learning tool. A.P., M.C.M., C.C.I., and
R.P.-S. analyzed the results. A.P., M.C.M., C.C.I., and R.P.-S. wrote the
paper. All authors commented on the manuscript.

\section*{Competing interests}

The authors declare no competing interests.


\clearpage

\renewcommand{\thepage}{\arabic{page}} 
\renewcommand{\theequation}{S\arabic{equation}} 
\renewcommand{\thesection}{\arabic{section}}  
\renewcommand{\thetable}{S\arabic{table}}  
\renewcommand{\thefigure}{S\arabic{figure}}
\renewcommand{\thevideo}{S\arabic{video}}


\setcounter{page}{1}
\setcounter{section}{0}
\setcounter{figure}{0}
\setcounter{equation}{0}
\onecolumngrid
\vspace*{\fill}
\begin{center}
\textbf{\large SUPPLEMENTARY INFORMATION\\~\\}
\end{center}
\vspace*{\fill}

\twocolumngrid

\newpage

\onecolumngrid
\newpage
\vspace*{\fill}
\begin{center}
\begin{video*}[h]
\caption{Rendering of the movement of fish overlaid with the digitized trajectories. We show habituated and non-habituated individuals in blue and red, respectively.}
\end{video*}\end{center}
\vspace*{\fill}
\newpage

\onecolumngrid
\newpage
\vspace*{\fill}
\begin{center}
\section*{SUPPLEMENTARY FIGURES}
\end{center}
\vspace*{\fill}
\newpage

\begin{figure}
	\centering
	\subfloat[]{%
  \includegraphics[width=0.25\textwidth]{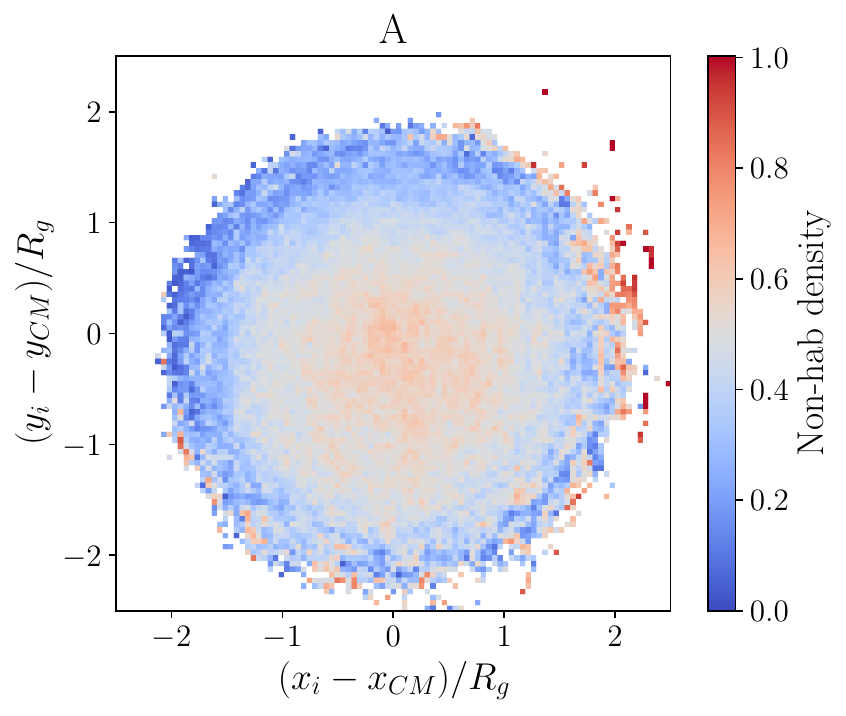}%
}
\subfloat[]{%
  \includegraphics[width=0.25\textwidth]{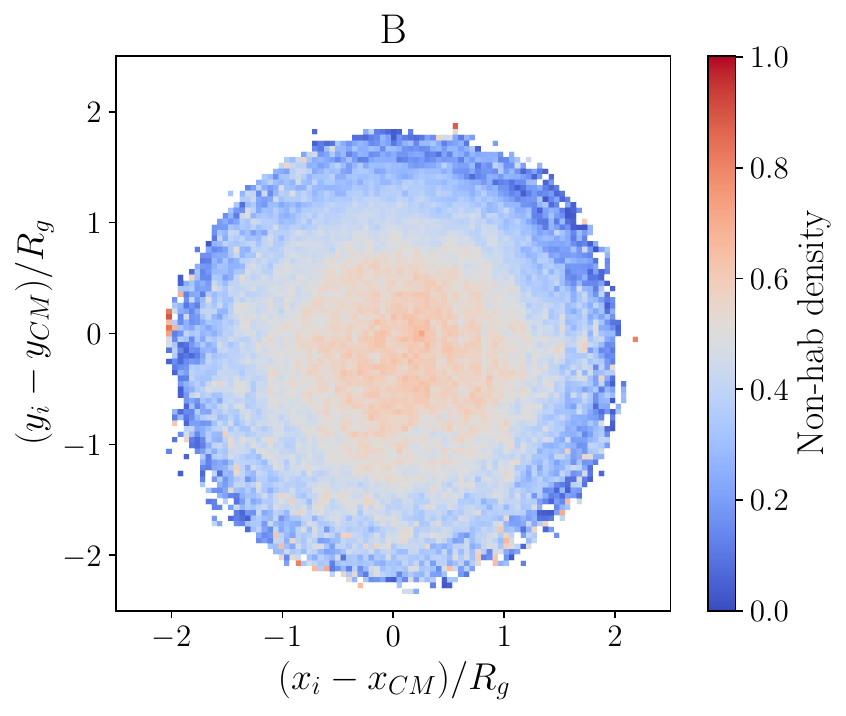}%
}
\subfloat[]{%
  \includegraphics[width=0.25\textwidth]{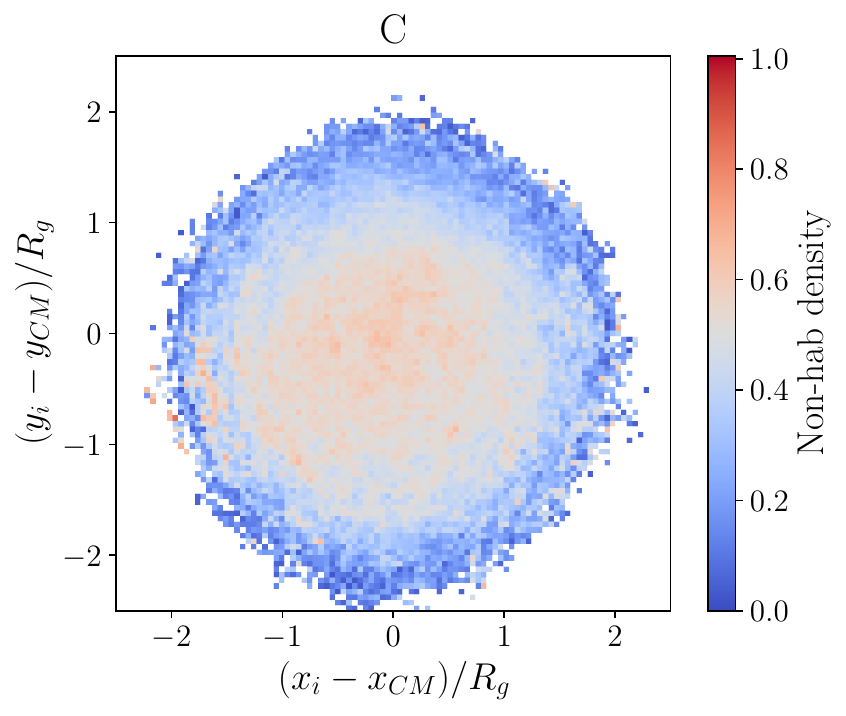}%
}

\subfloat[]{%
  \includegraphics[width=0.25\textwidth]{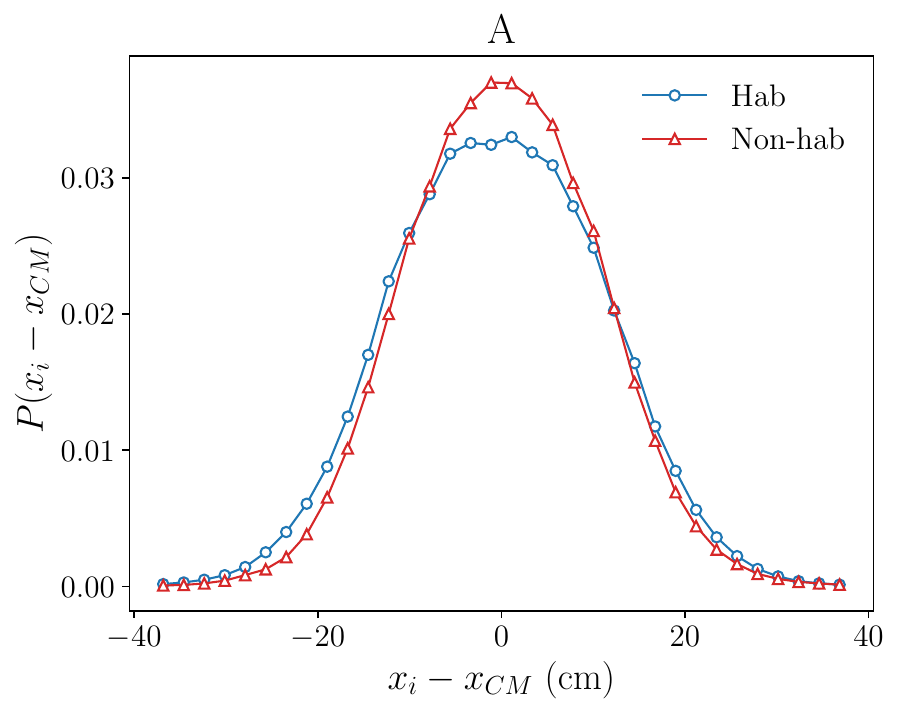}%
}
\subfloat[]{%
  \includegraphics[width=0.25\textwidth]{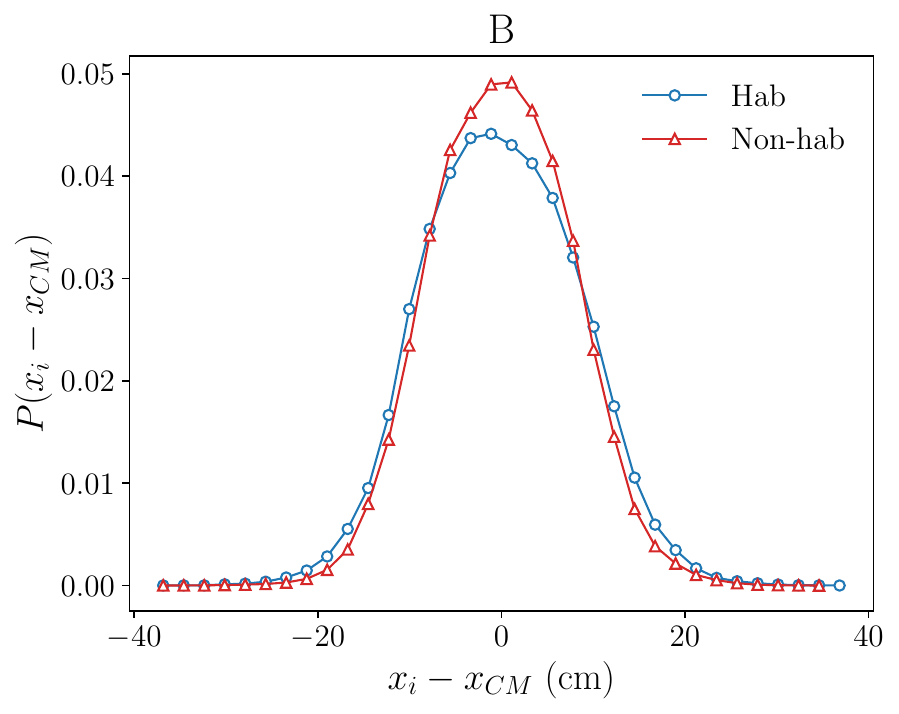}%
}
\subfloat[]{%
  \includegraphics[width=0.25\textwidth]{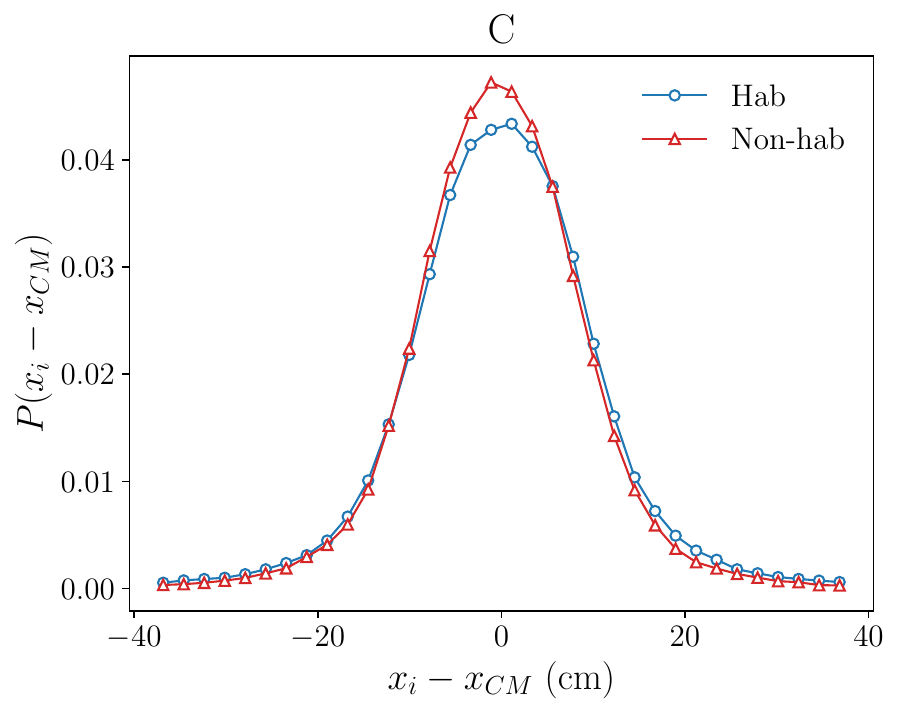}%
}

\subfloat[]{%
  \includegraphics[width=0.25\textwidth]{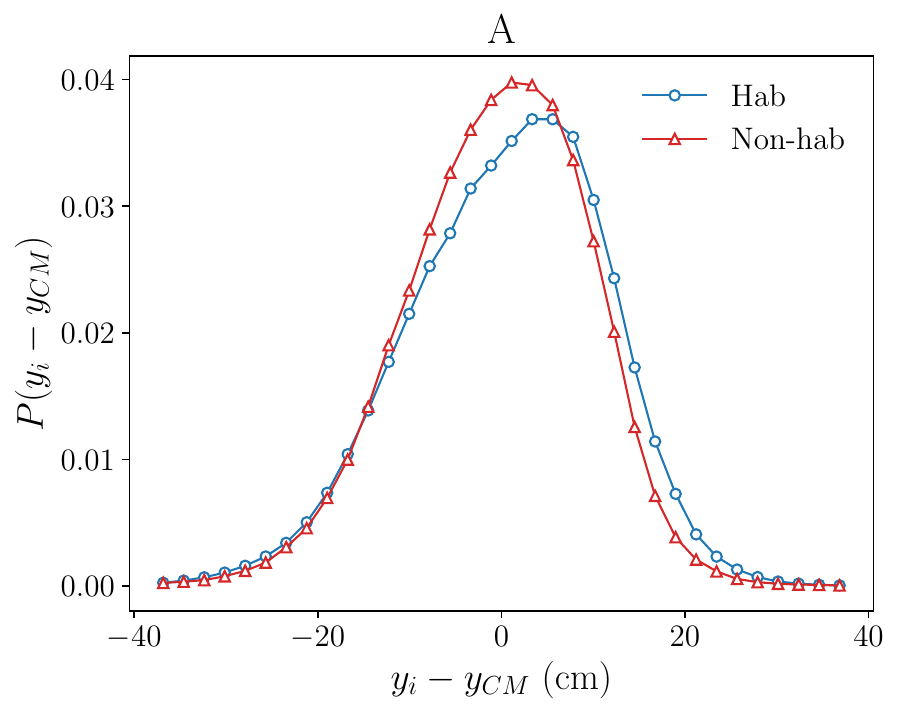}%
}
\subfloat[]{%
  \includegraphics[width=0.25\textwidth]{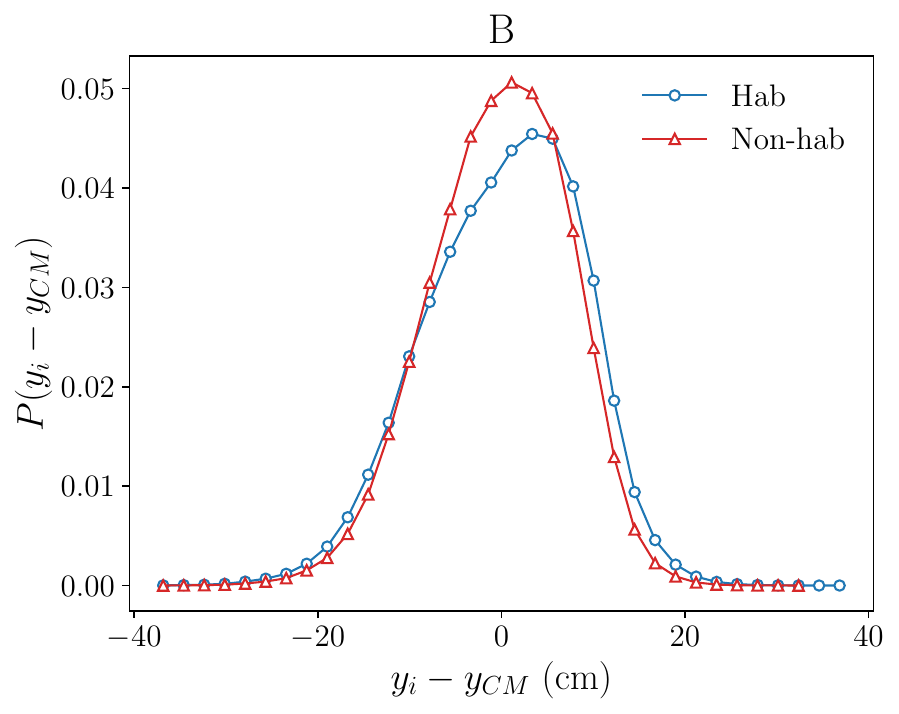}%
}
\subfloat[]{%
  \includegraphics[width=0.25\textwidth]{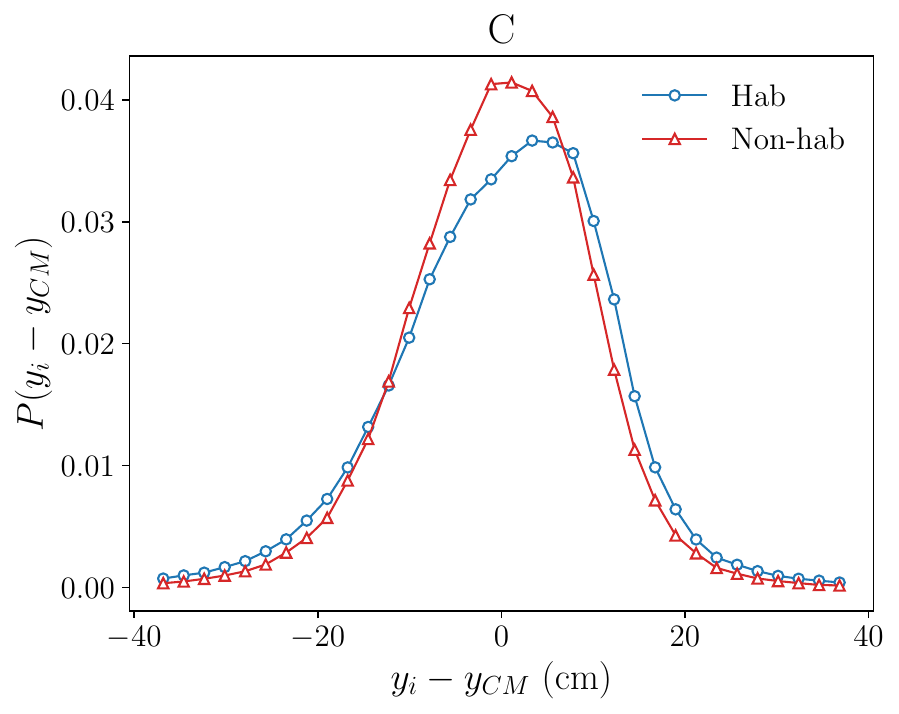}%
}

\subfloat[]{%
  \includegraphics[width=0.25\textwidth]{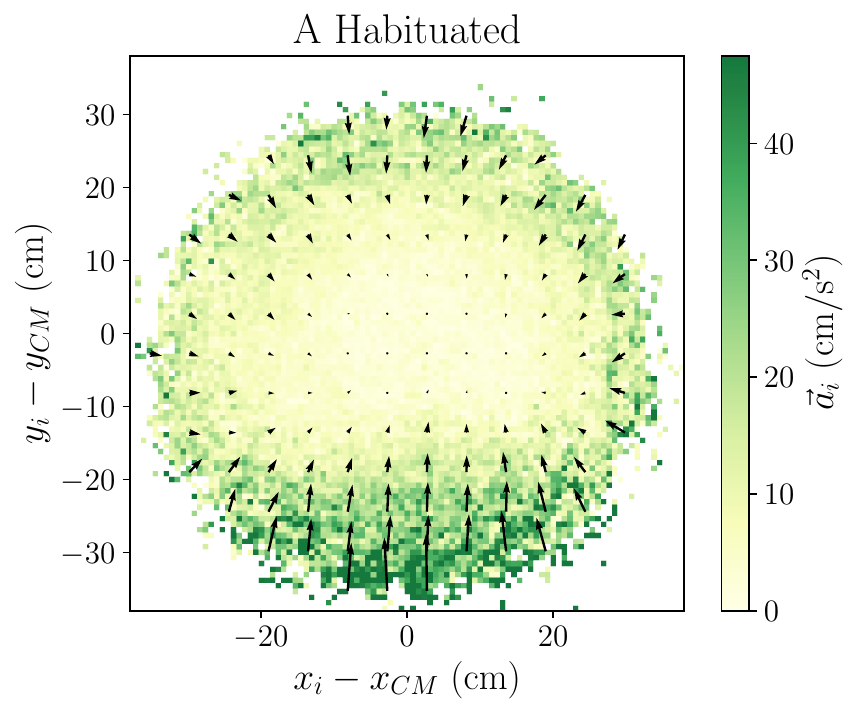}%
}
\subfloat[]{%
  \includegraphics[width=0.25\textwidth]{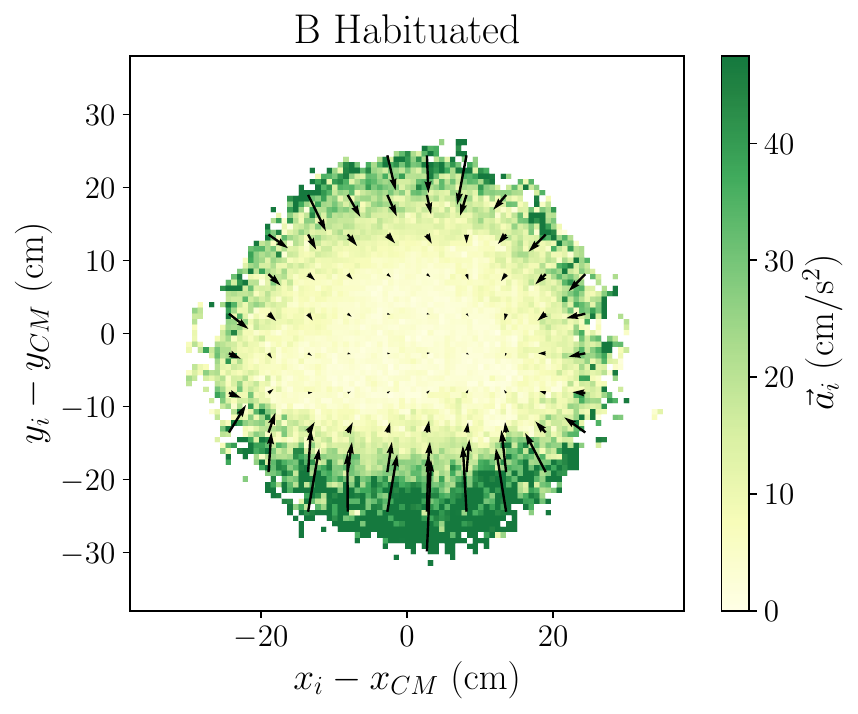}%
}
\subfloat[]{%
  \includegraphics[width=0.25\textwidth]{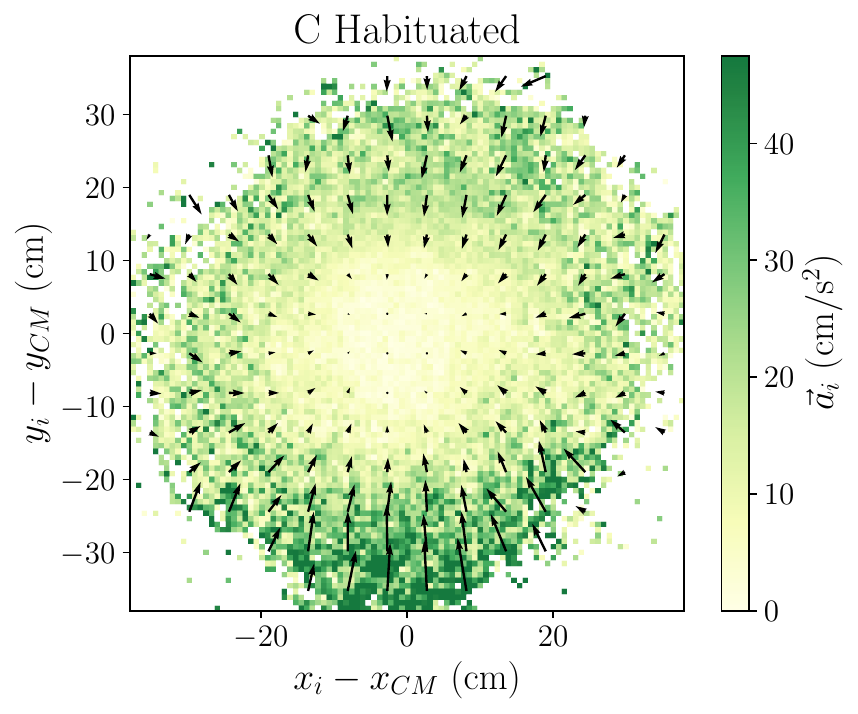}%
}

\subfloat[]{%
  \includegraphics[width=0.25\textwidth]{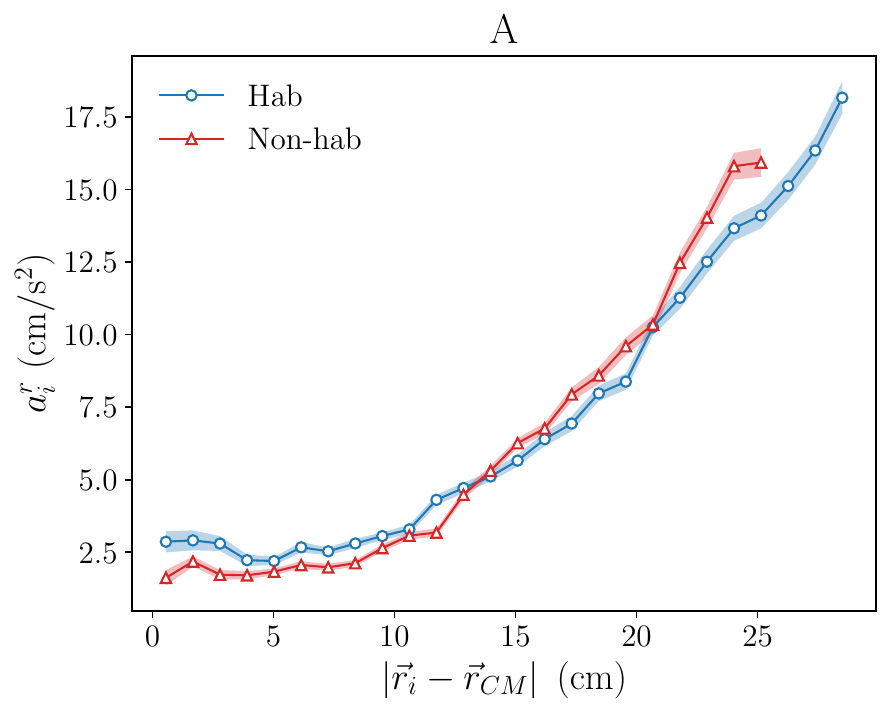}%
}
\subfloat[]{%
  \includegraphics[width=0.25\textwidth]{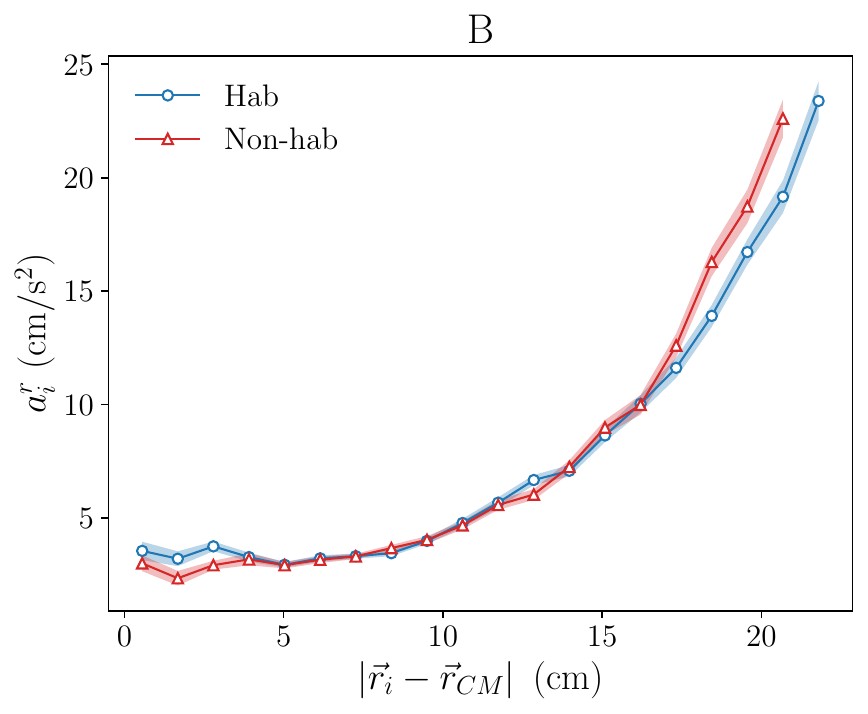}%
}
\subfloat[]{%
  \includegraphics[width=0.25\textwidth]{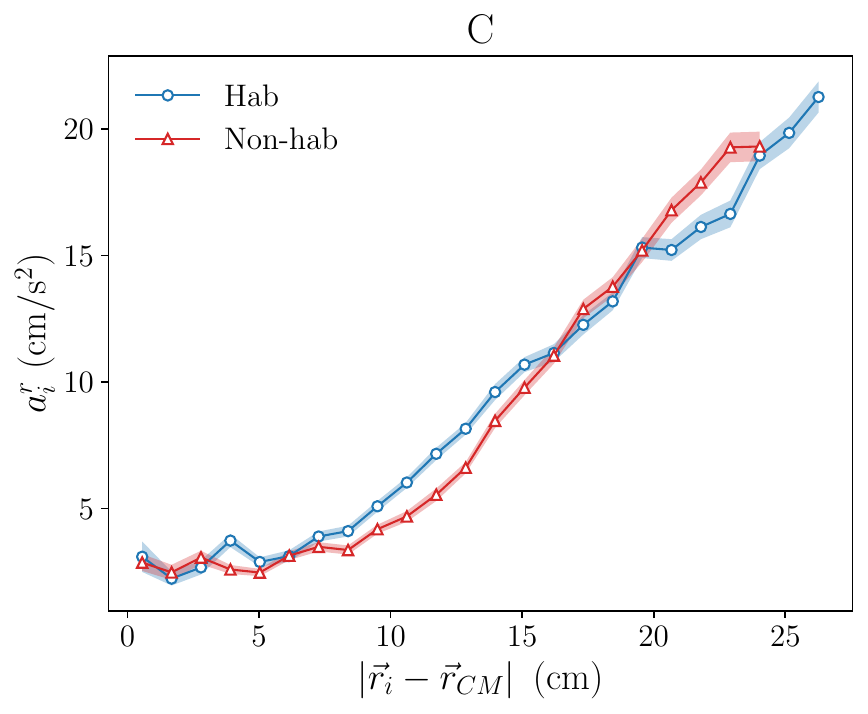}%
}
	\caption{(a)-(l) Group-level spatial interactions (see Fig.~\ref{fig:group_spatial}) for different series. (m)-(o) Average modulus of the acceleration in the radial direction of the force map, $a_i^r$, depending on the distance
    to the center-of-mass, $\left| \vec{r}_i - \vec{r}_{CM} \right|$. Their error bands are
    given by the standard deviation of the mean.}\label{supp:fig:group_spatial_series}
\end{figure}

\begin{figure}
	\centering
	\subfloat[]{%
  \includegraphics[width=0.25\textwidth]{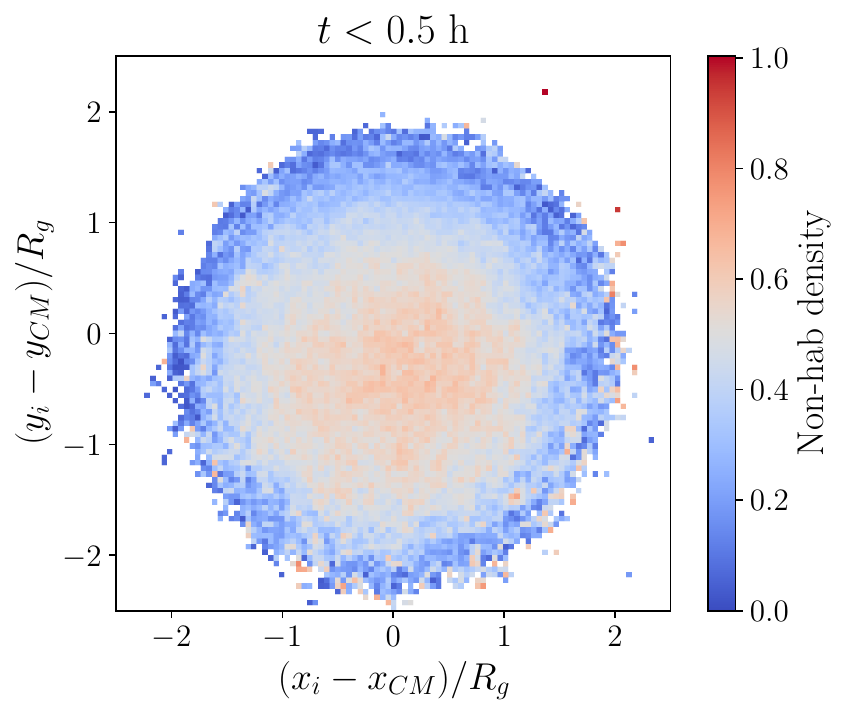}%
}
\subfloat[]{%
  \includegraphics[width=0.25\textwidth]{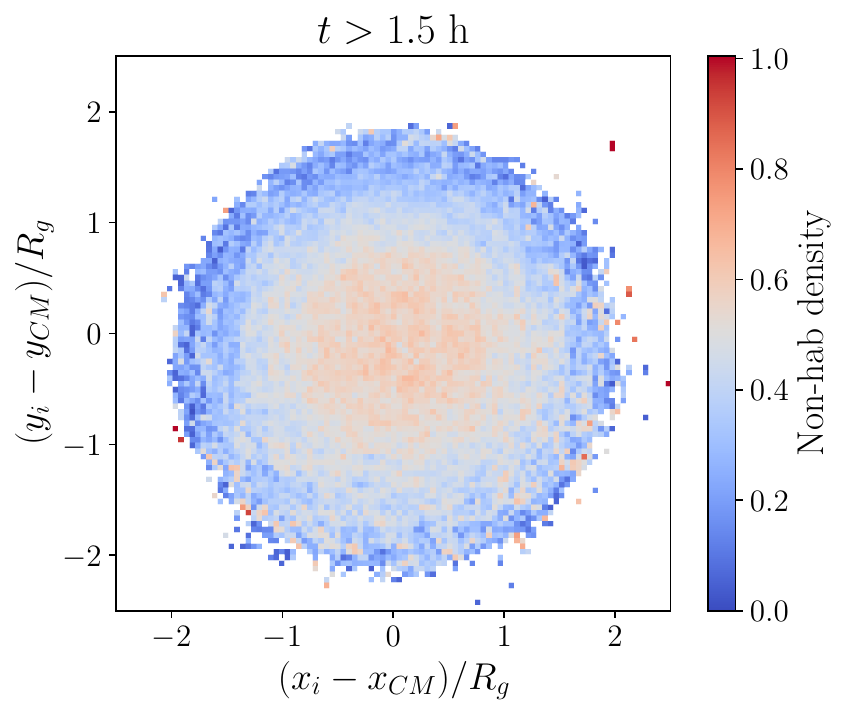}%
}

\subfloat[]{%
  \includegraphics[width=0.25\textwidth]{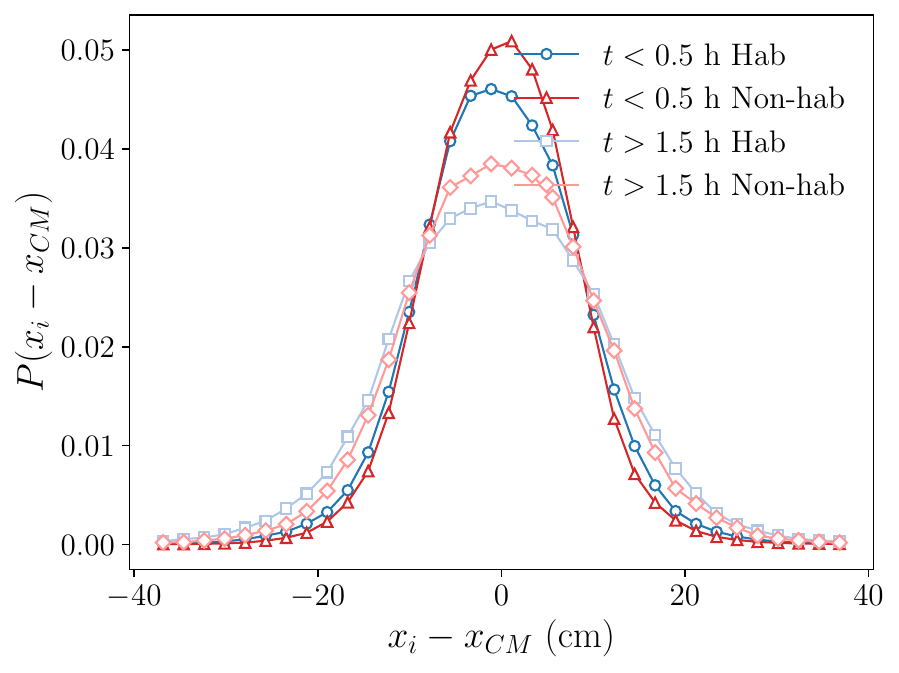}%
}
\subfloat[]{%
  \includegraphics[width=0.25\textwidth]{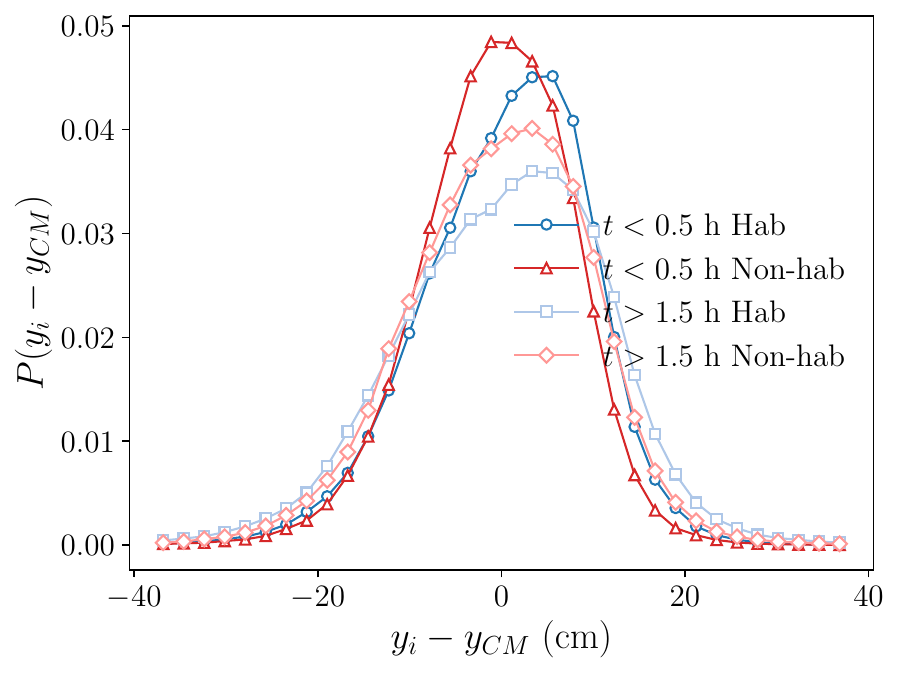}%
}
\subfloat[]{%
  \includegraphics[width=0.25\textwidth]{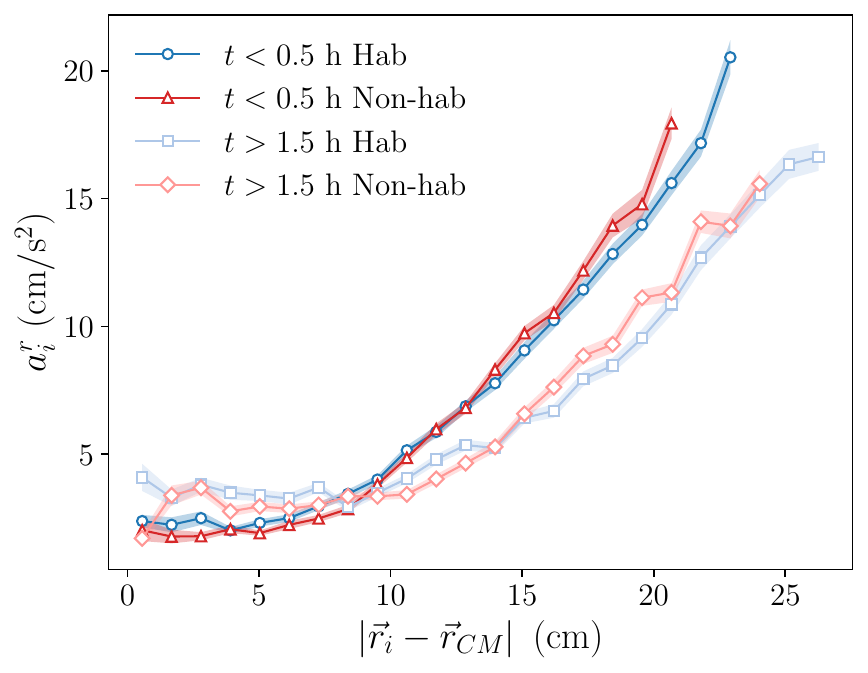}%
}
	\caption{(a)-(d) Temporal evolution of group-level spatial interactions (see Fig.~\ref{fig:group_spatial}). (e) Average modulus of the acceleration
    in the radial direction of the force map, $a_i^r$, depending on the distance
    to the center-of-mass, $\left| \vec{r}_i - \vec{r}_{CM} \right|$. Results are compared for the first half-hour and last half-hour of the experiment.}\label{supp:fig:group_spatial_tEvo}
\end{figure}

\begin{figure}
	\centering
	\subfloat[]{%
  \includegraphics[width=0.25\textwidth]{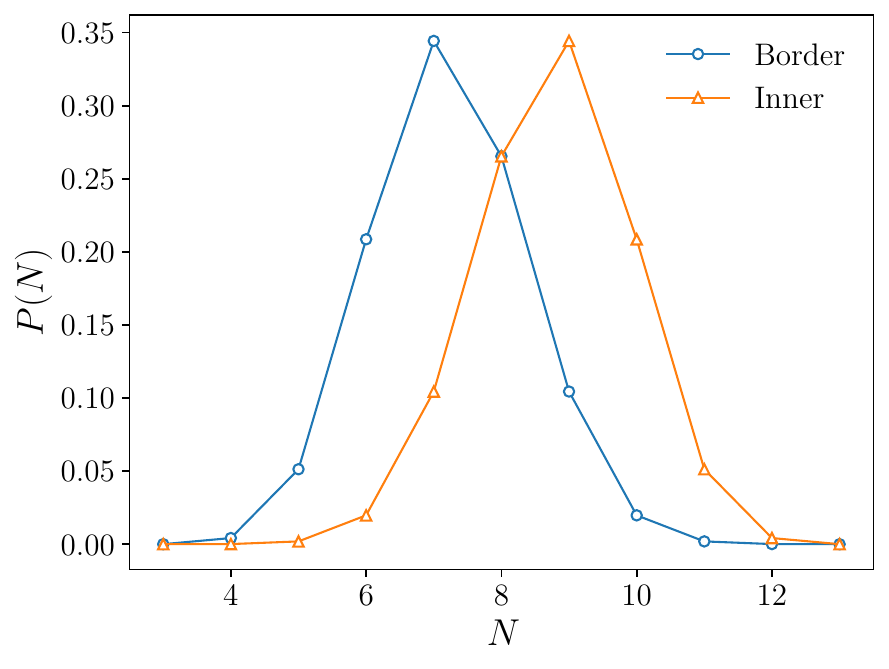}%
}
\subfloat[]{%
  \includegraphics[width=0.25\textwidth]{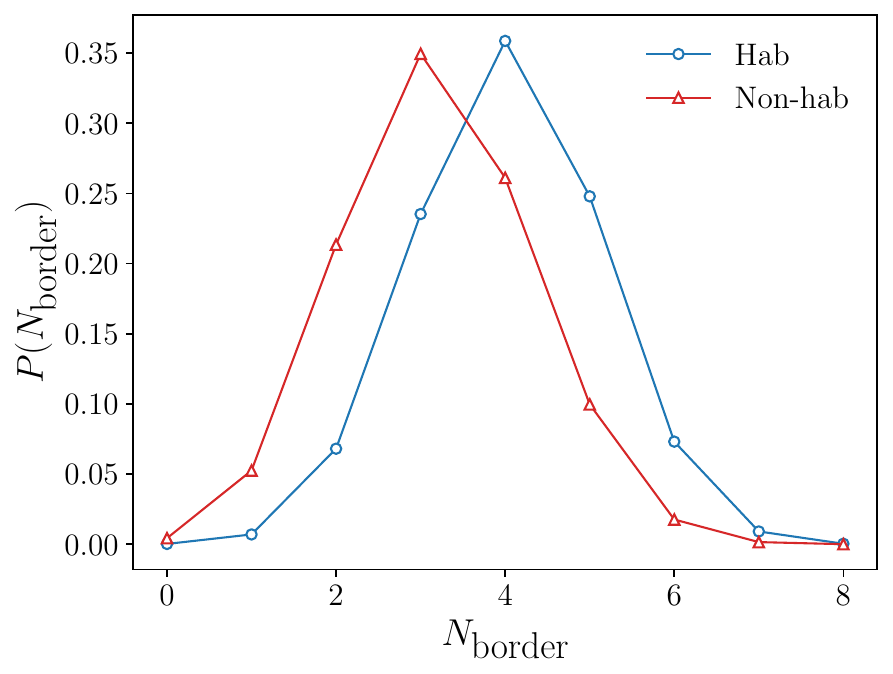}%
}
	\caption{PDF of (a) the number of individuals $N$ at the border and at the inner part of the group and (b) the number of individuals at the border $N_\textrm{border}$ for habituated and non-habituated individuals.}\label{supp:fig:border_definitions}
\end{figure}

\begin{figure}
	\centering
	\subfloat[]{%
  \includegraphics[width=0.25\textwidth]{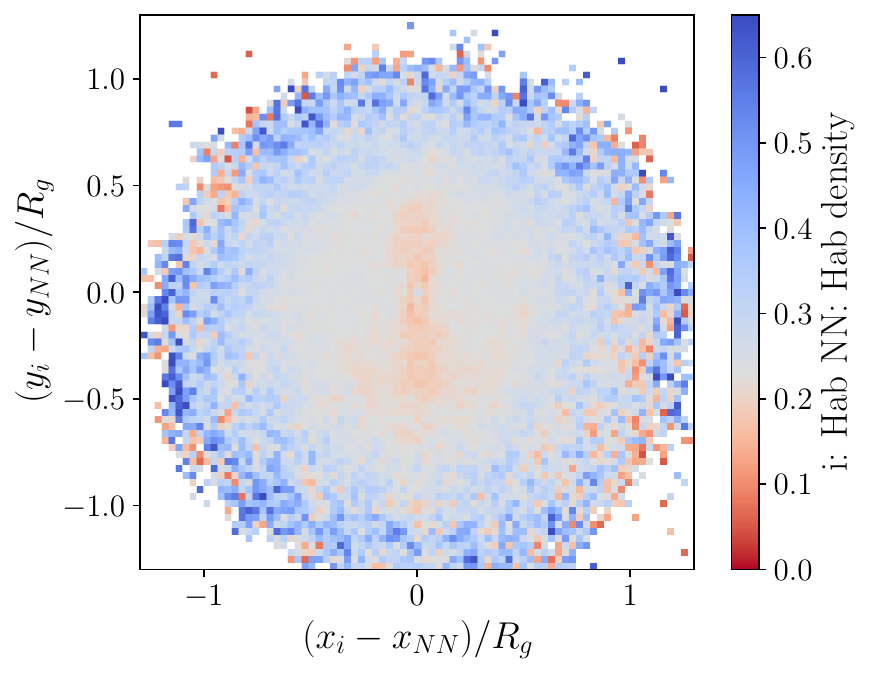}%
}
\subfloat[]{%
  \includegraphics[width=0.25\textwidth]{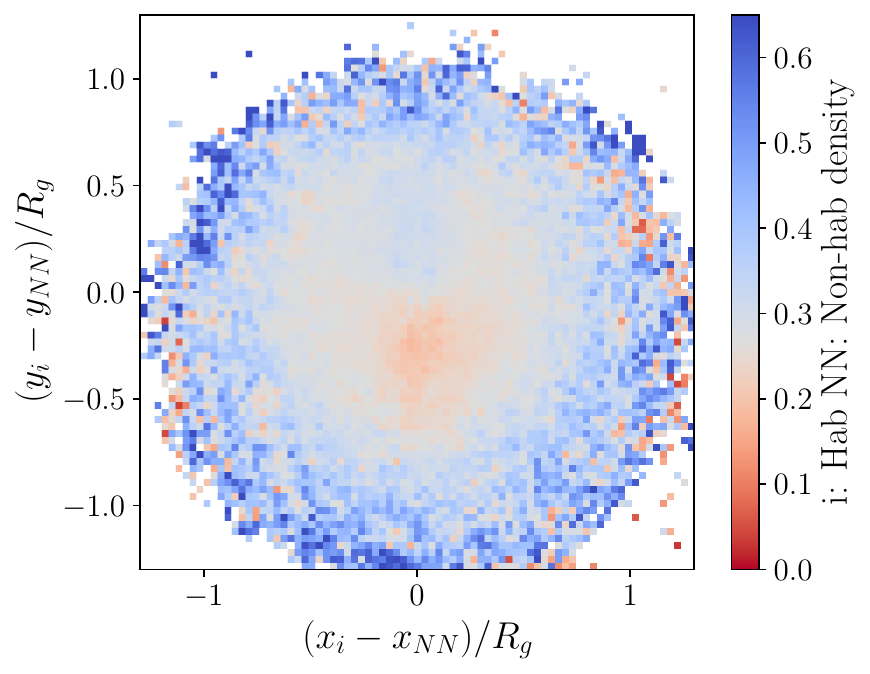}%
}
	\caption{(a) and (b) Density of habituated individuals $i$ with (a) habituated and (b) non-habituated nearest-neighbors $NN$ depending on the relative position of the individual with the nearest neighbor.}\label{supp:fig:density_habituated_positions}
\end{figure}

\begin{figure}
	\centering
	\subfloat[]{%
  \includegraphics[width=0.25\textwidth]{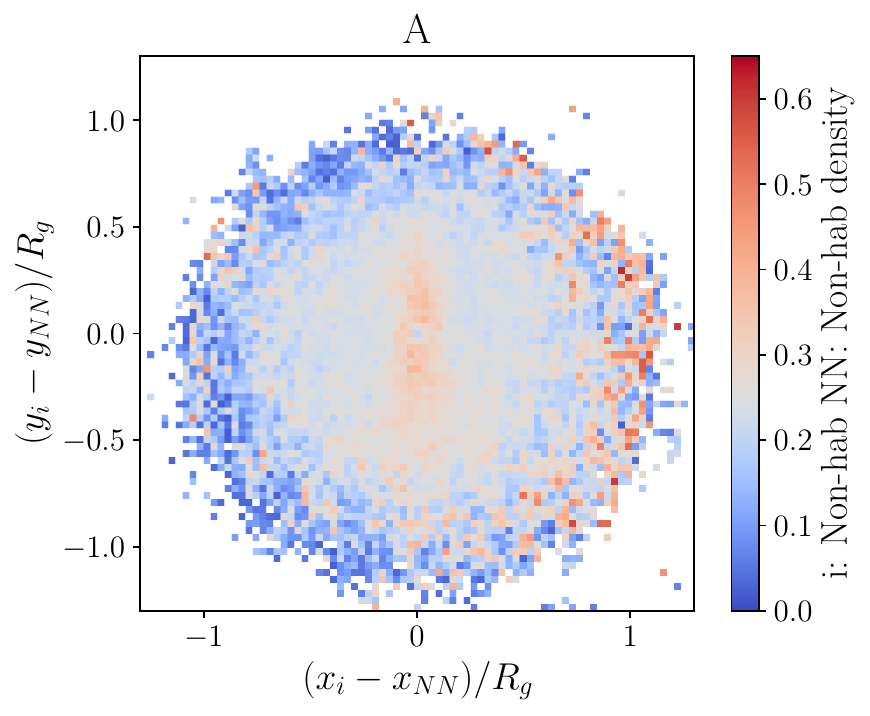}%
}
\subfloat[]{%
  \includegraphics[width=0.25\textwidth]{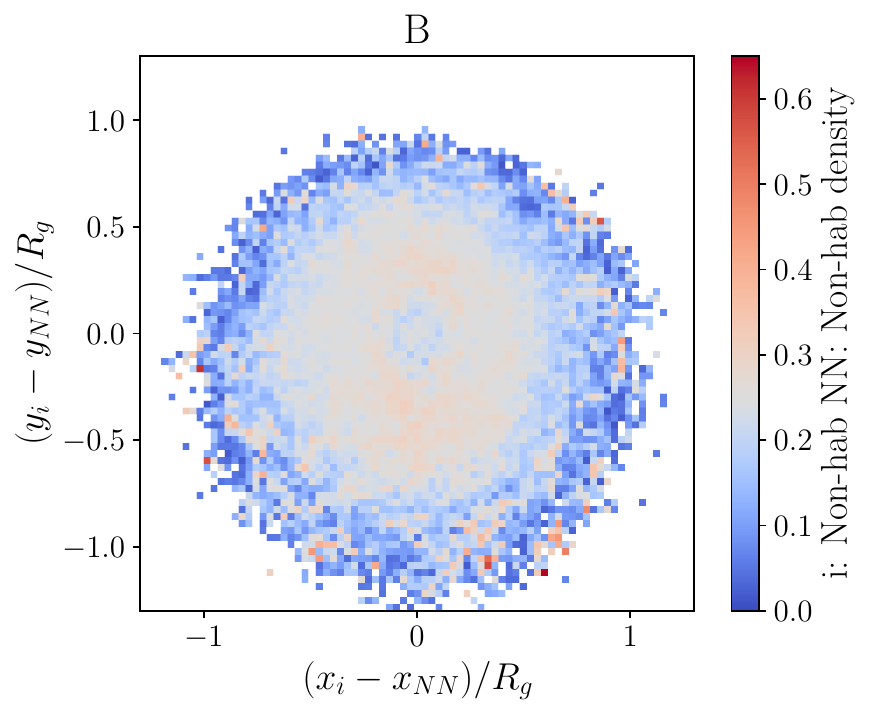}%
}
\subfloat[]{%
  \includegraphics[width=0.25\textwidth]{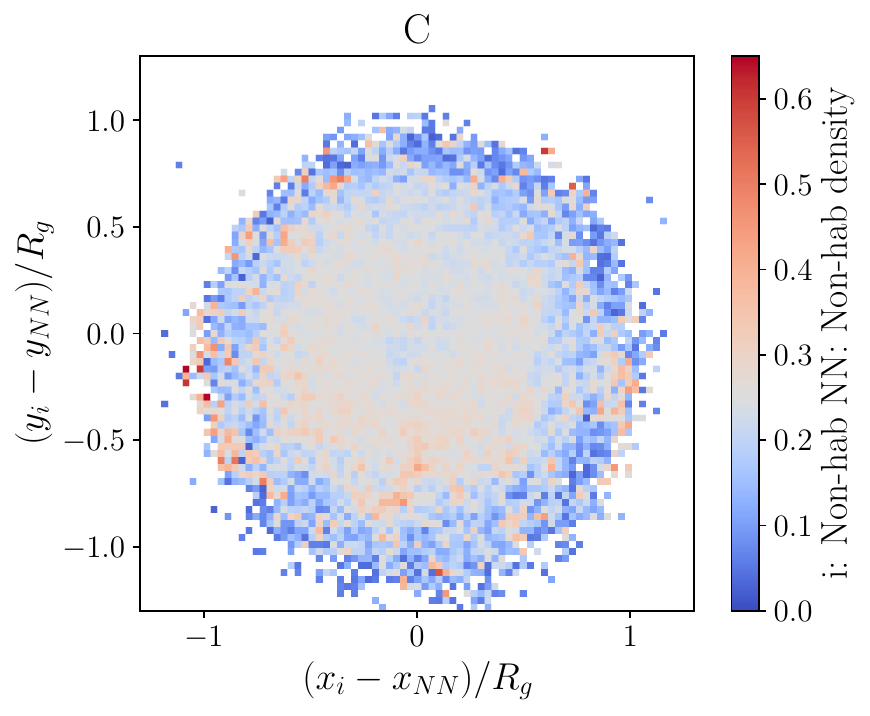}%
}

\subfloat[]{%
  \includegraphics[width=0.25\textwidth]{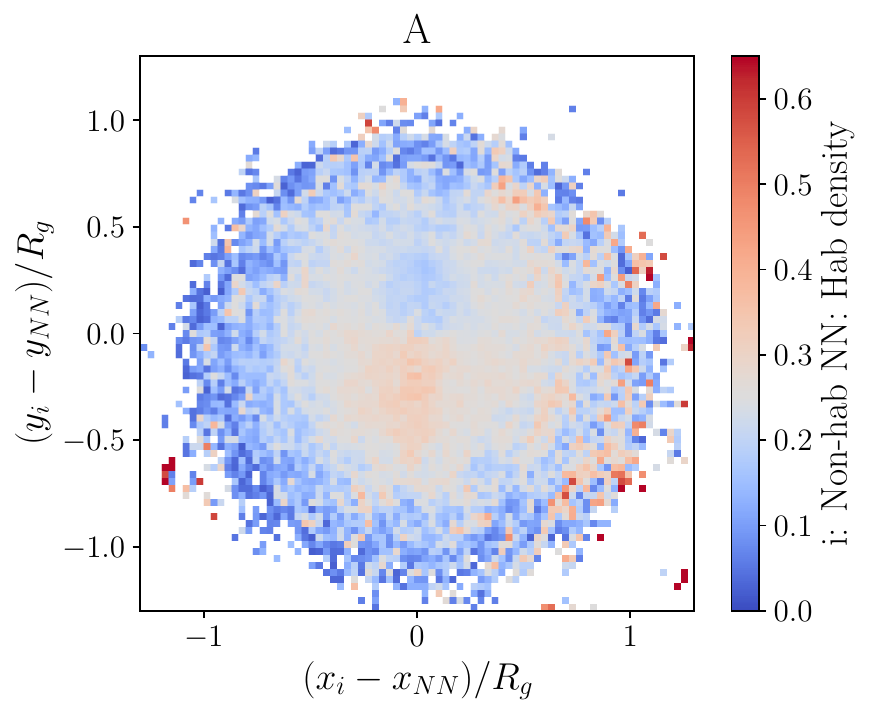}%
}
\subfloat[]{%
  \includegraphics[width=0.25\textwidth]{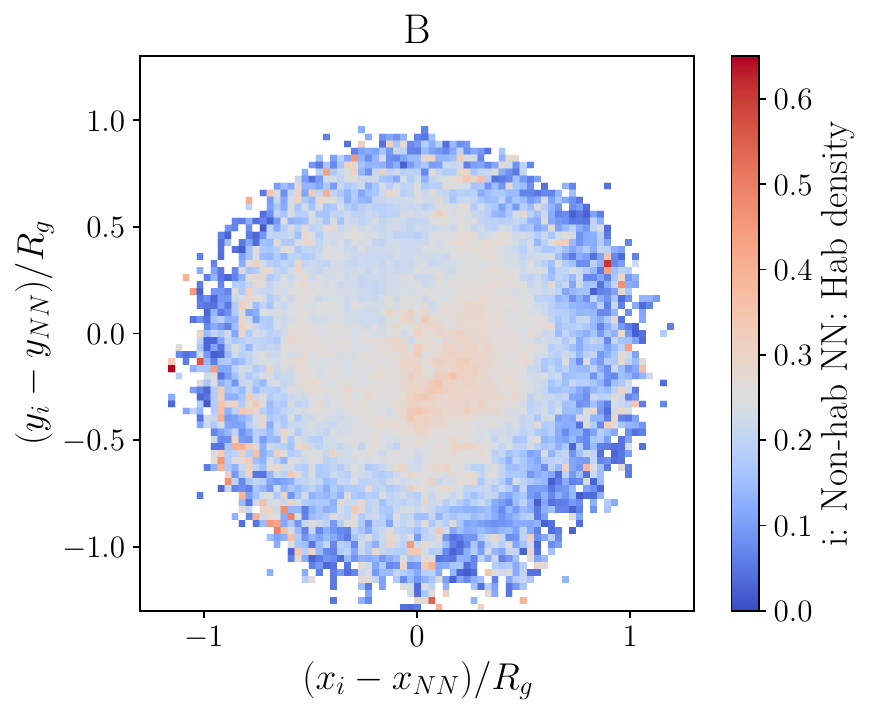}%
}
\subfloat[]{%
  \includegraphics[width=0.25\textwidth]{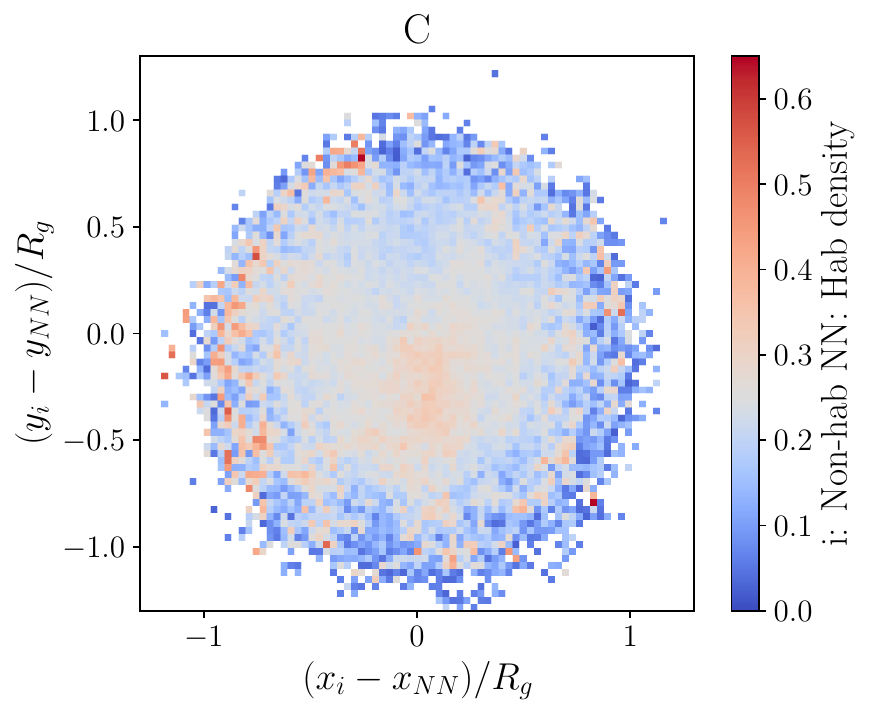}%
}

\subfloat[]{%
  \includegraphics[width=0.25\textwidth]{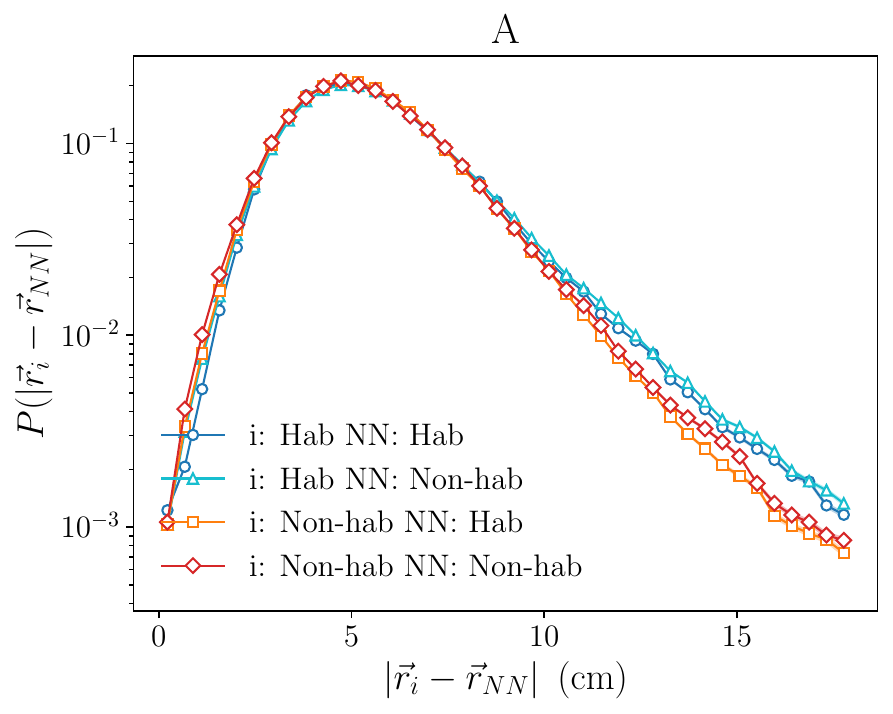}%
}
\subfloat[]{%
  \includegraphics[width=0.25\textwidth]{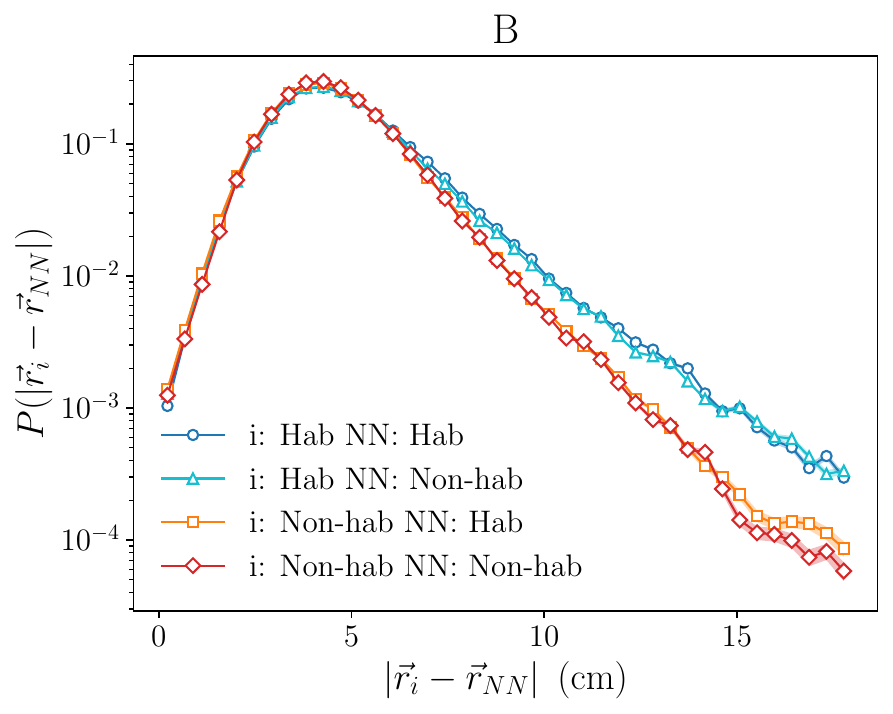}%
}
\subfloat[]{%
  \includegraphics[width=0.25\textwidth]{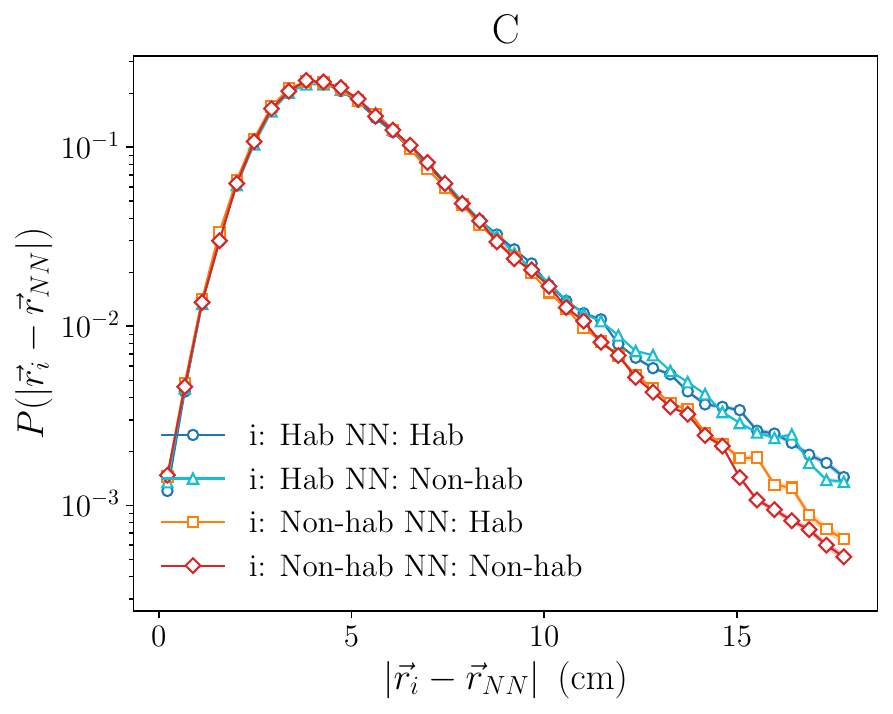}%
}

\subfloat[]{%
  \includegraphics[width=0.25\textwidth]{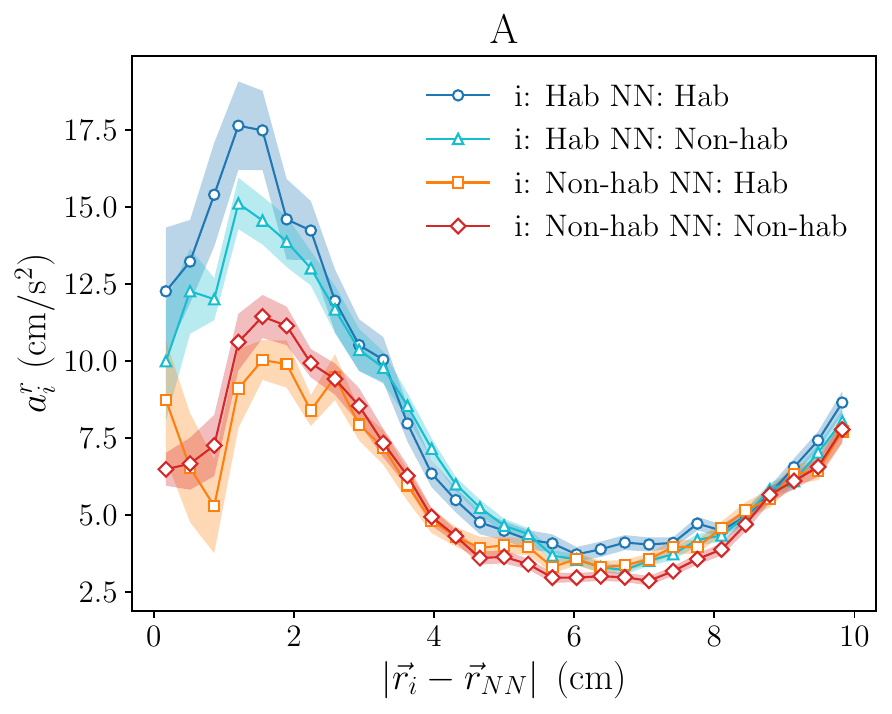}%
}
\subfloat[]{%
  \includegraphics[width=0.25\textwidth]{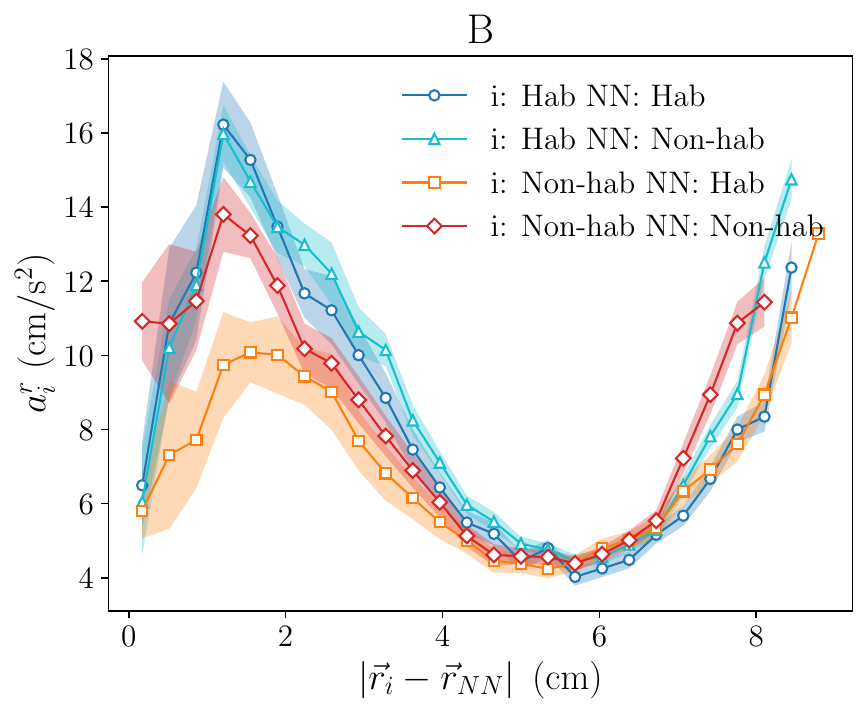}%
}
\subfloat[]{%
  \includegraphics[width=0.25\textwidth]{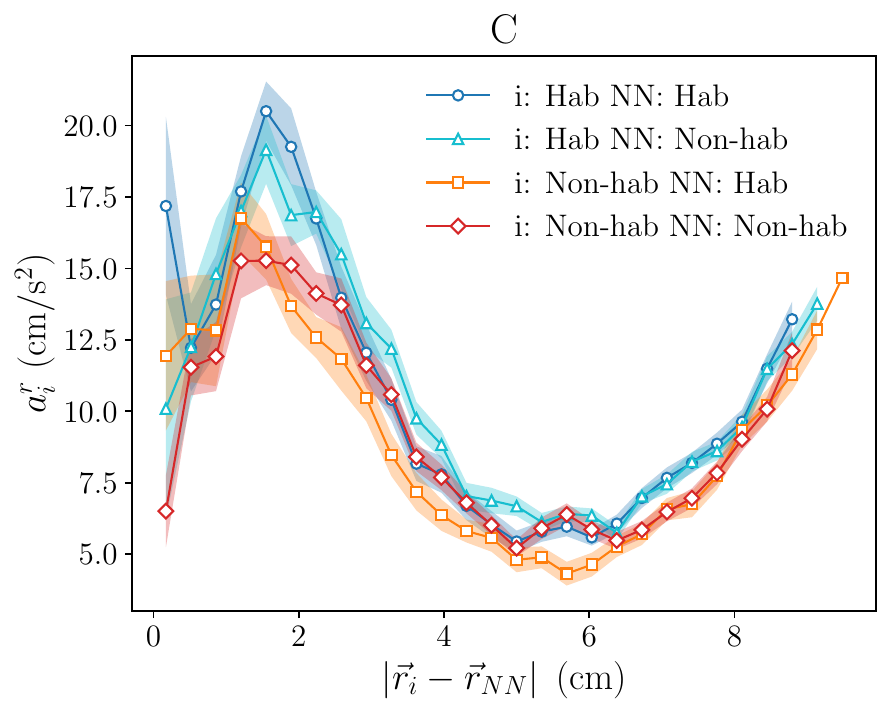}%
}
	\caption{(a)-(l) Local-level spatial interactions (see Fig.~\ref{fig:local_spatial}) for different series.}\label{supp:fig:local_spatial_series}
\end{figure}

\begin{figure}
	\centering
	\subfloat[]{%
  \includegraphics[width=0.25\textwidth]{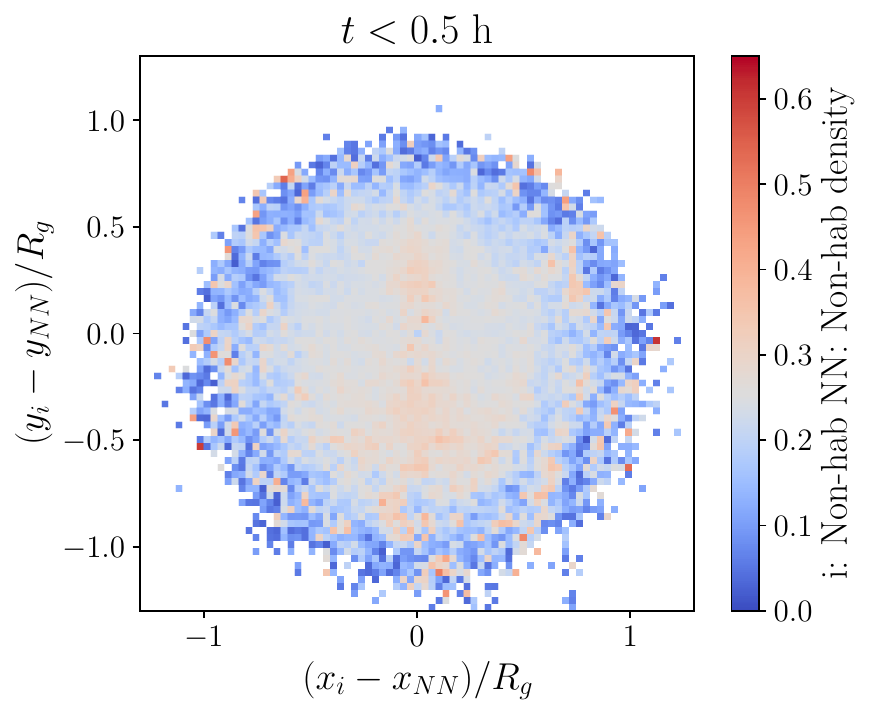}%
}
\subfloat[]{%
  \includegraphics[width=0.25\textwidth]{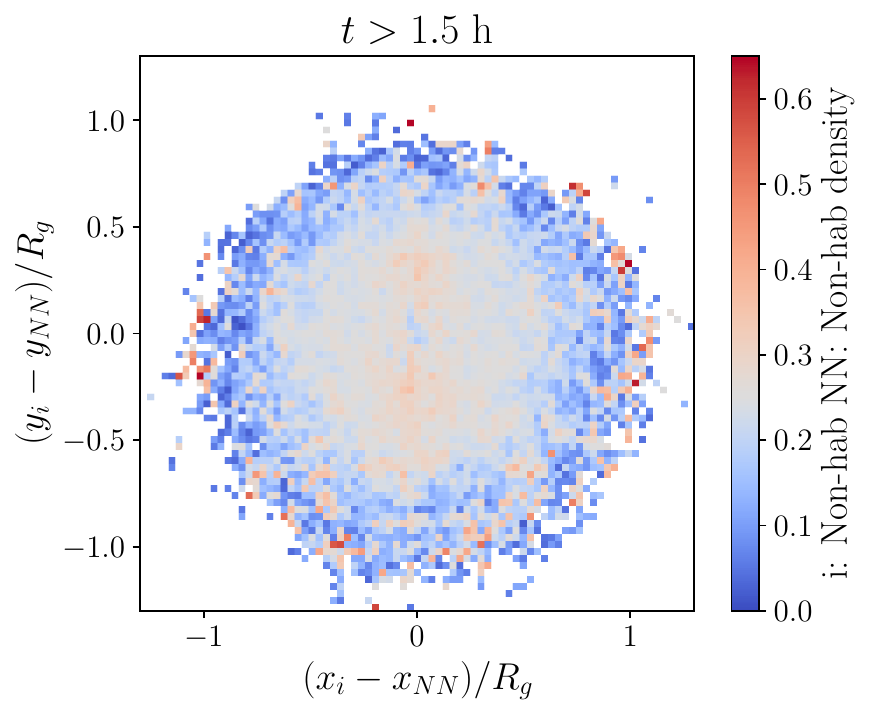}%
}
\subfloat[]{%
  \includegraphics[width=0.25\textwidth]{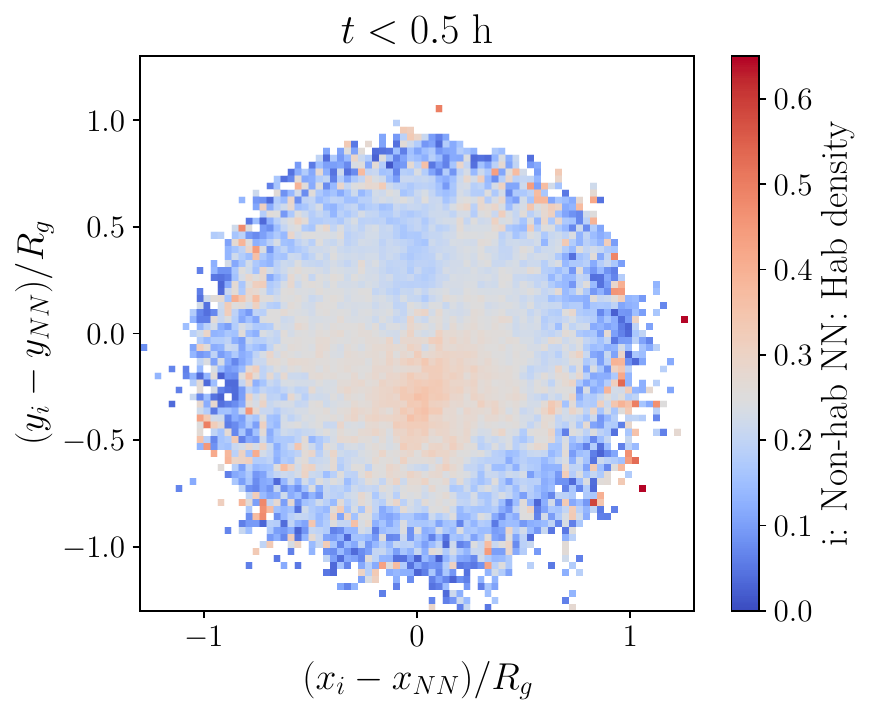}%
}
\subfloat[]{%
  \includegraphics[width=0.25\textwidth]{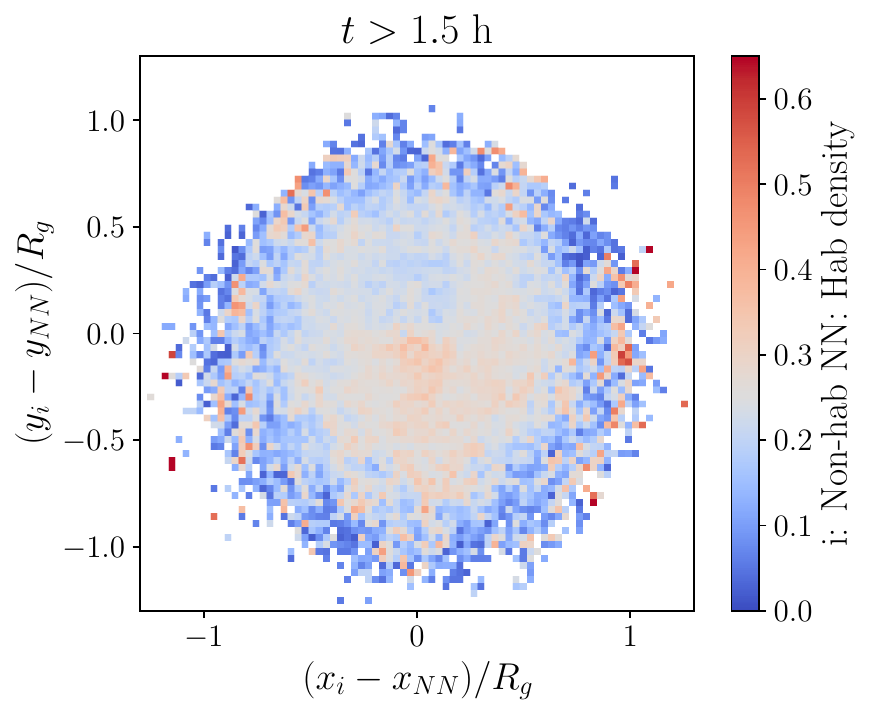}%
}

\subfloat[]{%
  \includegraphics[width=0.45\textwidth]{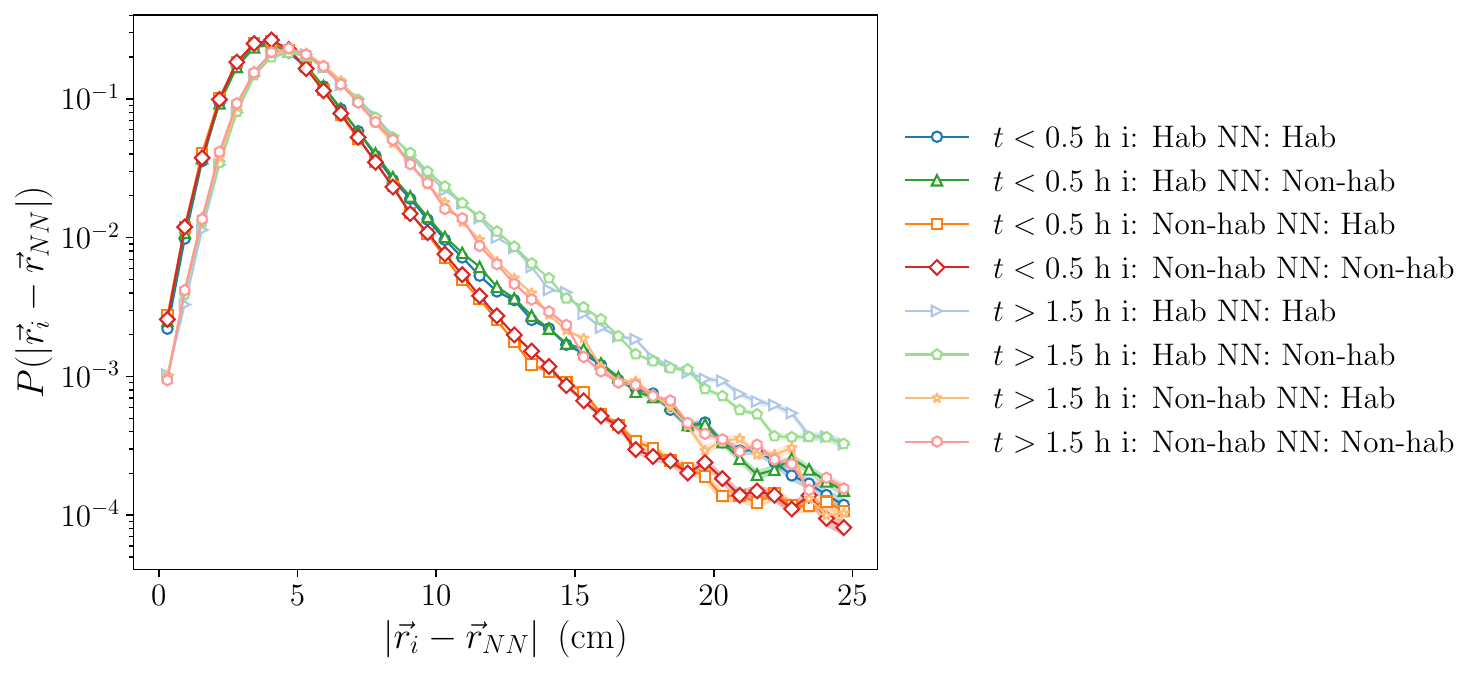}%
}
\subfloat[]{%
  \includegraphics[width=0.45\textwidth]{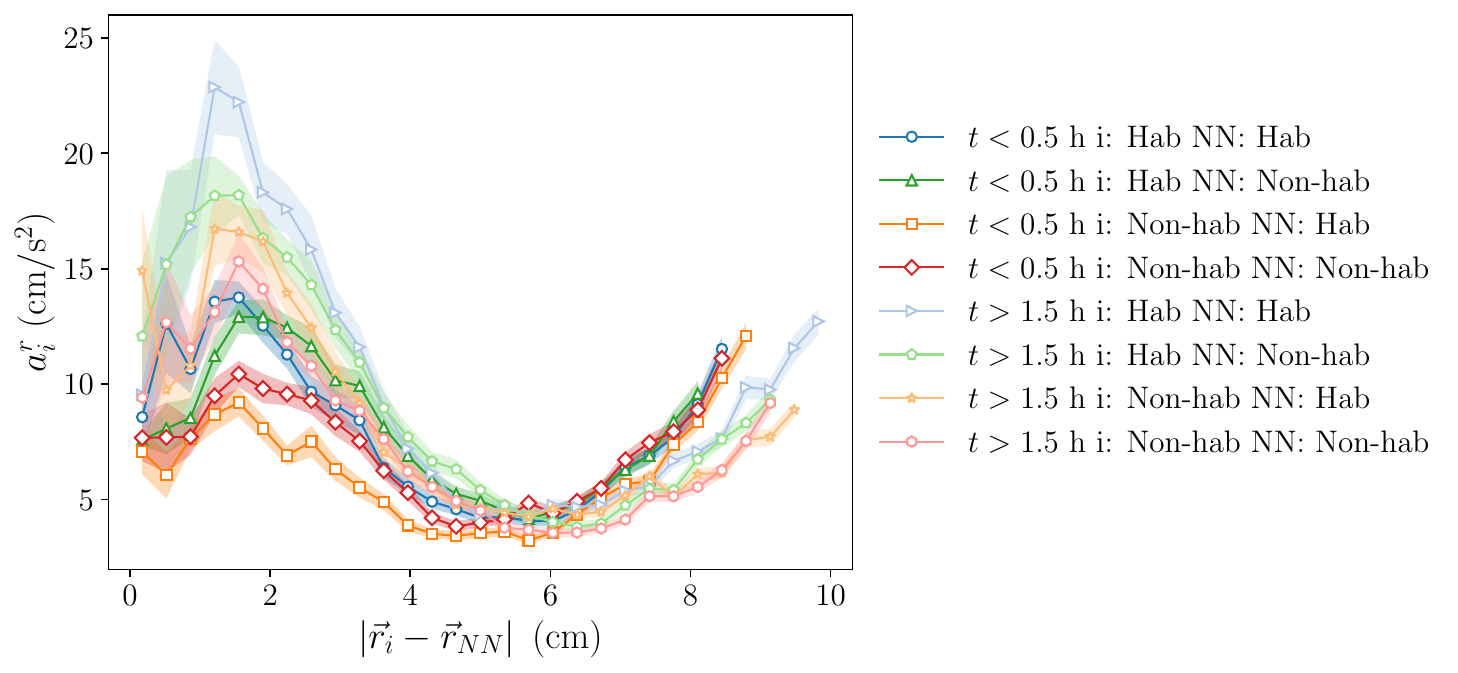}%
}
	\caption{(a)-(f) Temporal evolution of local-level spatial interactions (see Fig.~\ref{fig:local_spatial}).}\label{supp:fig:local_spatial_tEvo}
\end{figure}

\begin{figure}
	\centering
	\subfloat[]{%
  \includegraphics[width=0.25\textwidth]{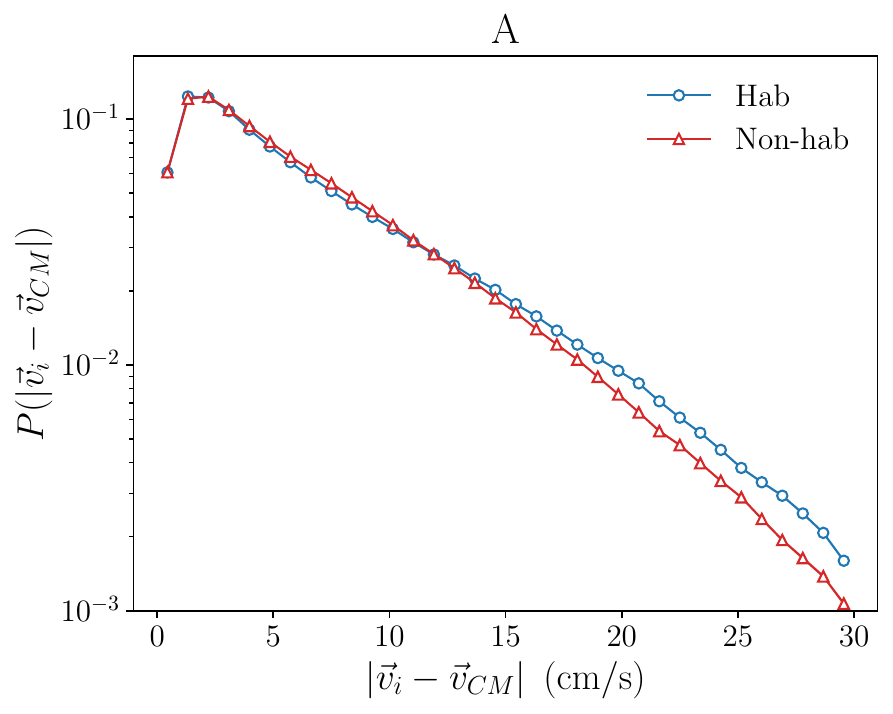}%
}
\subfloat[]{%
  \includegraphics[width=0.25\textwidth]{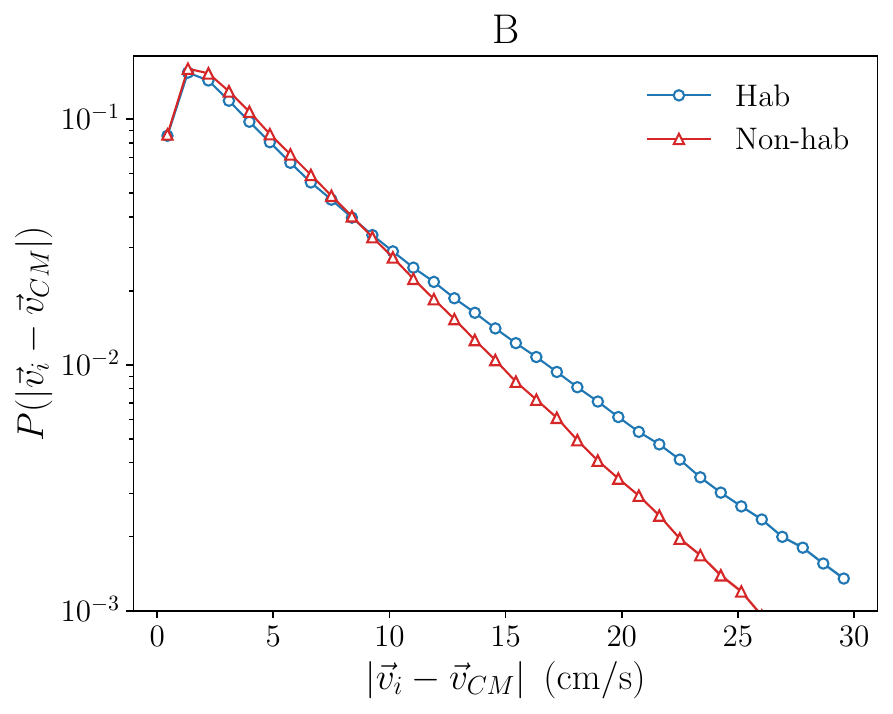}%
}
\subfloat[]{%
  \includegraphics[width=0.25\textwidth]{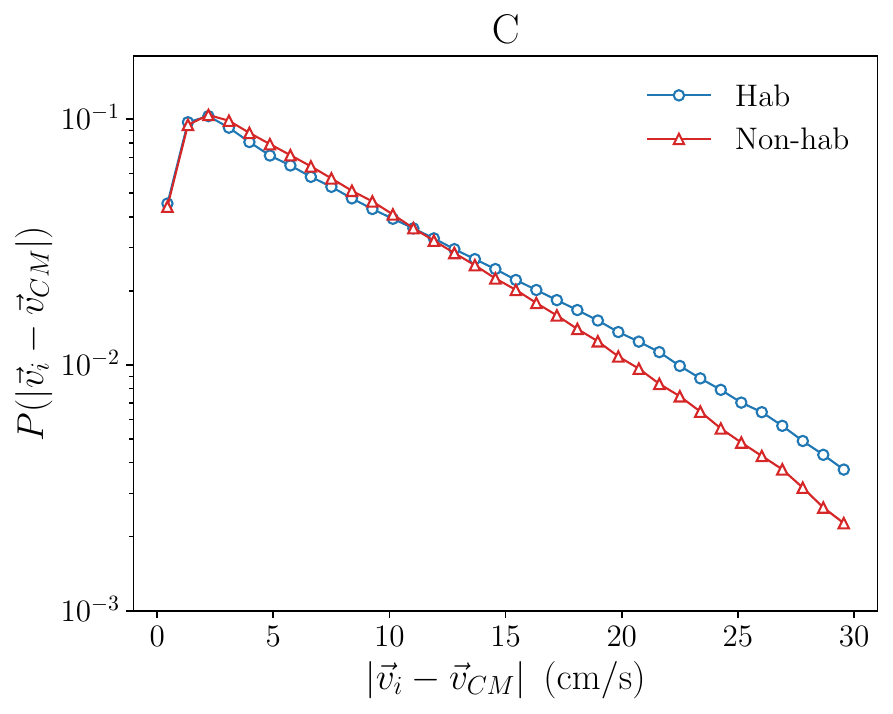}%
}

\subfloat[]{%
  \includegraphics[width=0.25\textwidth]{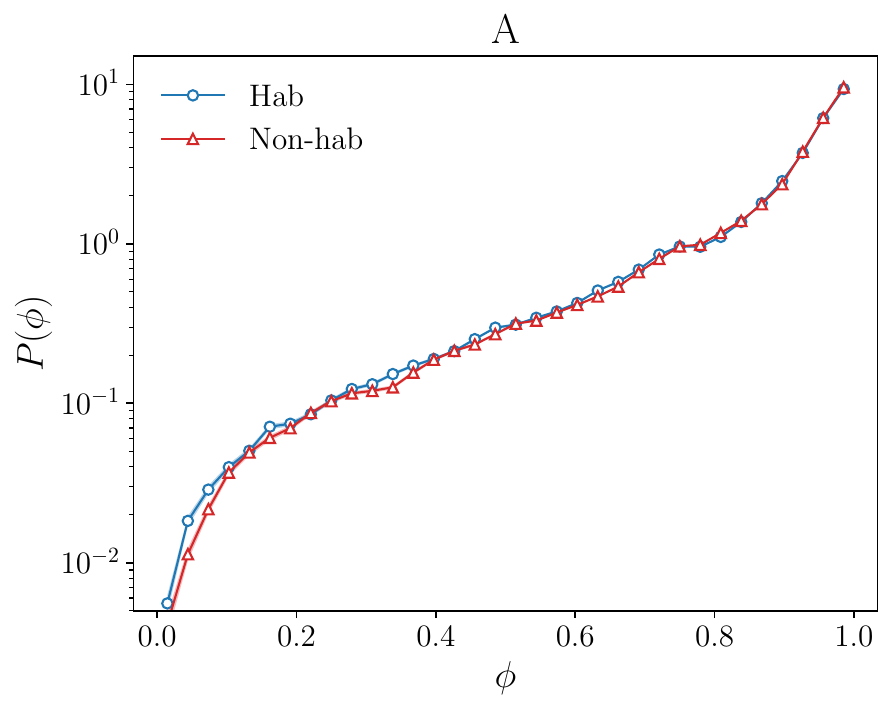}%
}
\subfloat[]{%
  \includegraphics[width=0.25\textwidth]{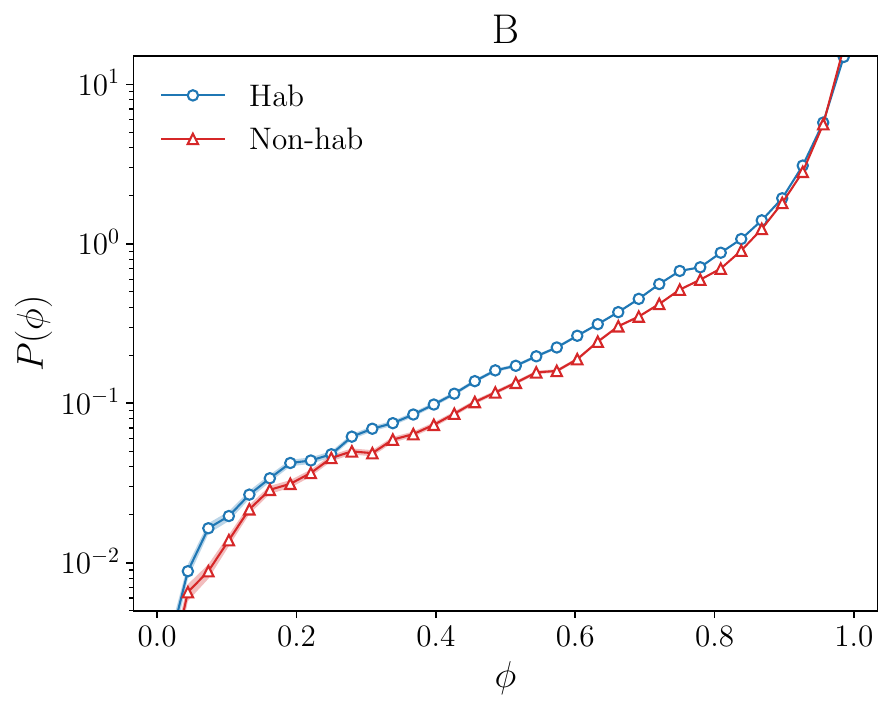}%
}
\subfloat[]{%
  \includegraphics[width=0.25\textwidth]{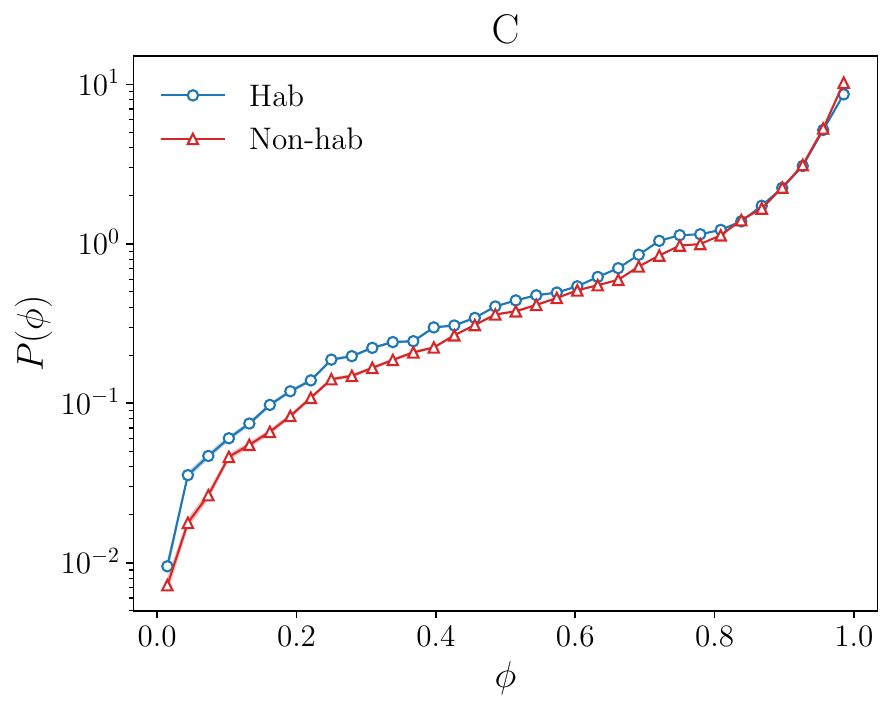}%
}

\subfloat[]{%
  \includegraphics[width=0.25\textwidth]{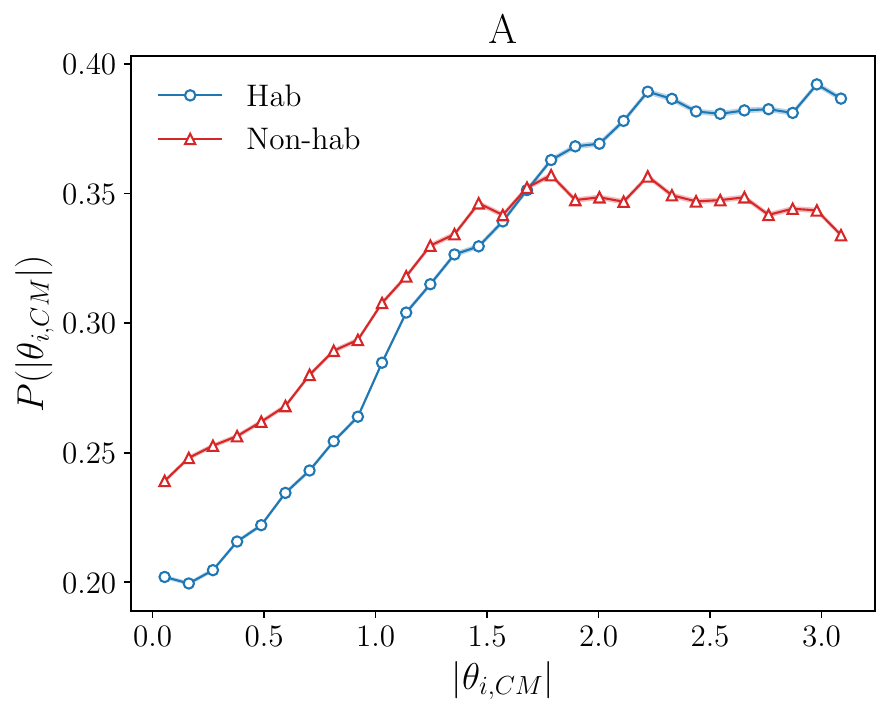}%
}
\subfloat[]{%
  \includegraphics[width=0.25\textwidth]{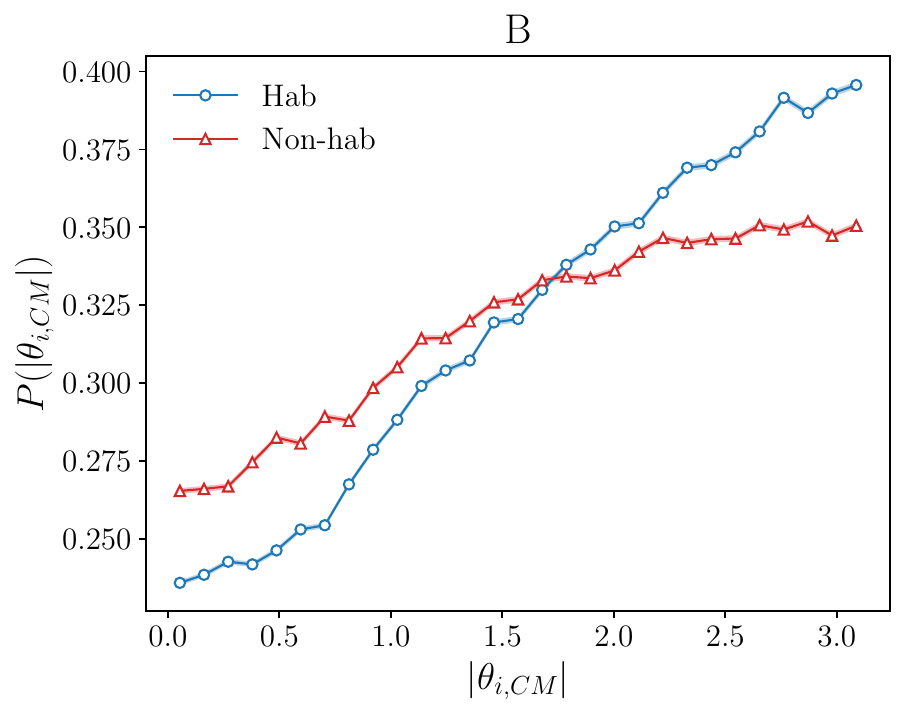}%
}
\subfloat[]{%
  \includegraphics[width=0.25\textwidth]{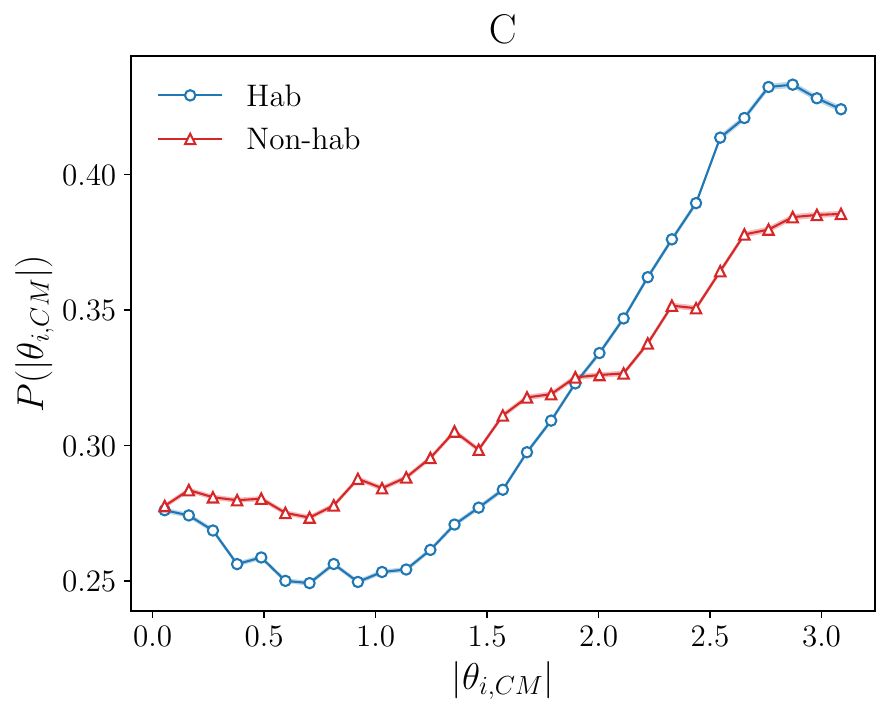}%
}
	\caption{(a)-(i) Group-level coordination (see Fig.~\ref{fig:group_velocity}) for different series.}\label{supp:fig:group_velocity_series}
\end{figure}

\begin{figure}
	\centering
	\subfloat[]{%
  \includegraphics[width=0.25\textwidth]{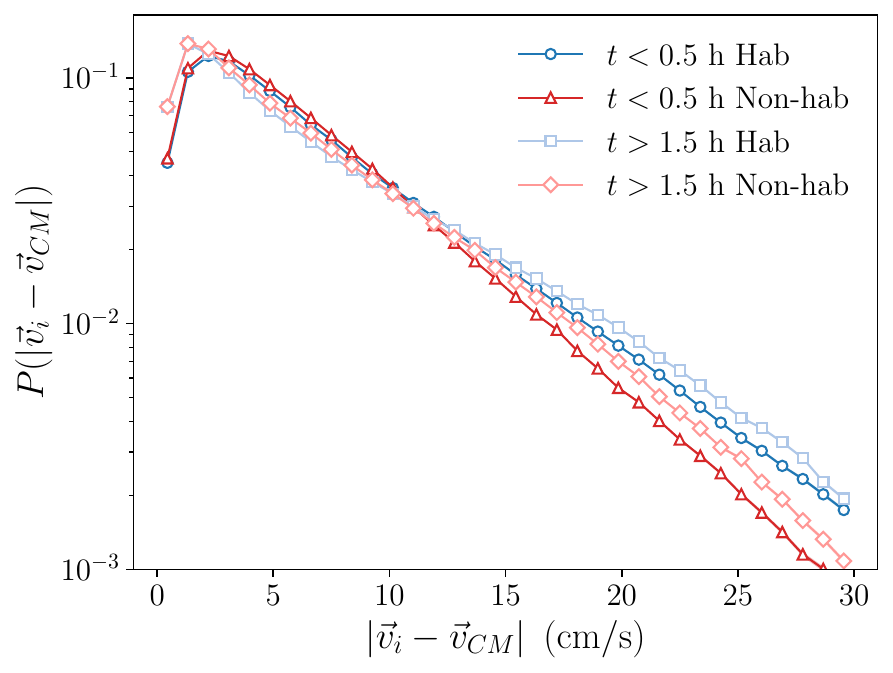}%
}
\subfloat[]{%
  \includegraphics[width=0.25\textwidth]{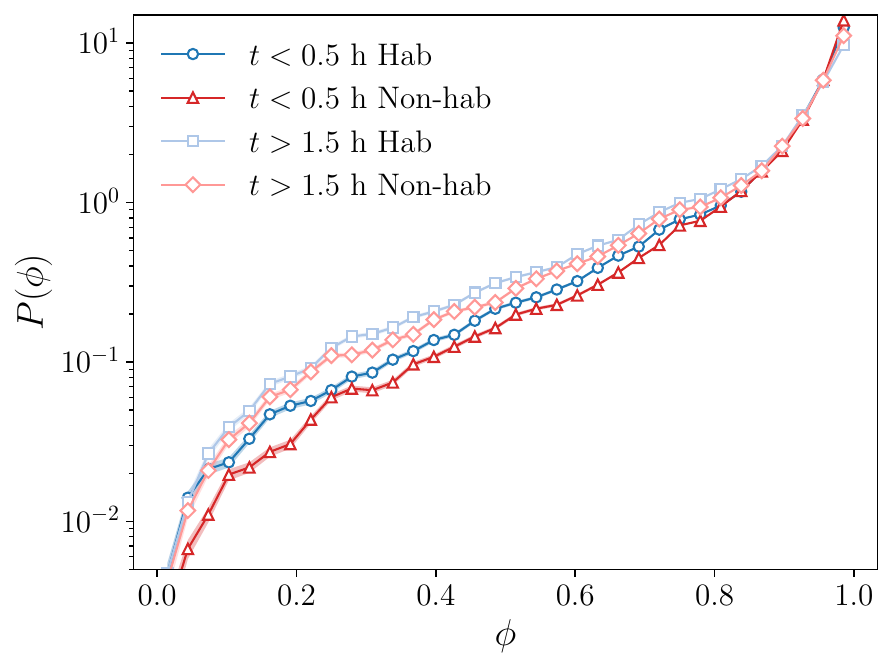}%
}
\subfloat[]{%
\includegraphics[width=0.25\textwidth]{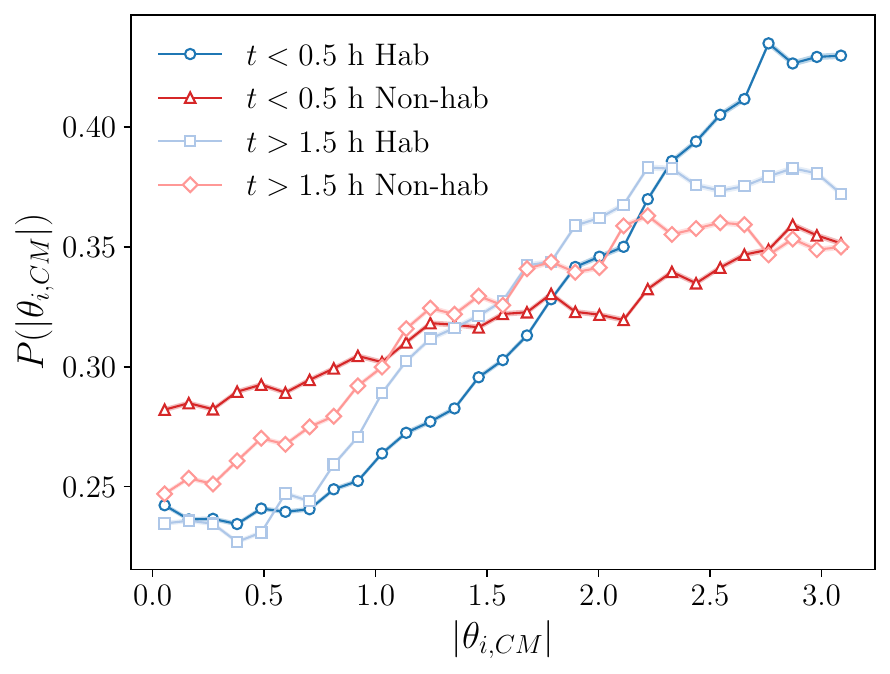}
}
	\caption{(a)-(c) Temporal evolution of group-level coordination (see Fig.~\ref{fig:group_velocity}).\label{supp:fig:group_velocity_tEvo}}
\end{figure}

\begin{figure}
	\centering
	\includegraphics[width=0.25\textwidth]{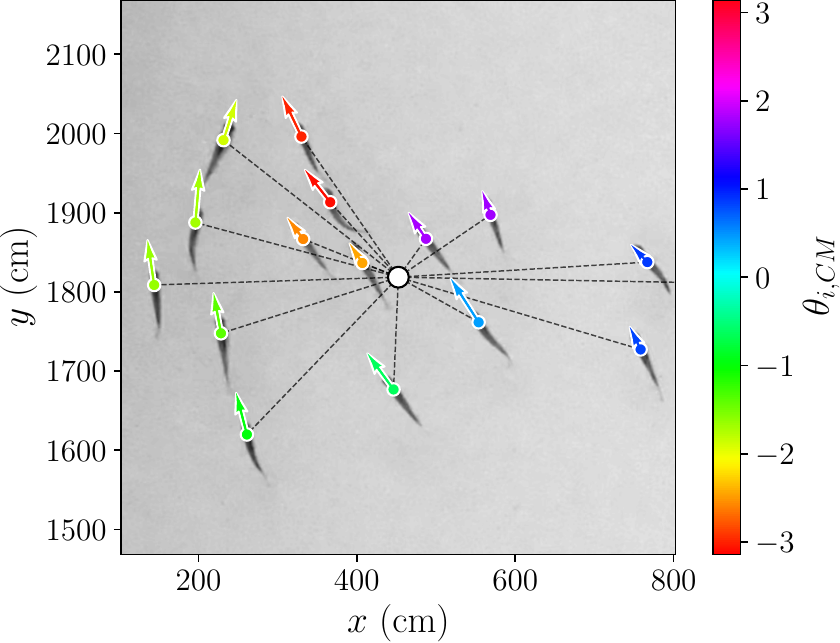}%
	\caption{Sample of the positions of individuals $\vec{r}_i$ color-coded by the individuals orientation relative to the center of mass (CM), $\theta_{i, CM}$. The white circle corresponds to the CM.} \label{supp:fig:theta_i_CM}
\end{figure}

\begin{figure}
	\centering
	\subfloat[]{%
  \includegraphics[width=0.25\textwidth]{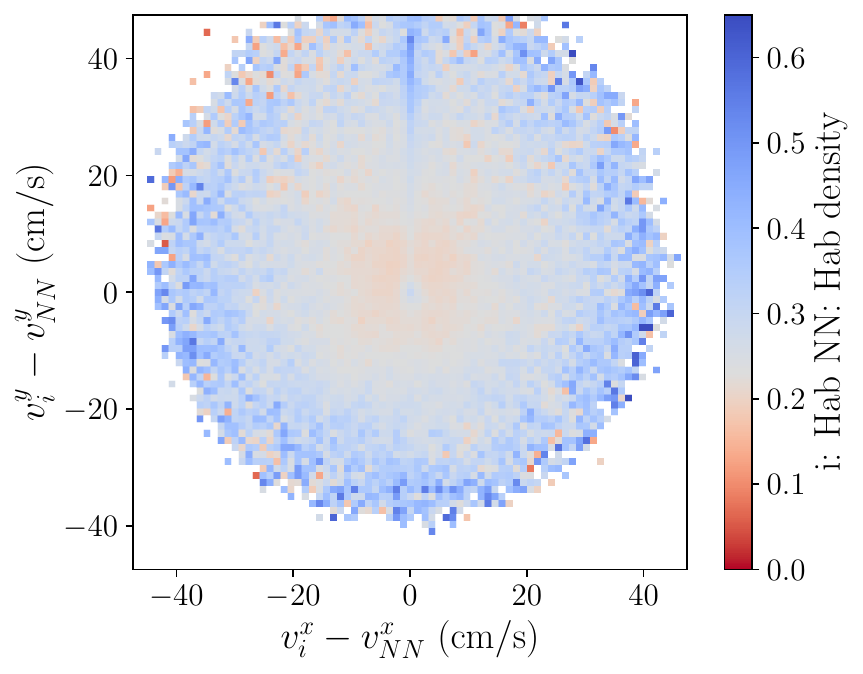}%
}
\subfloat[]{%
  \includegraphics[width=0.25\textwidth]{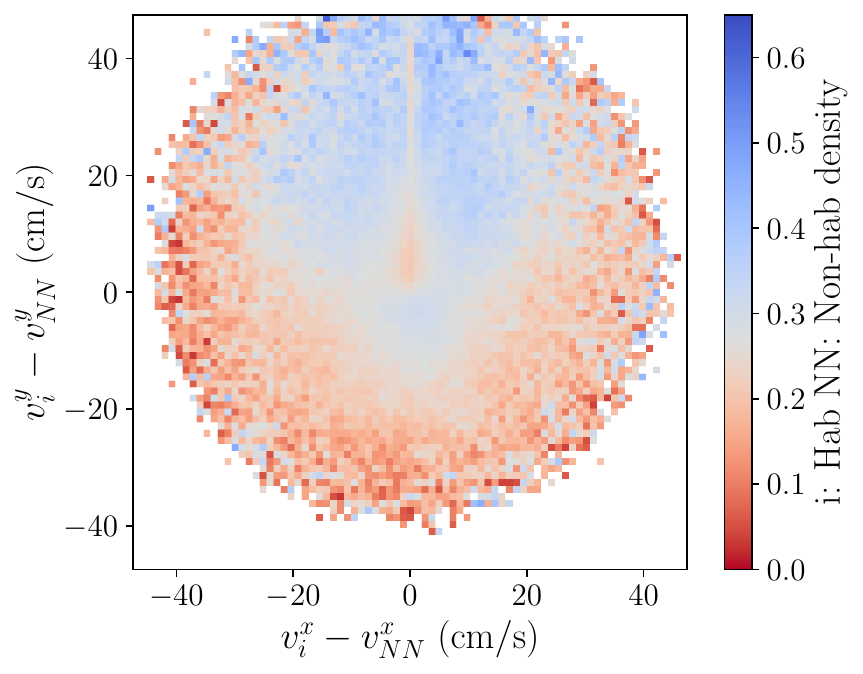}%
}
	\caption{(a) and (b) Density of habituated individuals $i$ with (a) habituated and (b) non-habituated nearest-neighbors $NN$ depending on the relative velocity of the individual with the nearest neighbor.}\label{supp:fig:density_habituated_velocities}
\end{figure}

\begin{figure}
	\centering
	\subfloat[]{%
  \includegraphics[width=0.25\textwidth]{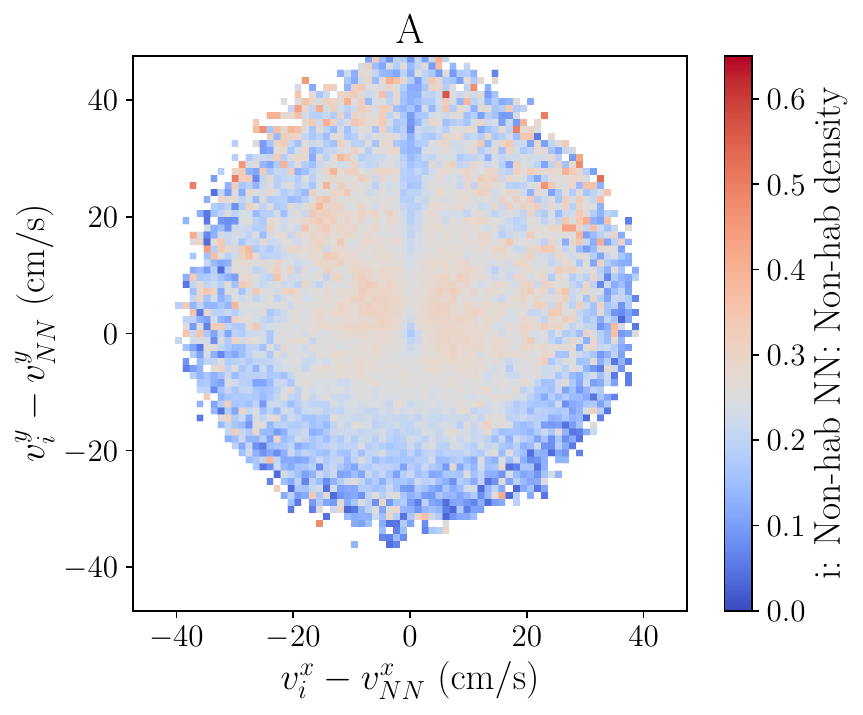}%
}
\subfloat[]{%
  \includegraphics[width=0.25\textwidth]{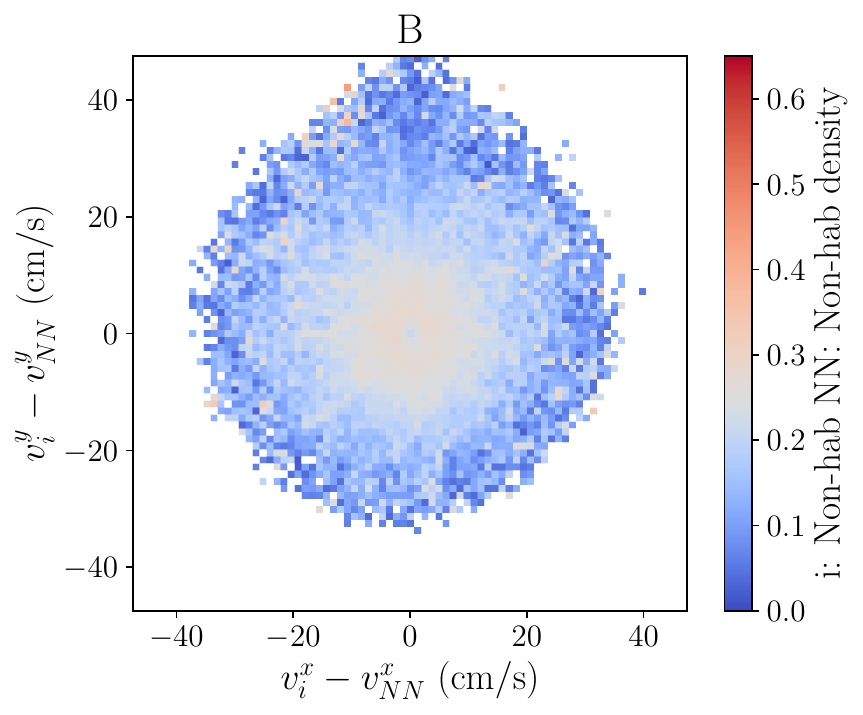}%
}
\subfloat[]{%
  \includegraphics[width=0.25\textwidth]{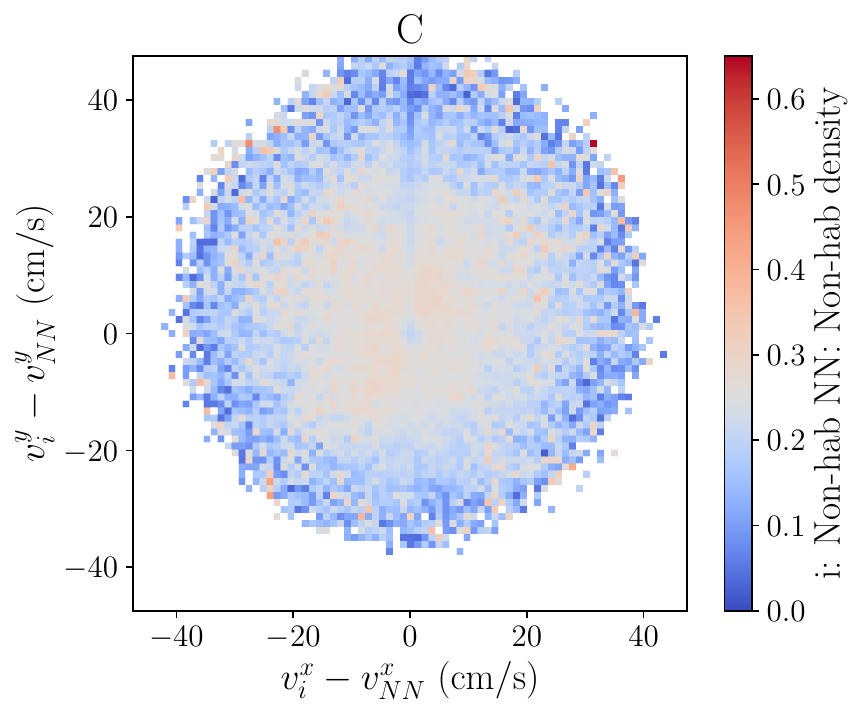}%
}

\subfloat[]{%
  \includegraphics[width=0.25\textwidth]{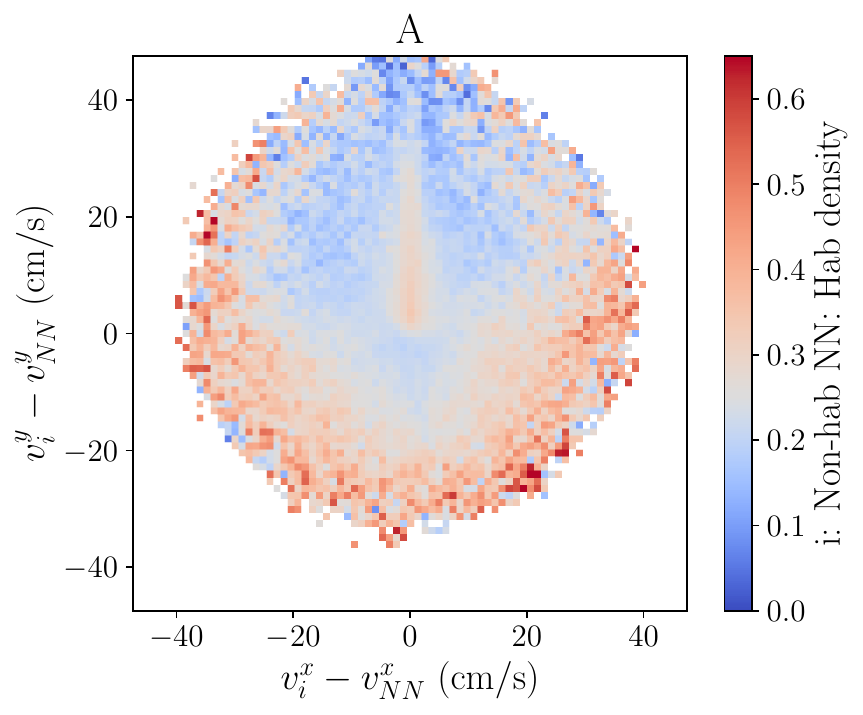}%
}
\subfloat[]{%
  \includegraphics[width=0.25\textwidth]{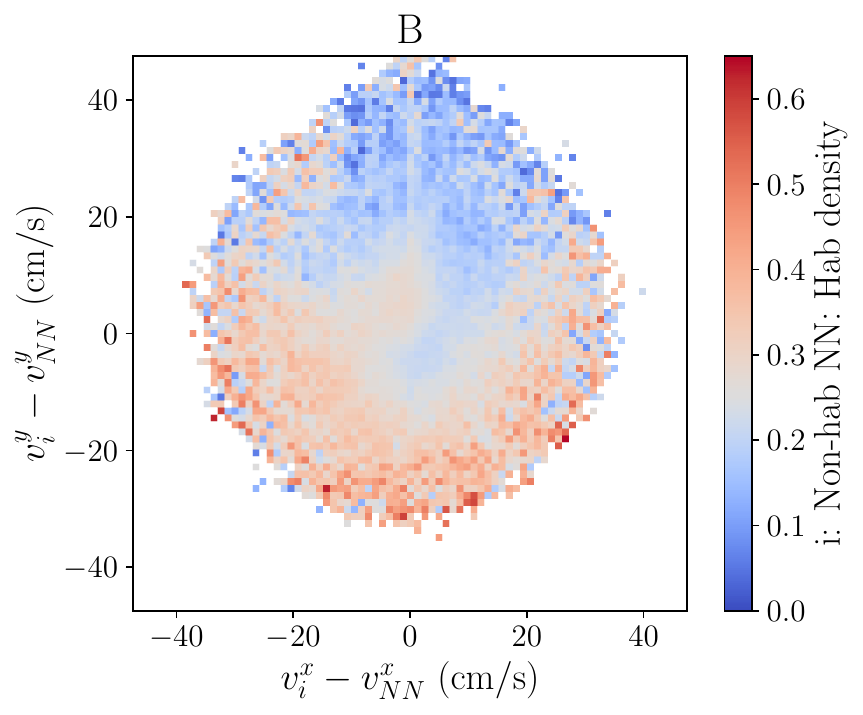}%
}
\subfloat[]{%
  \includegraphics[width=0.25\textwidth]{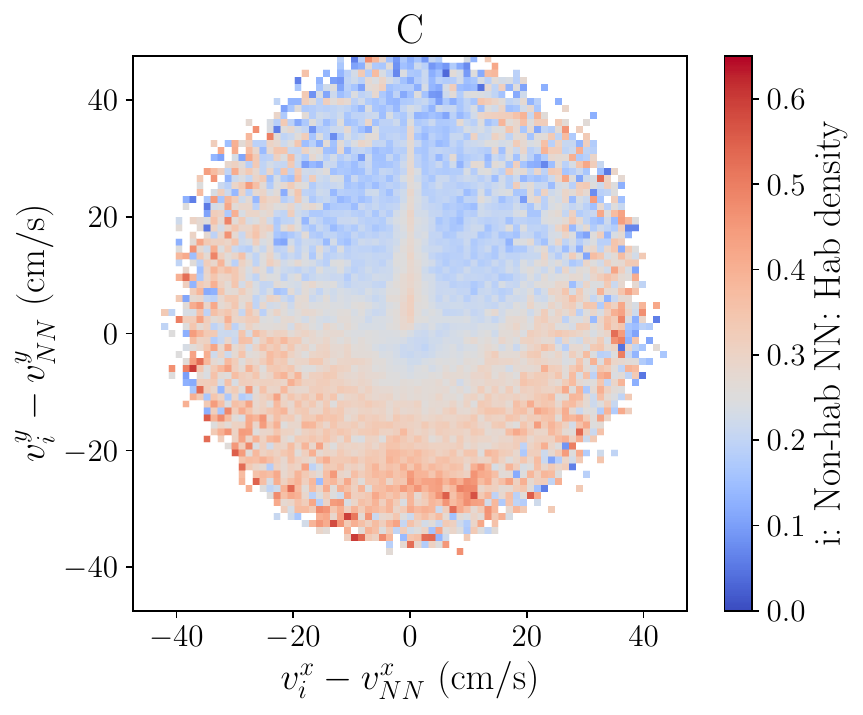}%
}

\subfloat[]{%
  \includegraphics[width=0.25\textwidth]{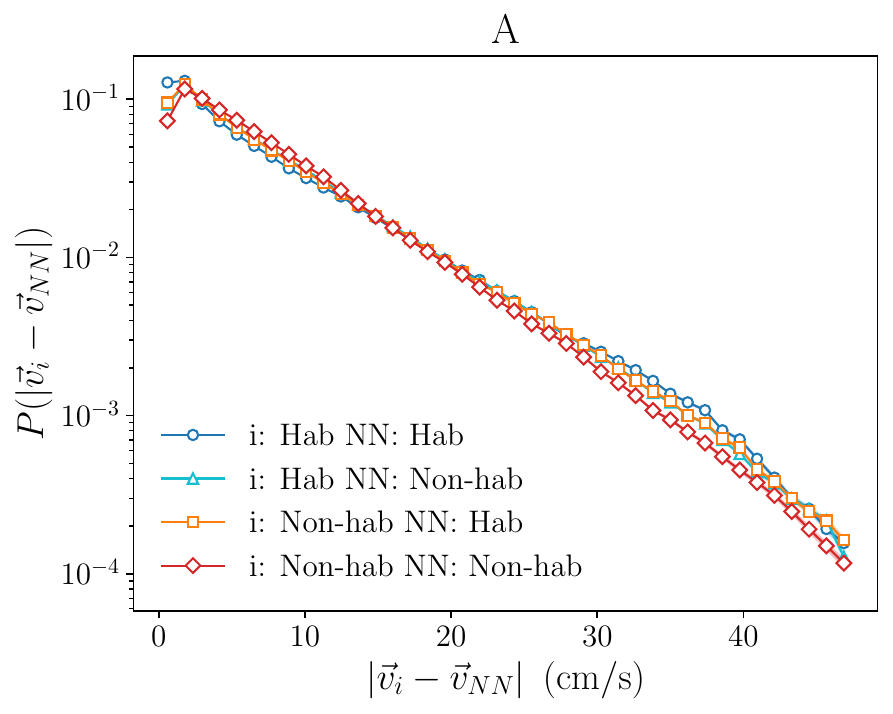}%
}
\subfloat[]{%
  \includegraphics[width=0.25\textwidth]{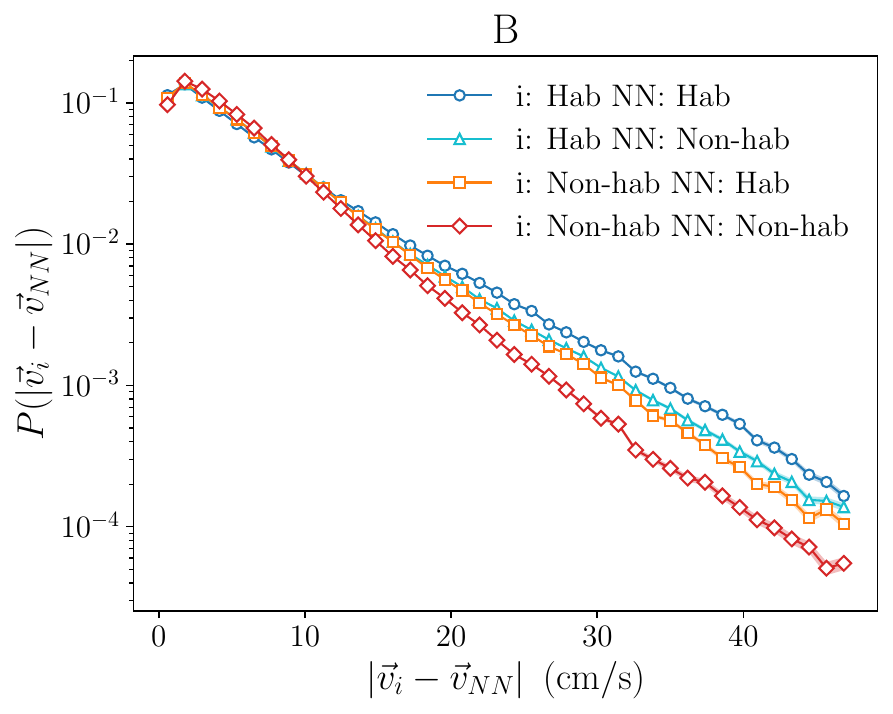}%
}
\subfloat[]{%
  \includegraphics[width=0.25\textwidth]{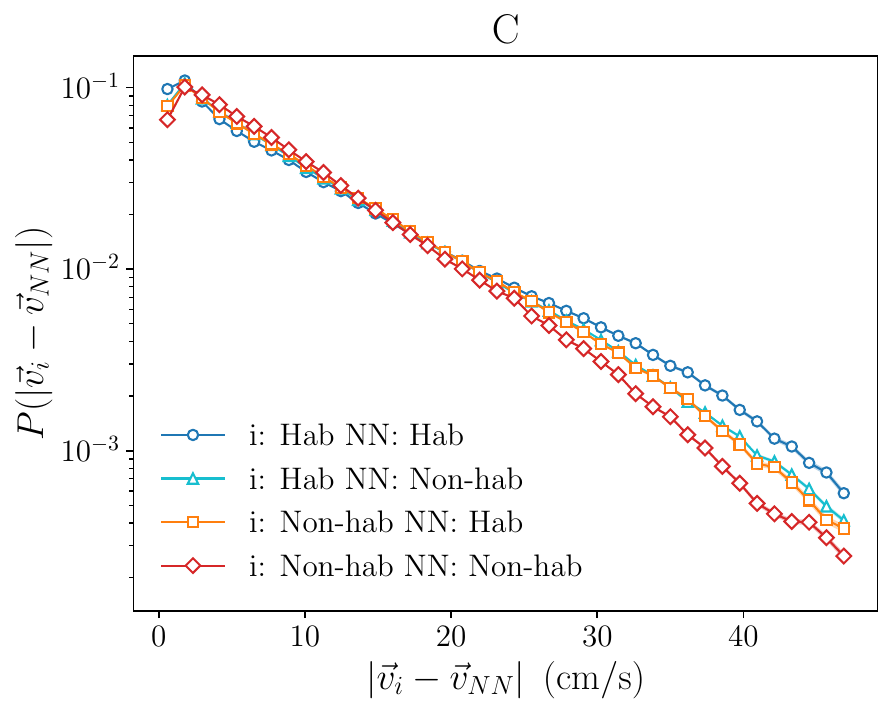}%
}
	\caption{(a)-(i) Local-level coordination (see Fig.~\ref{fig:local_velocity}) for different series.} \label{supp:fig:local_velocity_series}
\end{figure}

\begin{figure}
	\centering
	\subfloat[]{%
  \includegraphics[width=0.25\textwidth]{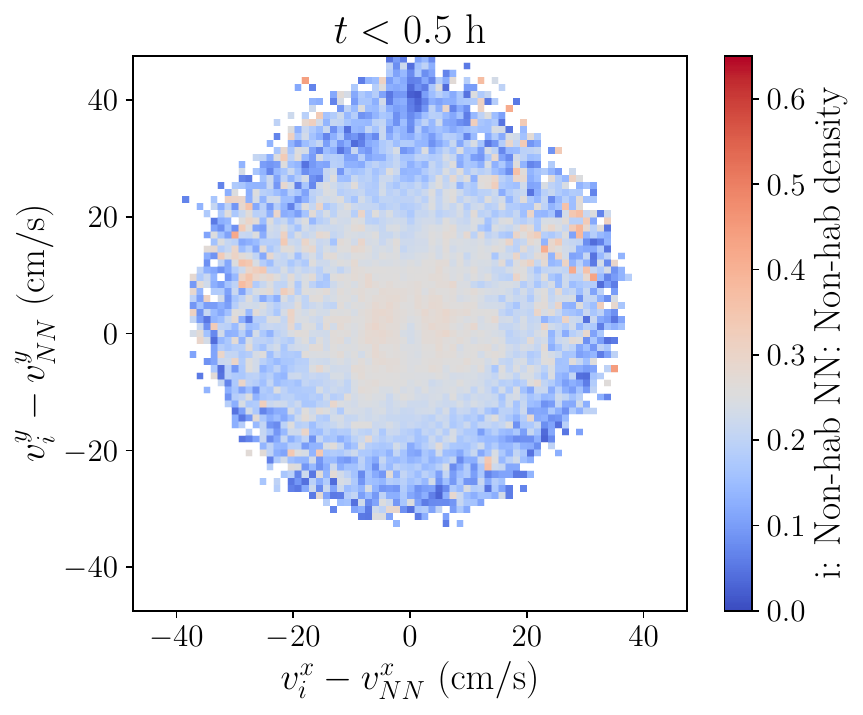}%
}
\subfloat[]{%
  \includegraphics[width=0.25\textwidth]{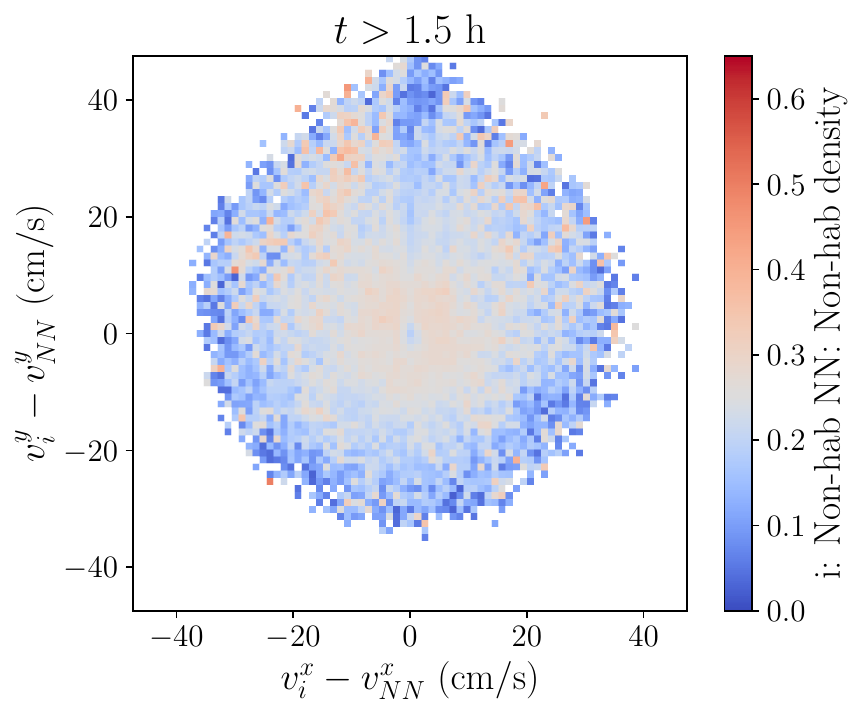}%
}

\subfloat[]{%
  \includegraphics[width=0.25\textwidth]{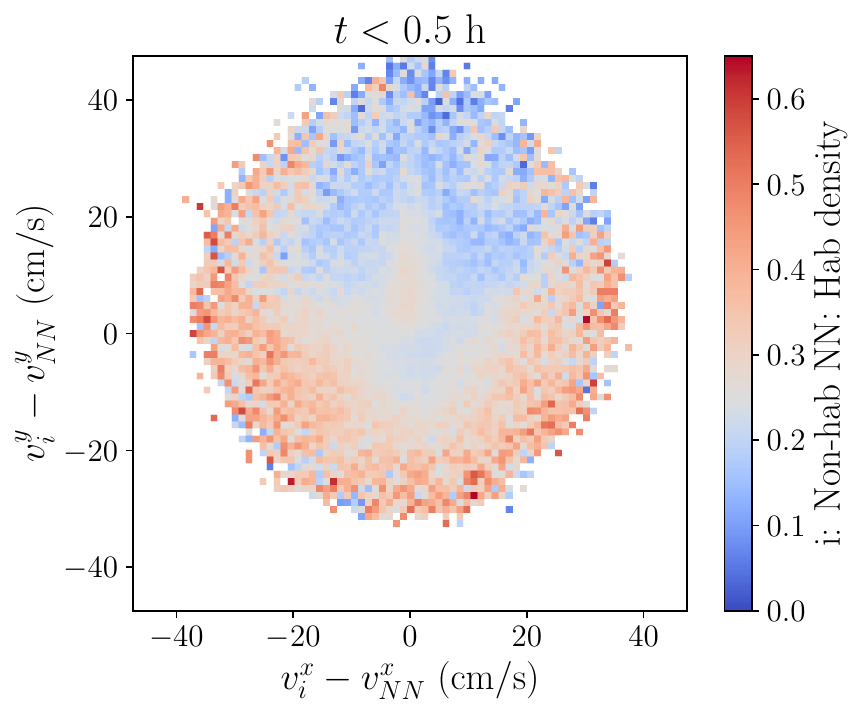}%
}
\subfloat[]{%
  \includegraphics[width=0.25\textwidth]{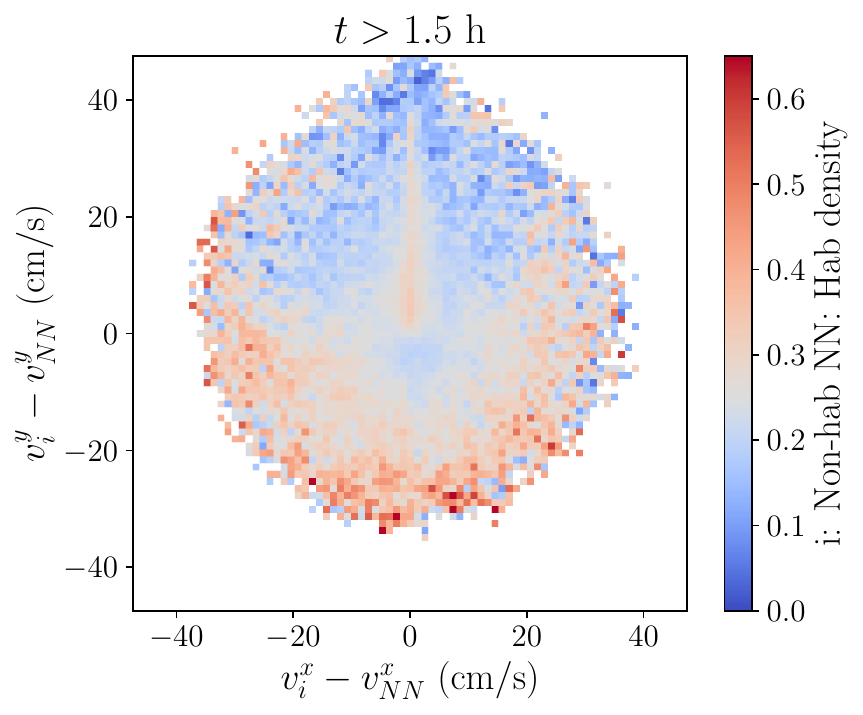}%
}

\subfloat[]{%
  \includegraphics[width=0.45\textwidth]{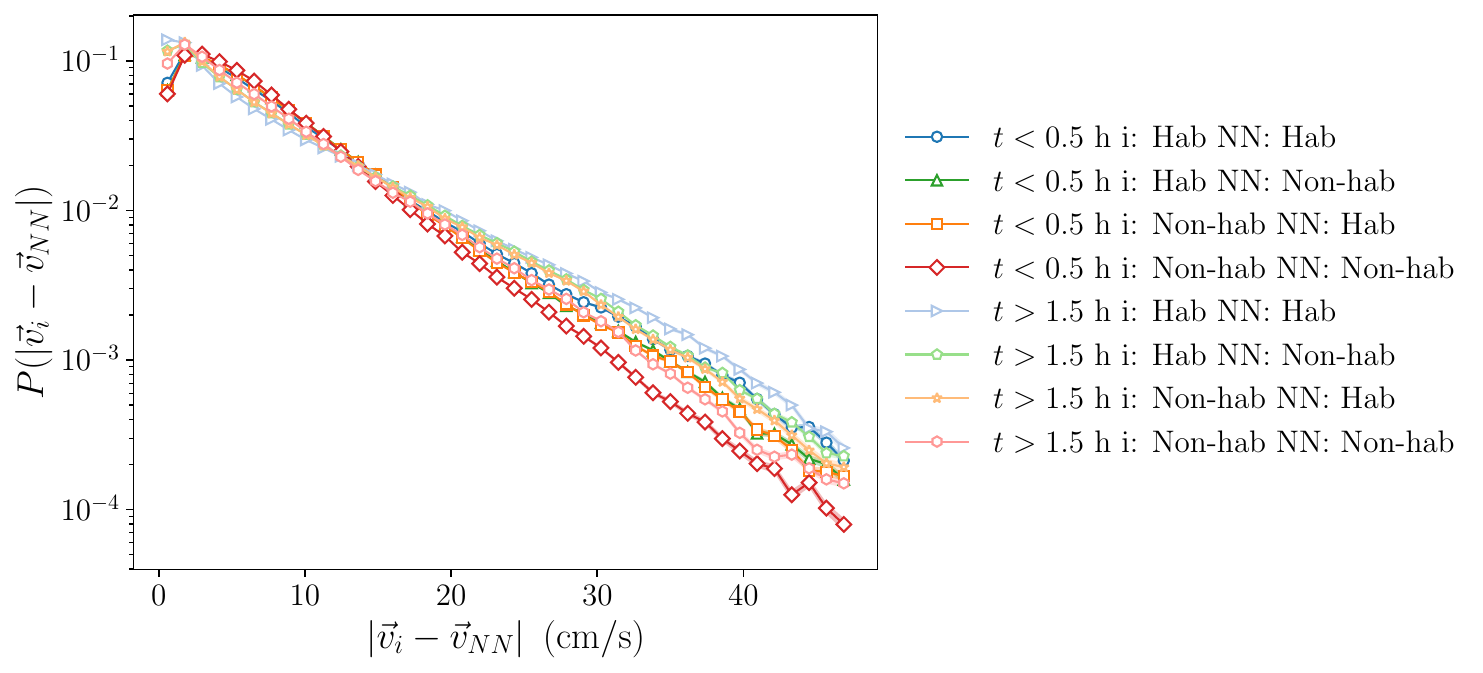}%
}
	\caption{(a)-(e) Temporal evolution of local-level coordination (see Fig.~\ref{fig:local_velocity}).} \label{supp:fig:local_velocity_tEvo}
\end{figure}

\begin{figure}
	\centering
	\subfloat[]{%
  \includegraphics[width=0.25\textwidth]{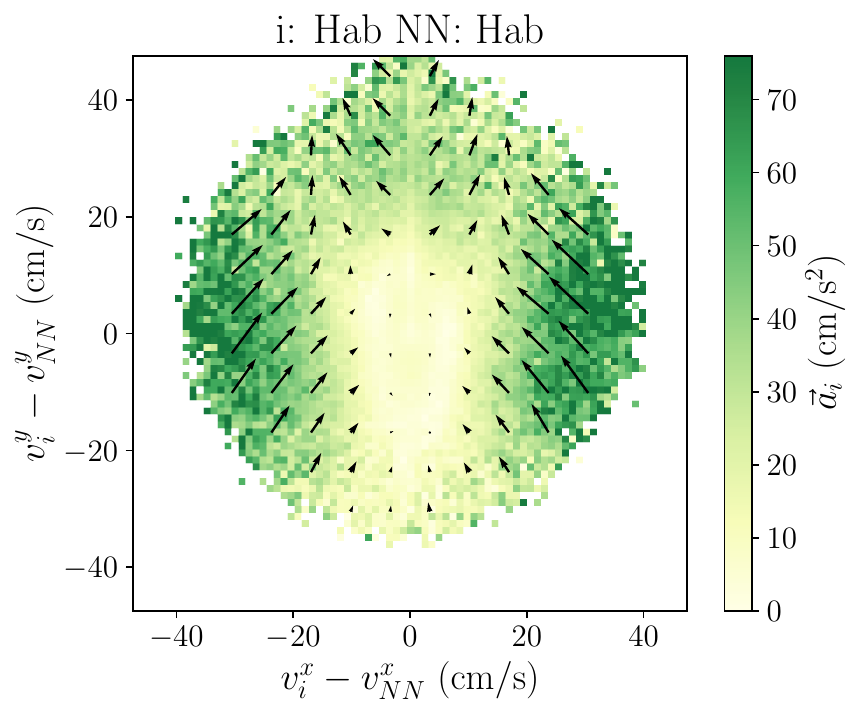}%
}
\subfloat[]{%
  \includegraphics[width=0.25\textwidth]{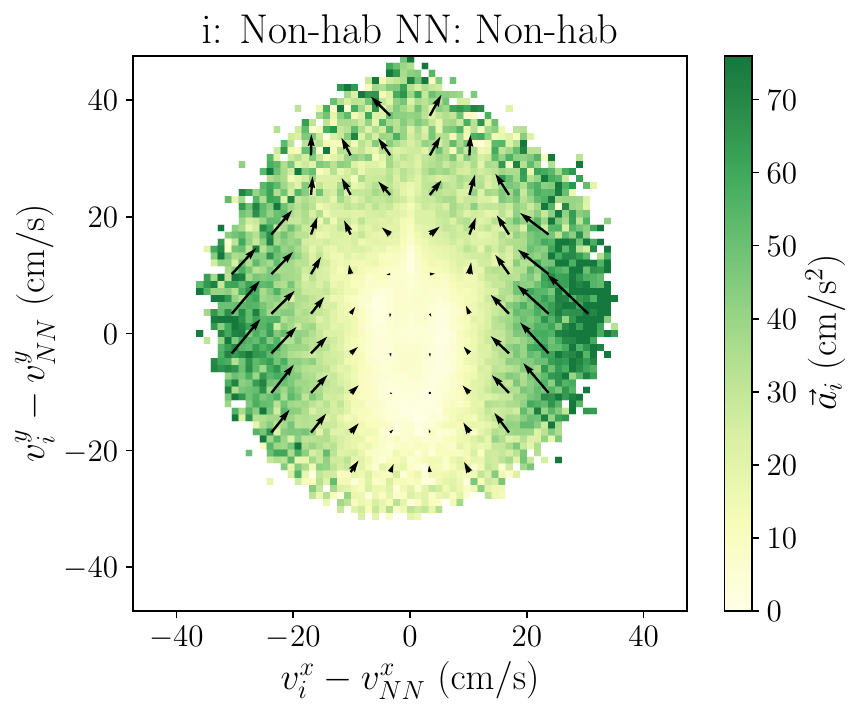}%
}
\subfloat[]{%
  \includegraphics[width=0.25\textwidth]{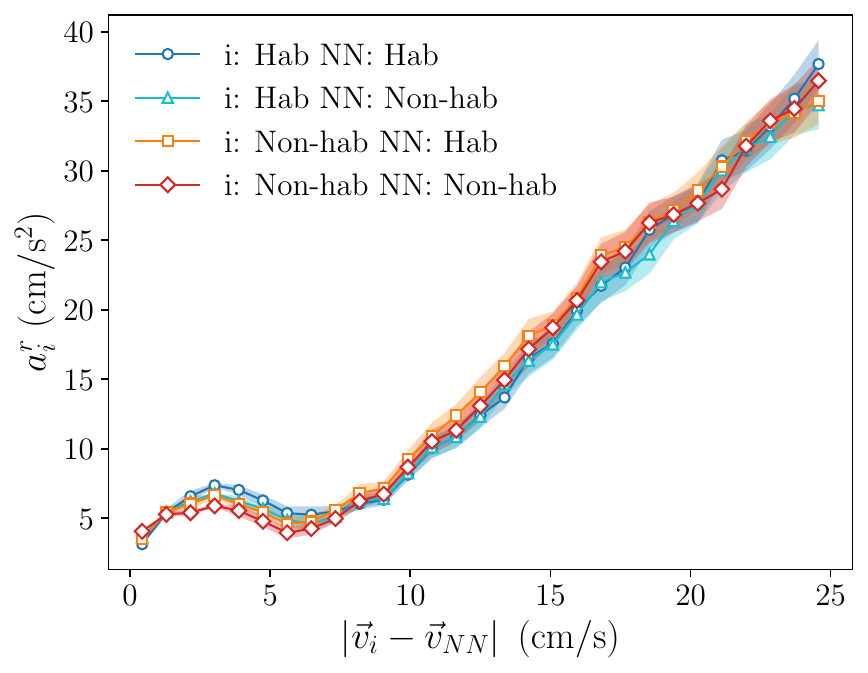}%
}
	\caption{Local-level alignment force map. (a) and (b) Average acceleration (alignment force map) of individual $i$ depending on the relative velocity with the nearest neighbor, $\vec{v}_i - \vec{v}_{NN}$, between (a) habituated and (b) non-habituated individuals, and (c) average projection of the acceleration in the radial direction of the alignment force map, $\left|\vec{v}_i - \vec{v}_{NN}\right|$.} \label{supp:fig:local_aligForceMap}
\end{figure}

\begin{figure}
	\centering
	\subfloat[]{%
  \includegraphics[width=0.25\textwidth]{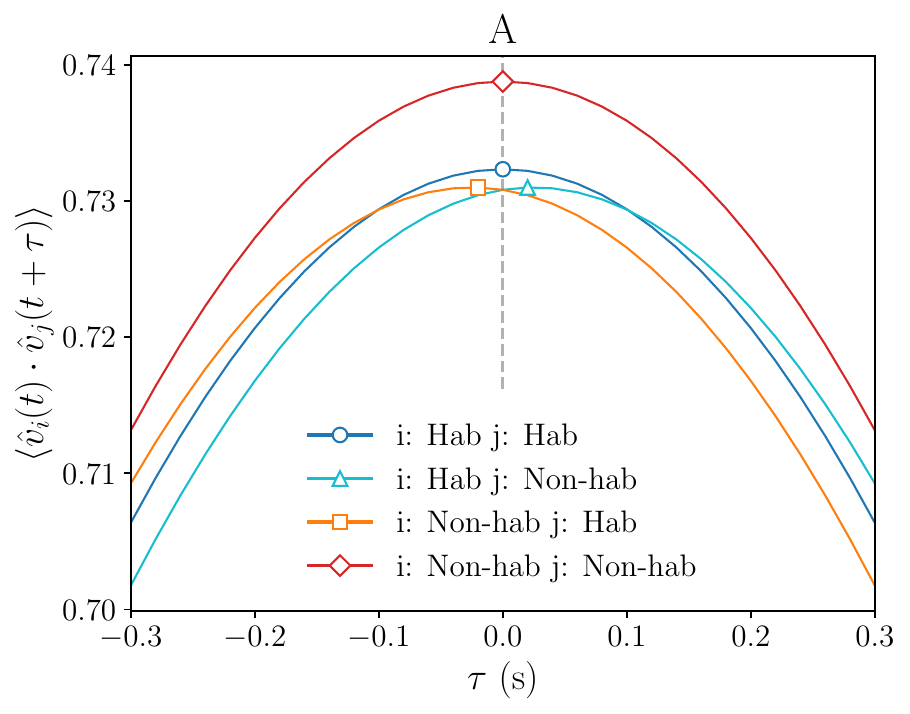}%
}
\subfloat[]{%
  \includegraphics[width=0.25\textwidth]{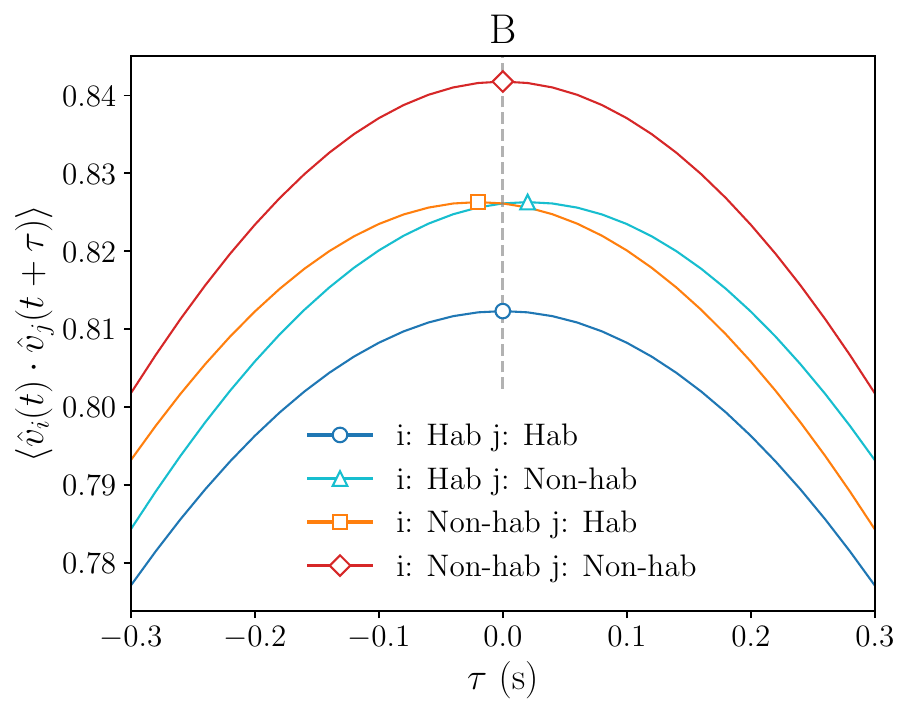}%
}
\subfloat[]{%
  \includegraphics[width=0.25\textwidth]{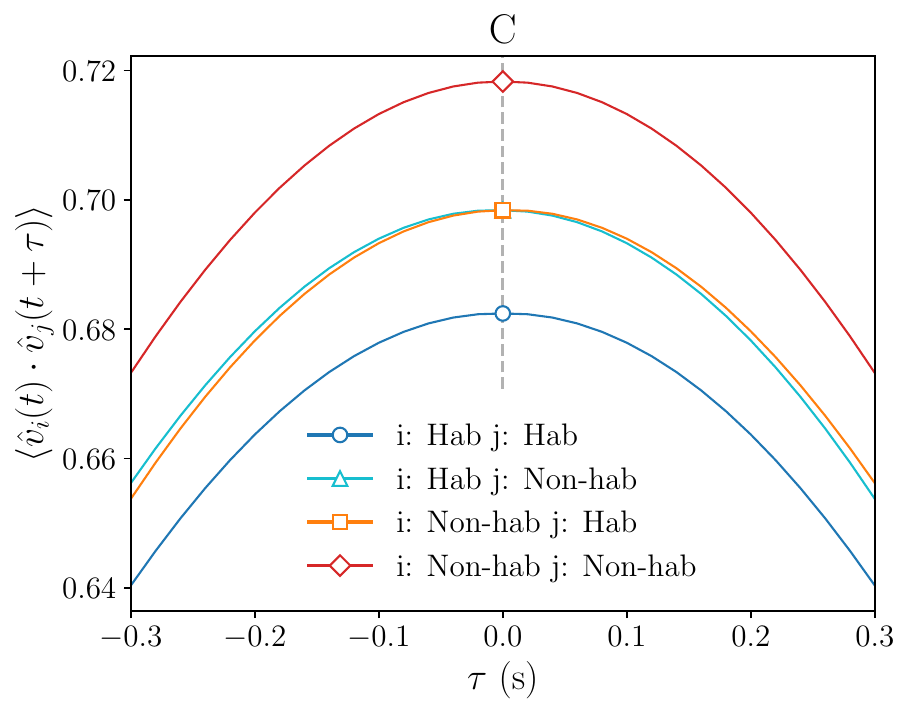}%
}

\subfloat[]{%
  \includegraphics[width=0.25\textwidth]{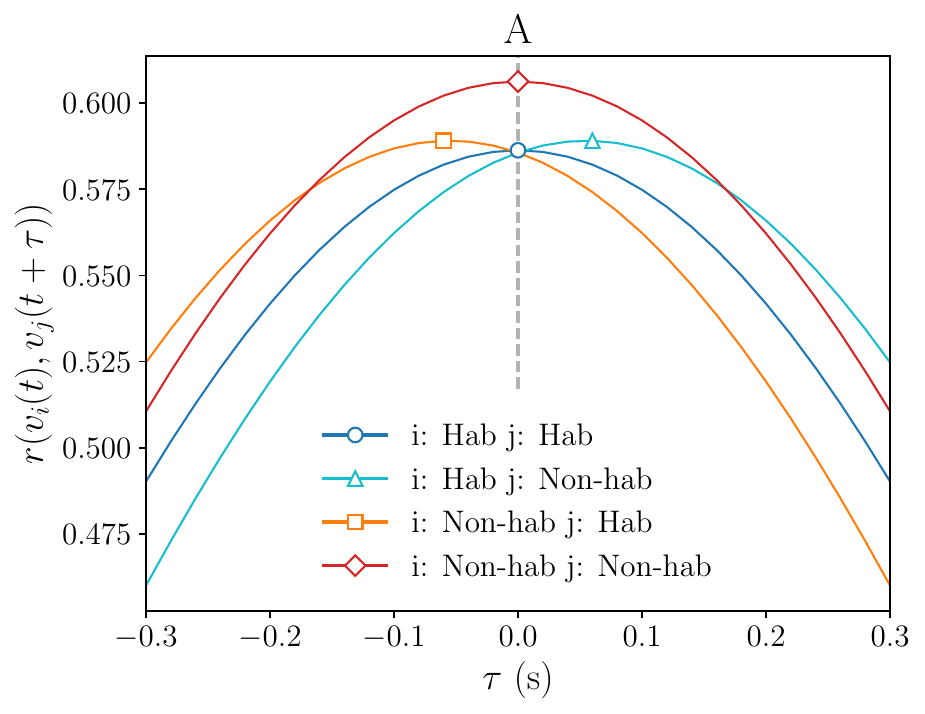}%
}
\subfloat[]{%
  \includegraphics[width=0.25\textwidth]{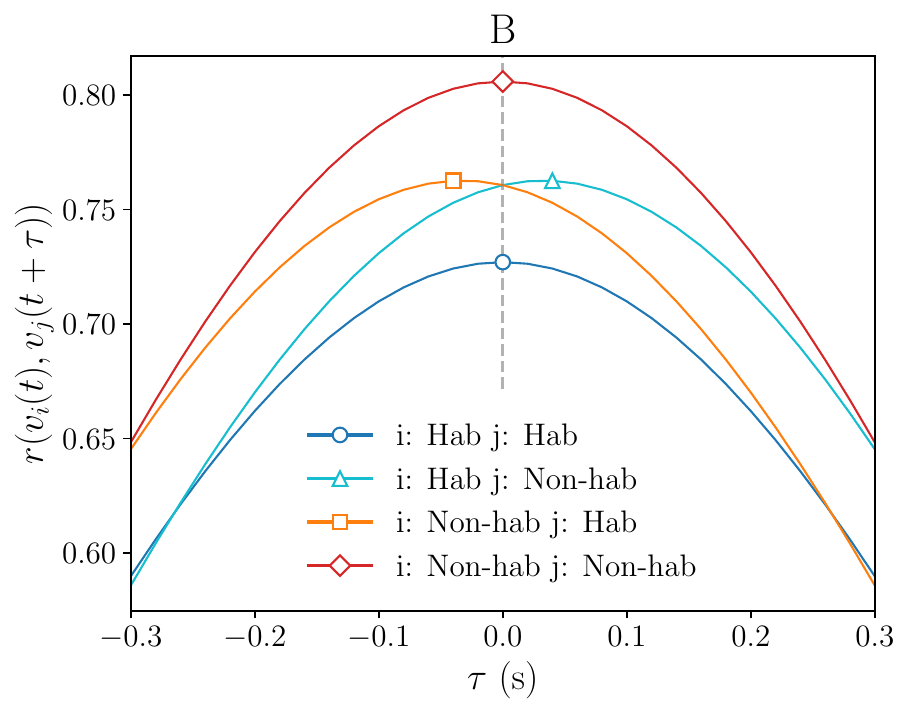}%
}
\subfloat[]{%
  \includegraphics[width=0.25\textwidth]{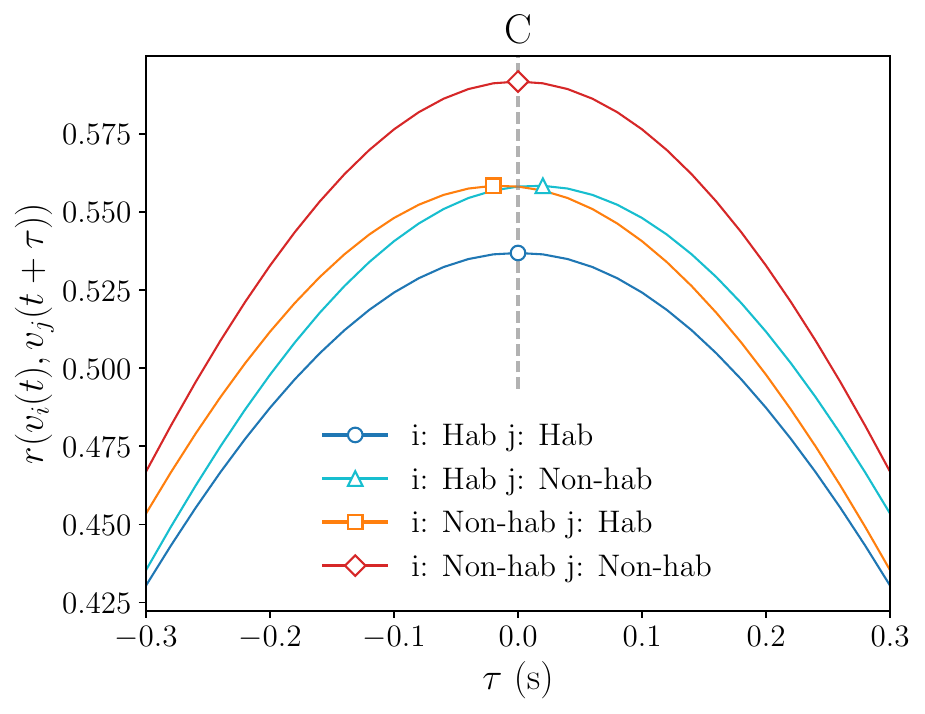}%
}
	\caption{(a)-(f) Leader-follower relationships (see Fig.~\ref{fig:leader-follower_relationships}) for different series.} \label{supp:fig:leader_follower_series}
\end{figure}

\begin{figure}
	\centering
	\subfloat[]{%
  \includegraphics[width=0.45\textwidth]{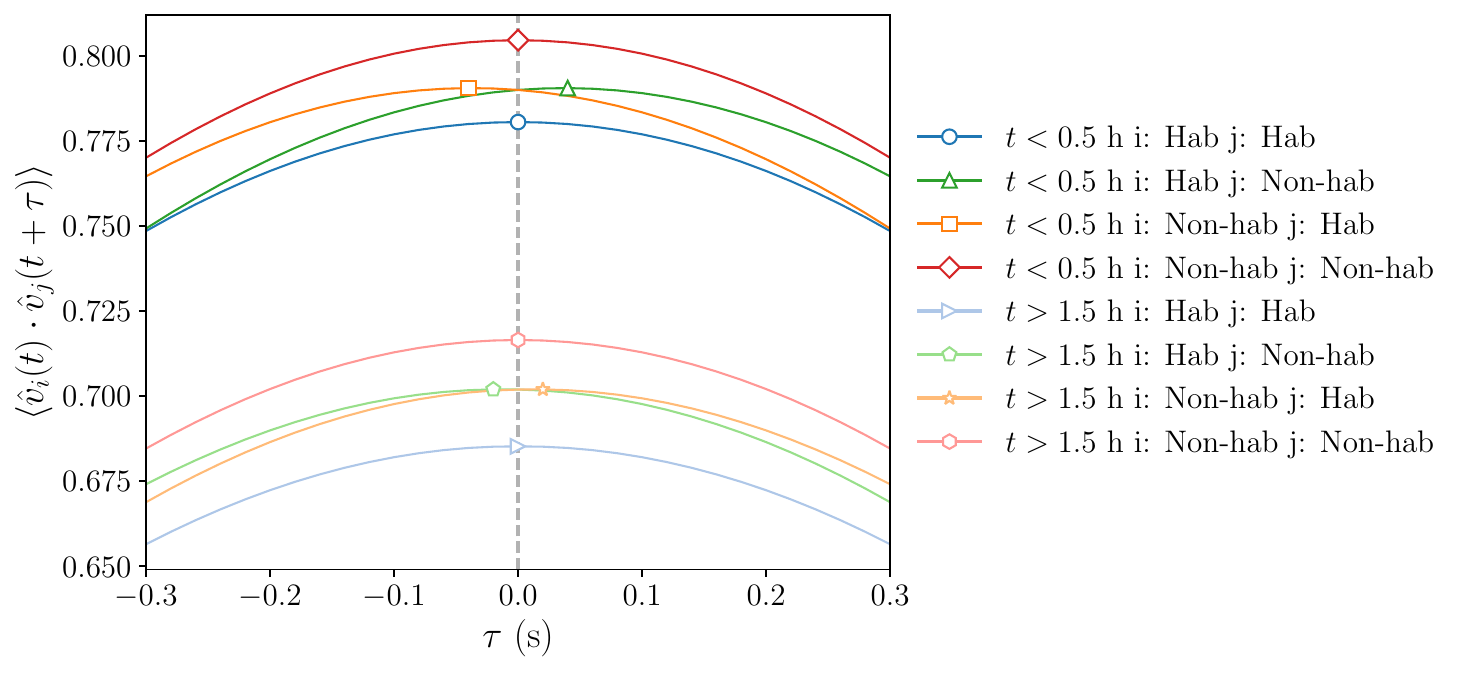}%
}
\subfloat[]{%
  \includegraphics[width=0.45\textwidth]{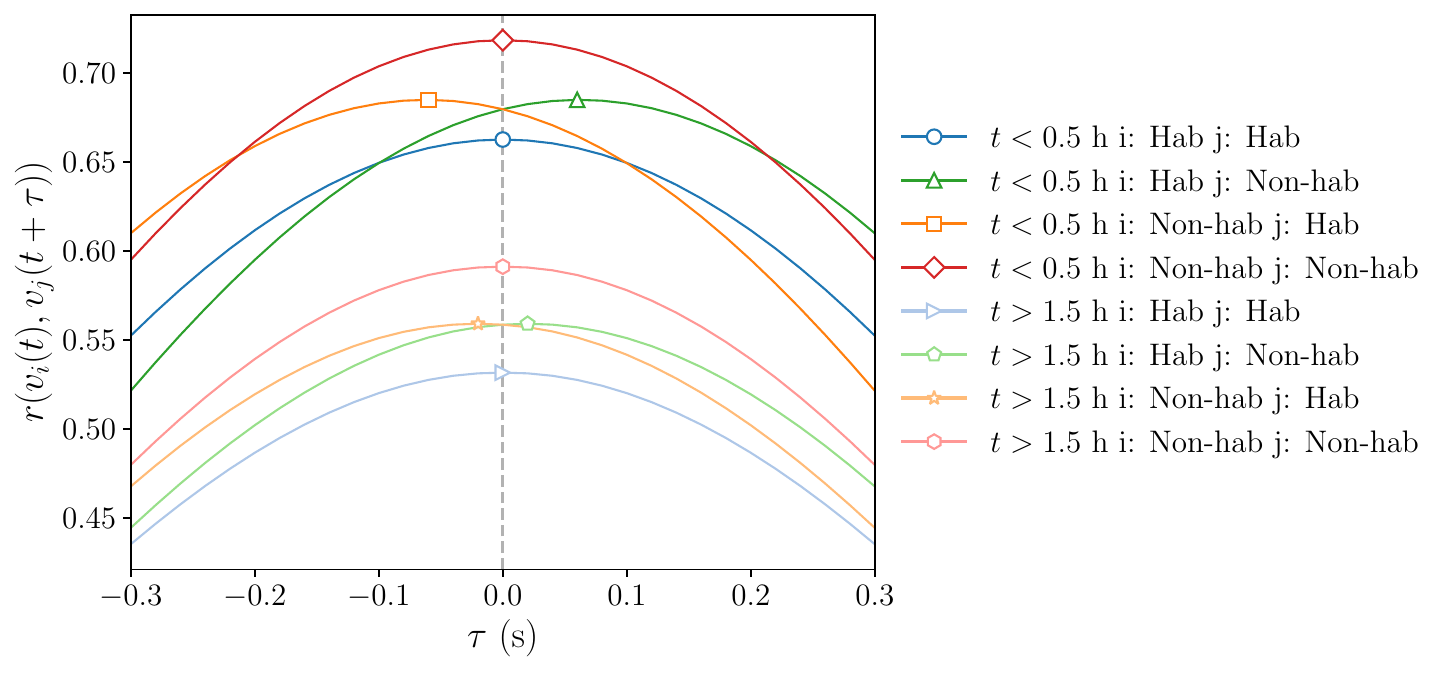}%
}
	\caption{(a) and (b) Temporal evolution of leader-follower relationships (see Fig.~\ref{fig:leader-follower_relationships}).} \label{supp:fig:leader_follower_tEvo}
\end{figure}

\begin{figure}
	\centering
	\subfloat[]{%
  \includegraphics[width=0.25\textwidth]{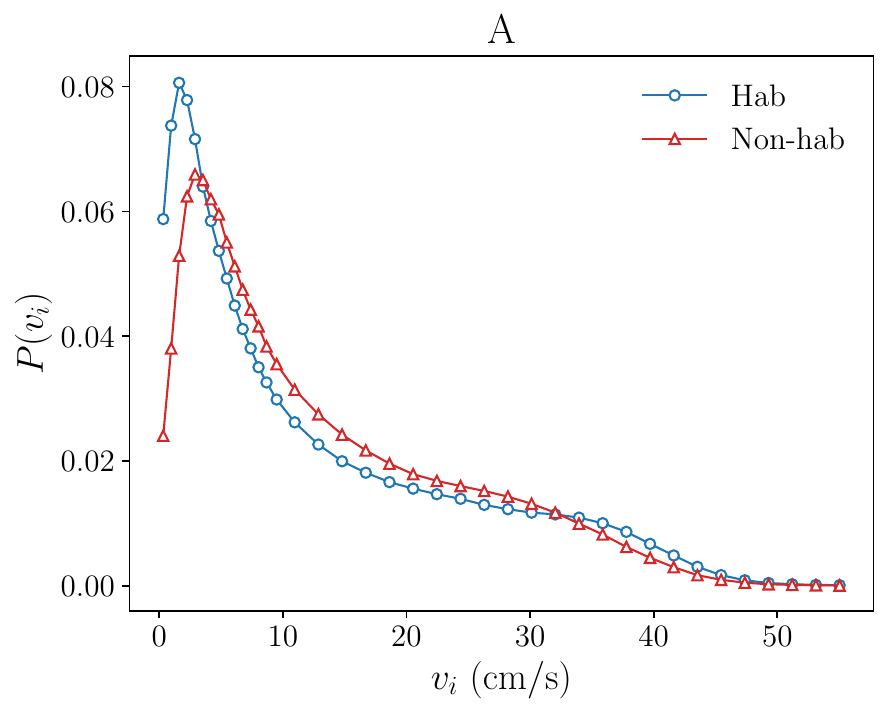}%
}
\subfloat[]{%
  \includegraphics[width=0.25\textwidth]{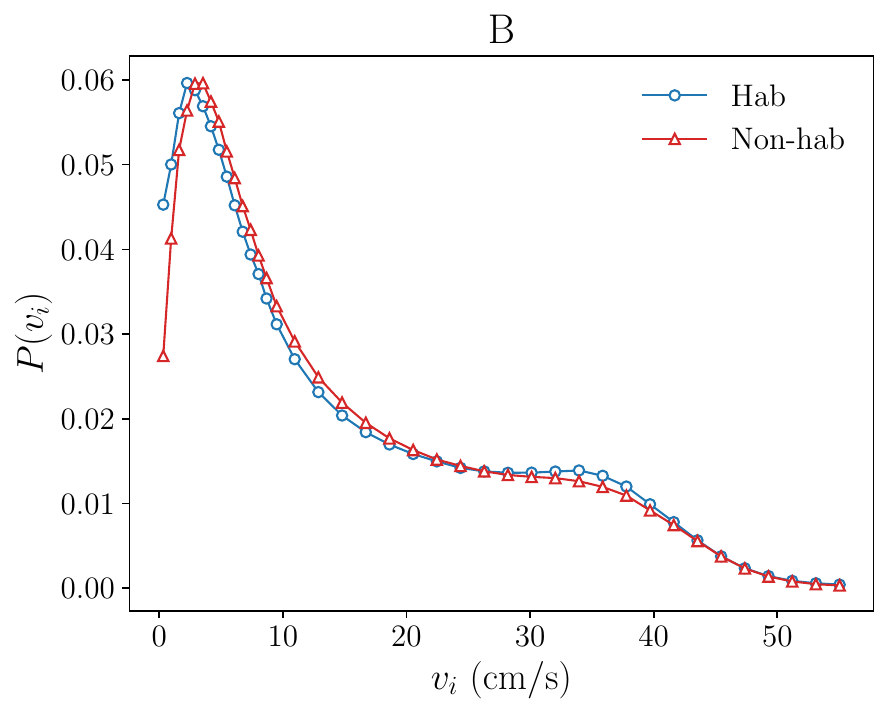}%
}
\subfloat[]{%
  \includegraphics[width=0.25\textwidth]{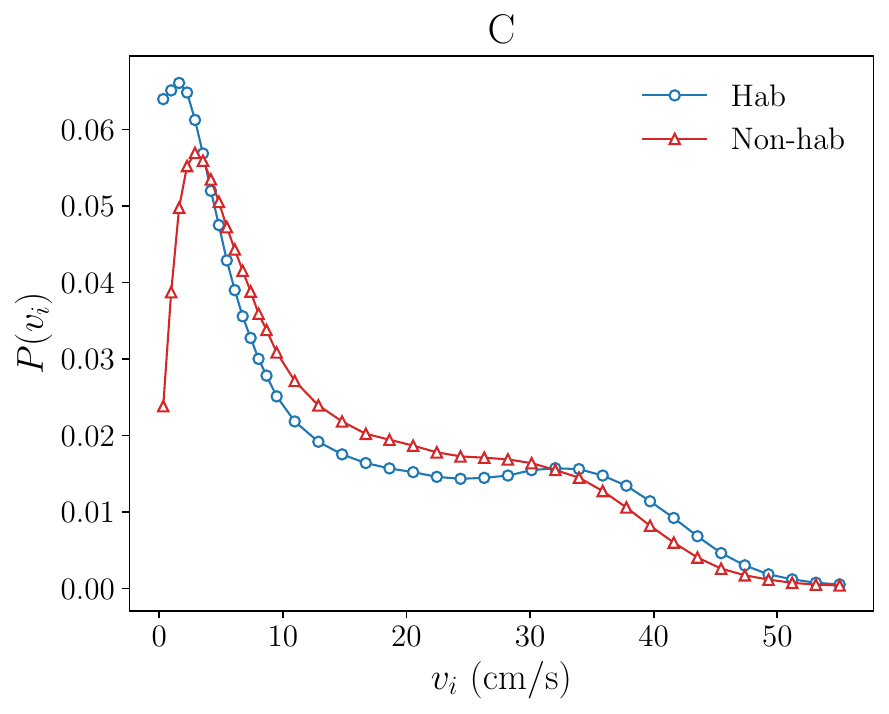}%
}

\subfloat[]{%
  \includegraphics[width=0.25\textwidth]{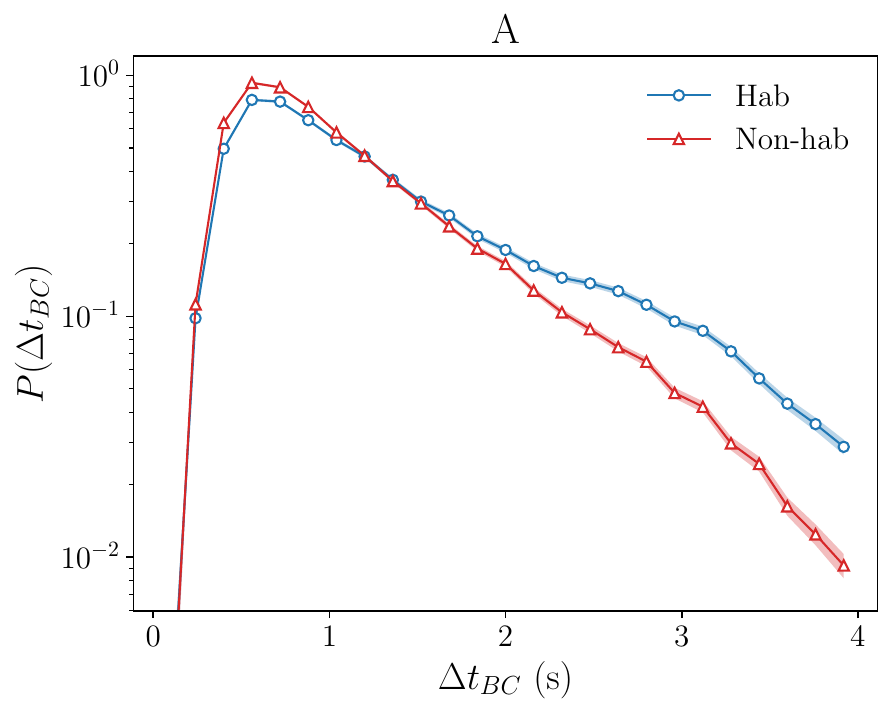}%
}
\subfloat[]{%
  \includegraphics[width=0.25\textwidth]{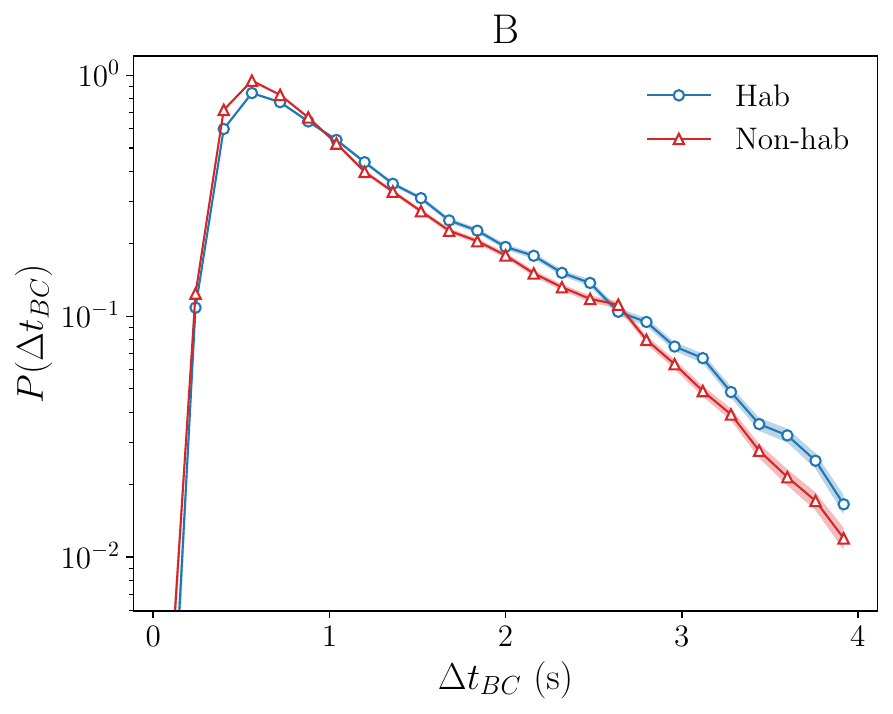}%
}
\subfloat[]{%
  \includegraphics[width=0.25\textwidth]{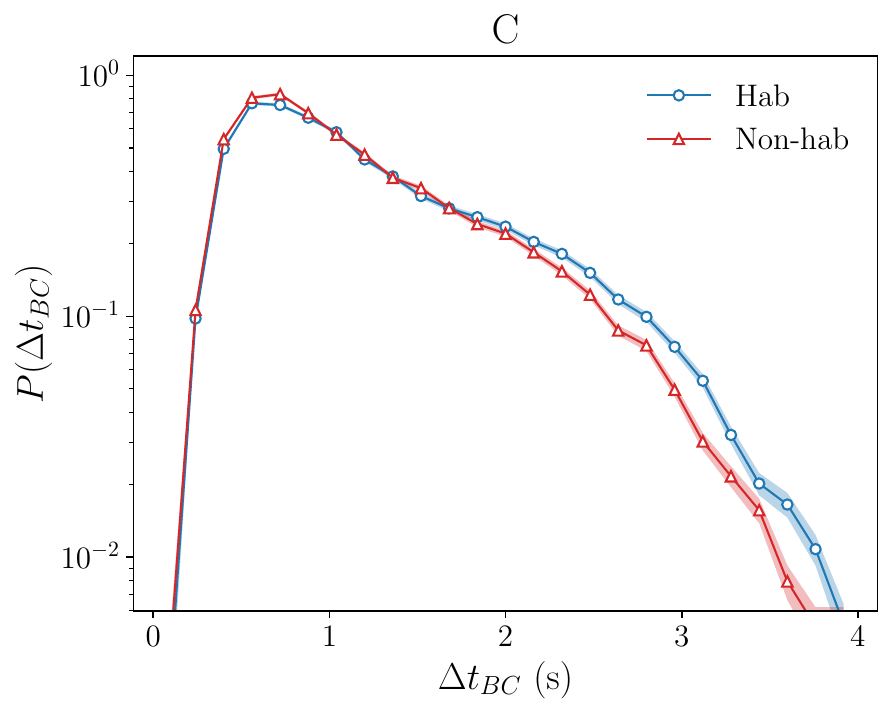}%
}

\subfloat[]{%
  \includegraphics[width=0.25\textwidth]{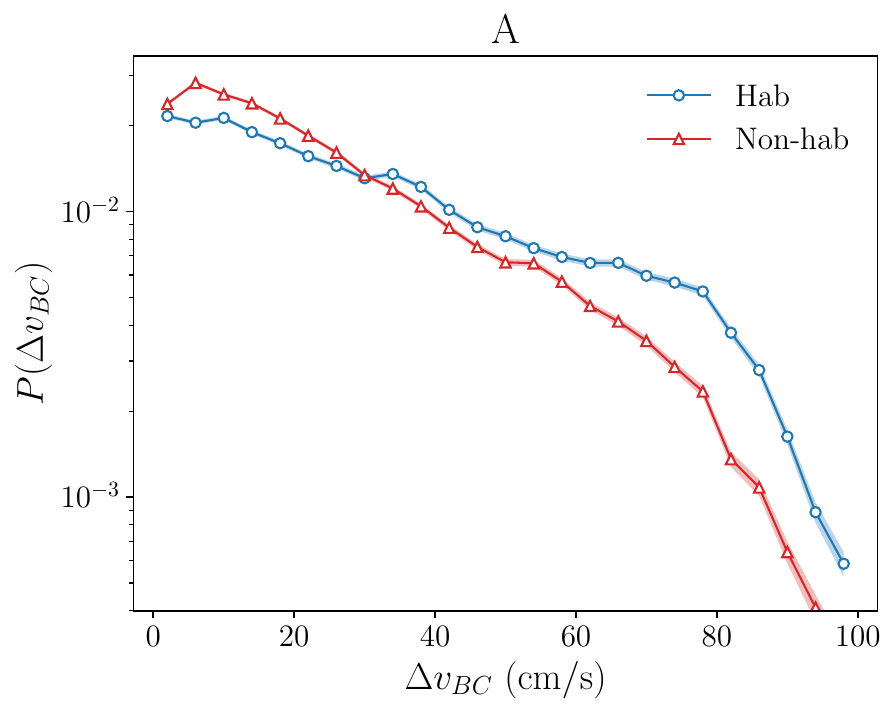}%
}
\subfloat[]{%
  \includegraphics[width=0.25\textwidth]{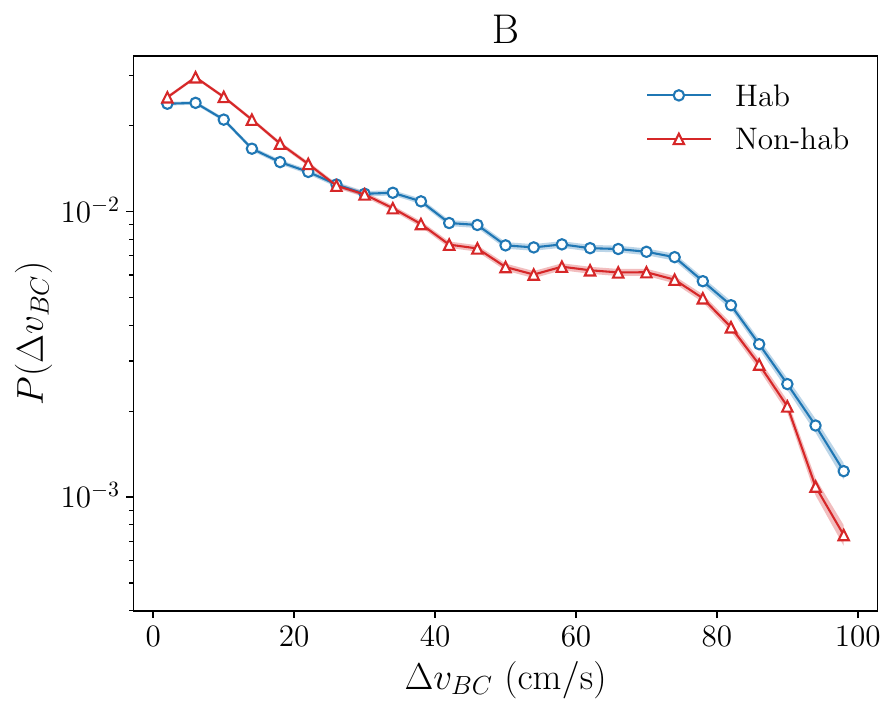}%
}
\subfloat[]{%
  \includegraphics[width=0.25\textwidth]{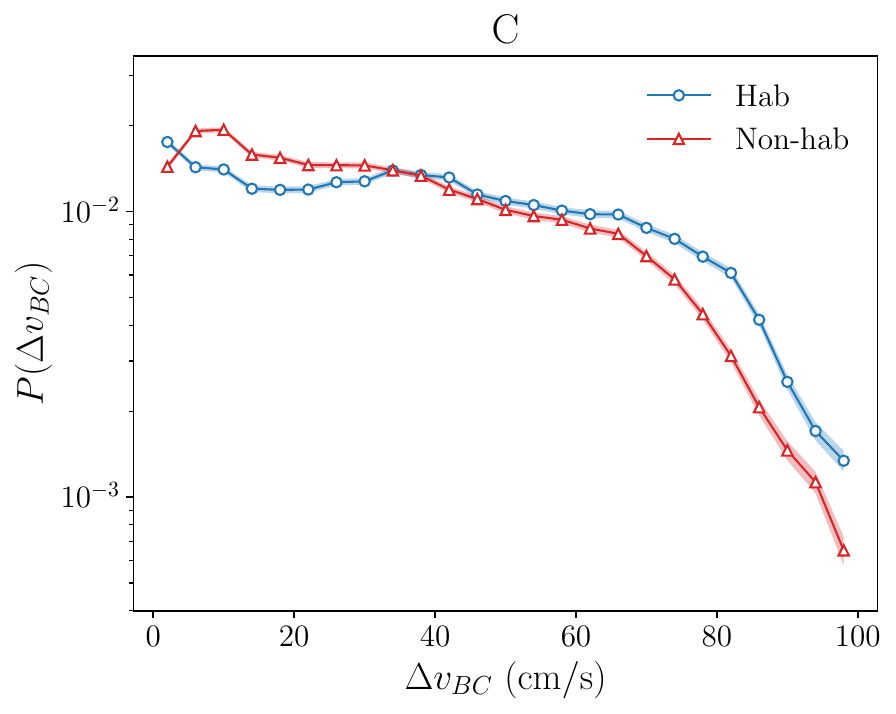}%
}
	\caption{(a)-(i) Burst-and-coast dynamics (see Fig.~\ref{fig:self-propulsion}) for different series.}\label{supp:fig:self_propulsion_series}
\end{figure}

\begin{figure}
	\centering
	\subfloat[]{%
  \includegraphics[width=0.25\textwidth]{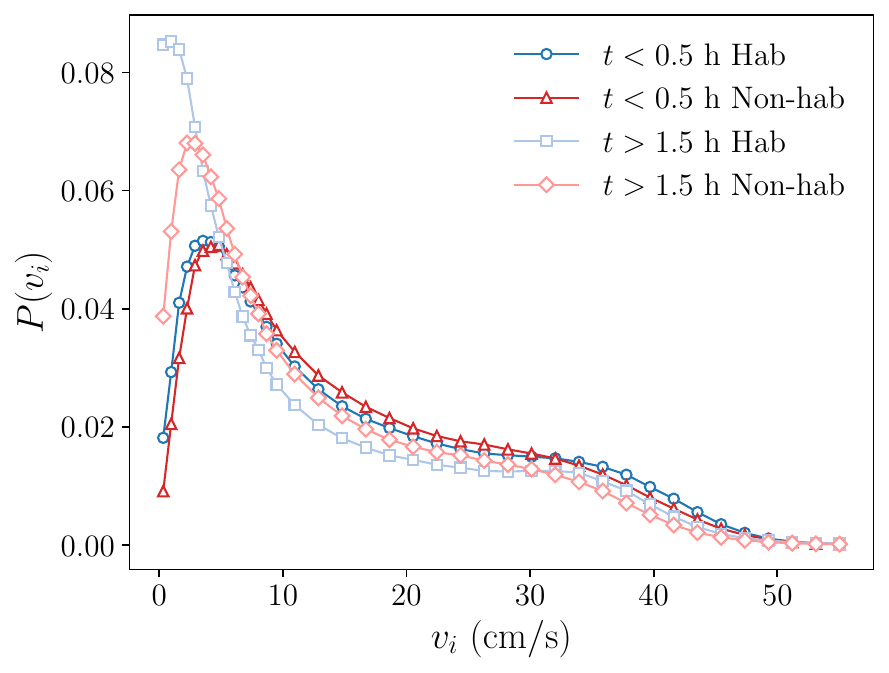}%
}
\subfloat[]{%
  \includegraphics[width=0.25\textwidth]{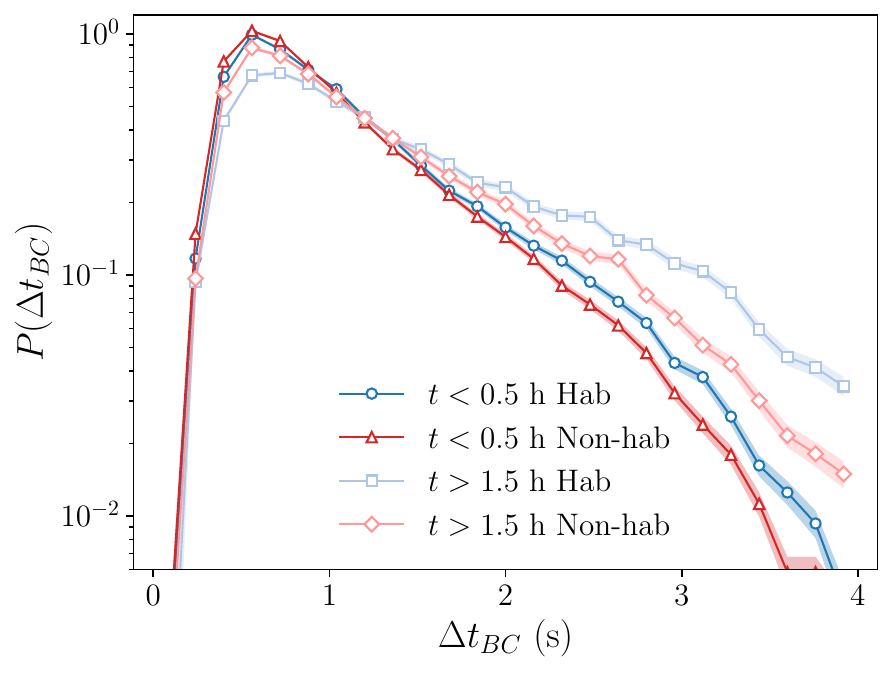}%
}
\subfloat[]{%
  \includegraphics[width=0.25\textwidth]{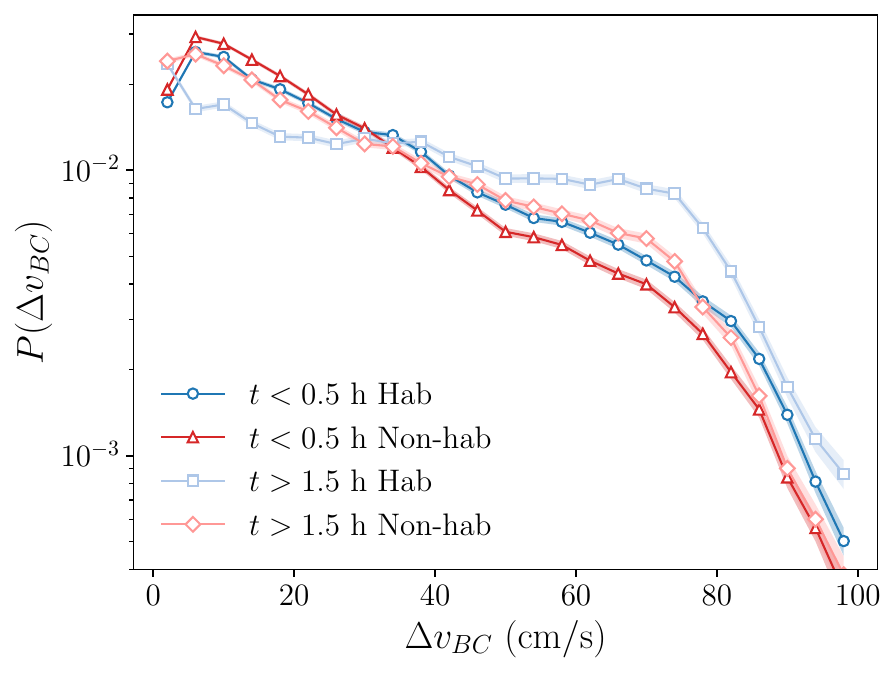}%
}
	\caption{(a)-(c) Temporal evolution of burst-and-coast dynamics (see Fig.~\ref{fig:self-propulsion})}\label{supp:fig:self_propulsion_tEvo}
\end{figure}

\begin{figure}
	\centering
	\subfloat[]{%
  \includegraphics[width=0.35\textwidth]{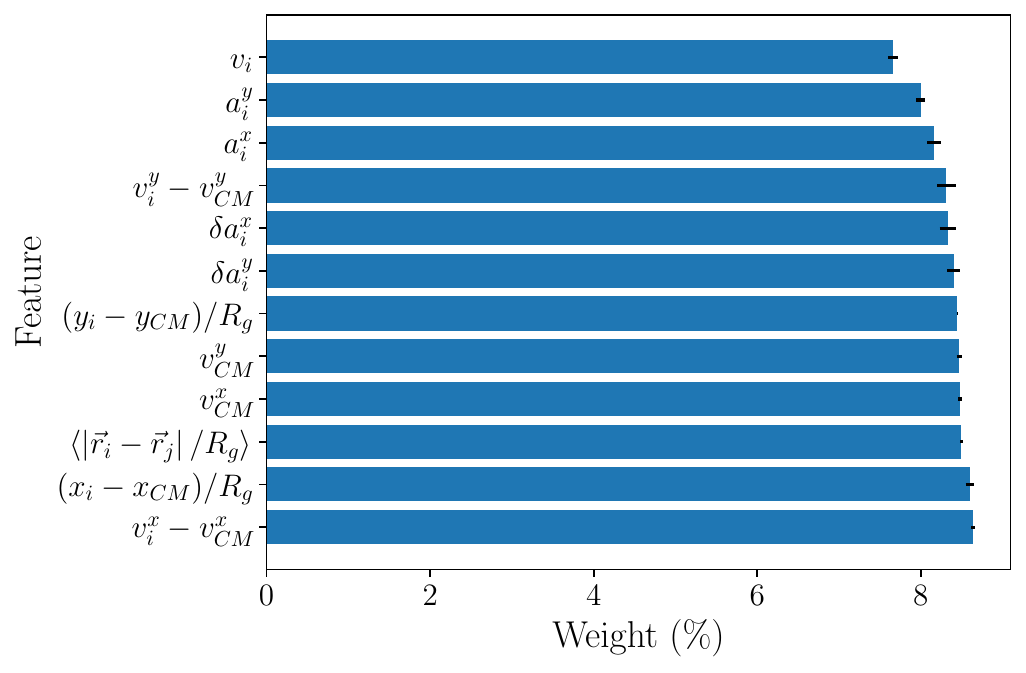}%
}
\subfloat[]{%
  \includegraphics[width=0.35\textwidth]{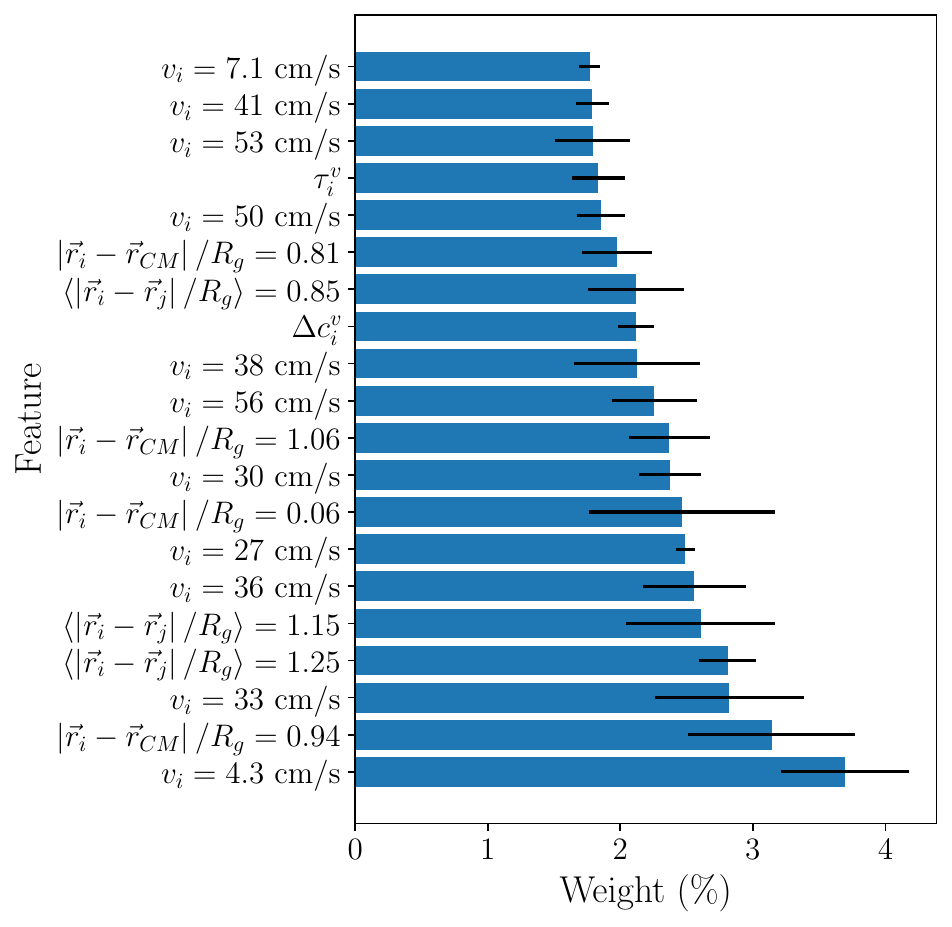}%
}
	\caption{Feature weights for (a) the instantaneous and (b) the temporal window machine learning tools. They are averaged across leave-one-series-out cross-validation folds. In (b), only the 20 highest-weight features (out of 124) are shown; labels denote either individual histogram bins or quantities derived from the leader–follower correlations (see~\ref{supp:fig:machineLearning_plots_tWindow}).}\label{supp:fig:machineLearning_weights}
\end{figure}

\begin{figure}
	\centering
\subfloat[]{%
  \includegraphics[width=0.3\textwidth]{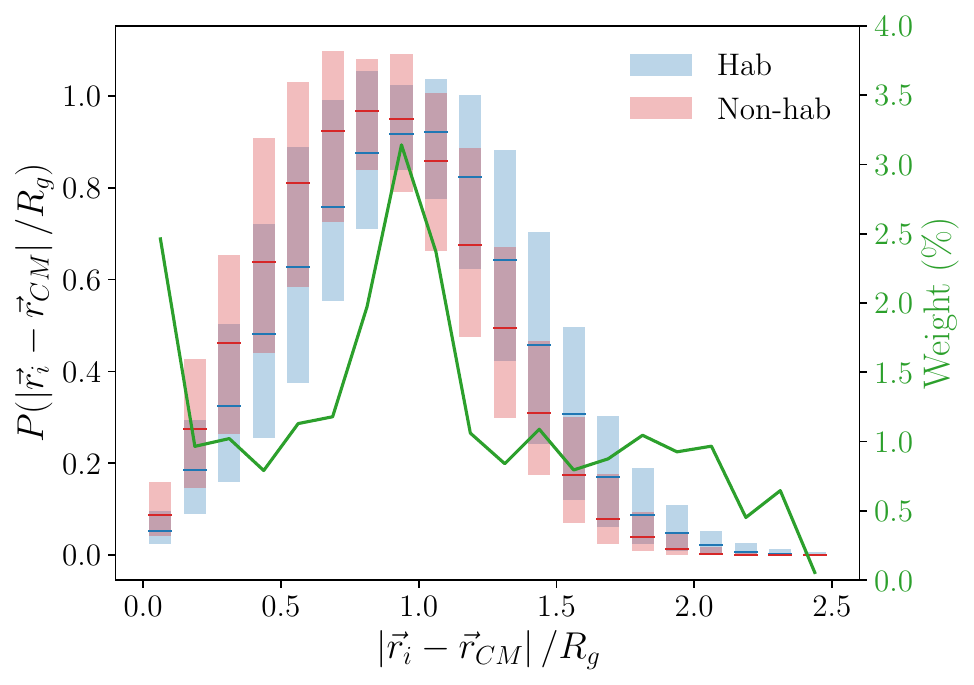}%
}
\subfloat[]{%
  \includegraphics[width=0.3\textwidth]{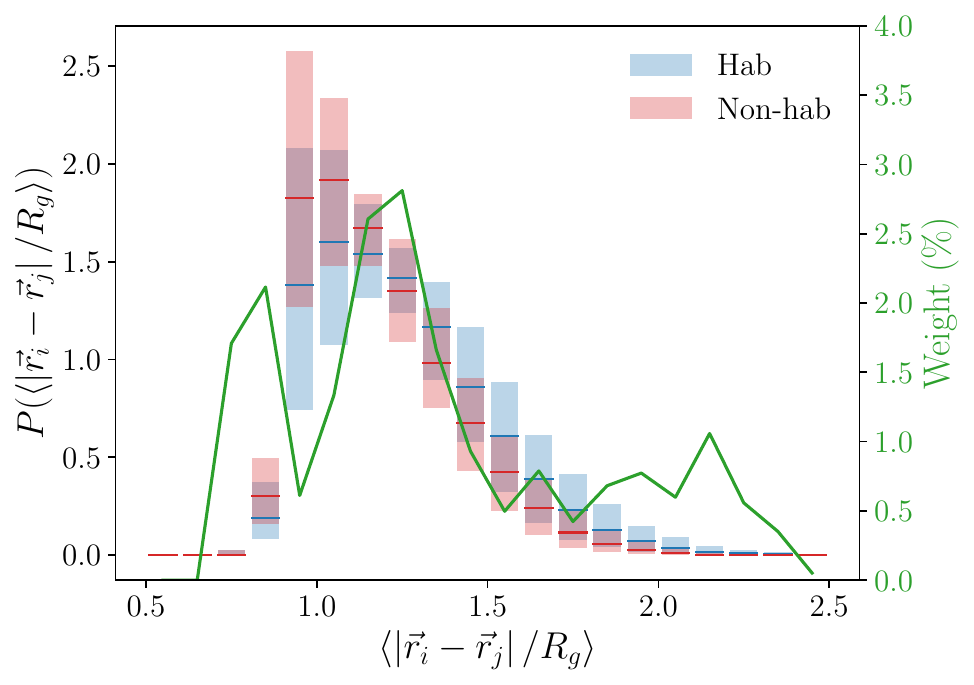}%
}
\subfloat[]{%
  \includegraphics[width=0.3\textwidth]{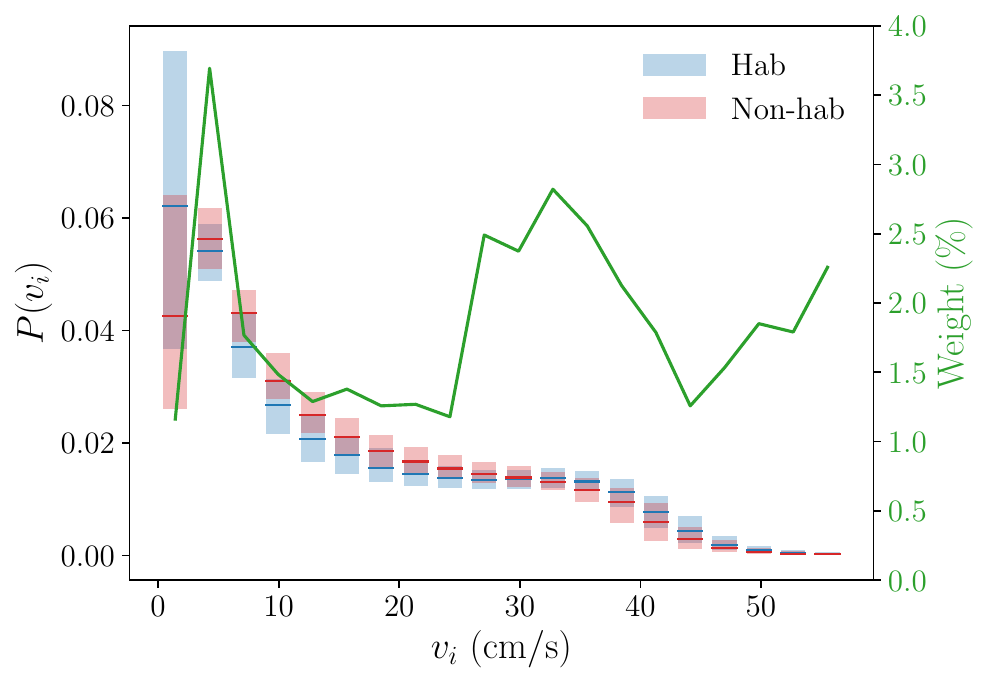}%
}

\subfloat[]{%
  \includegraphics[width=0.3\textwidth]{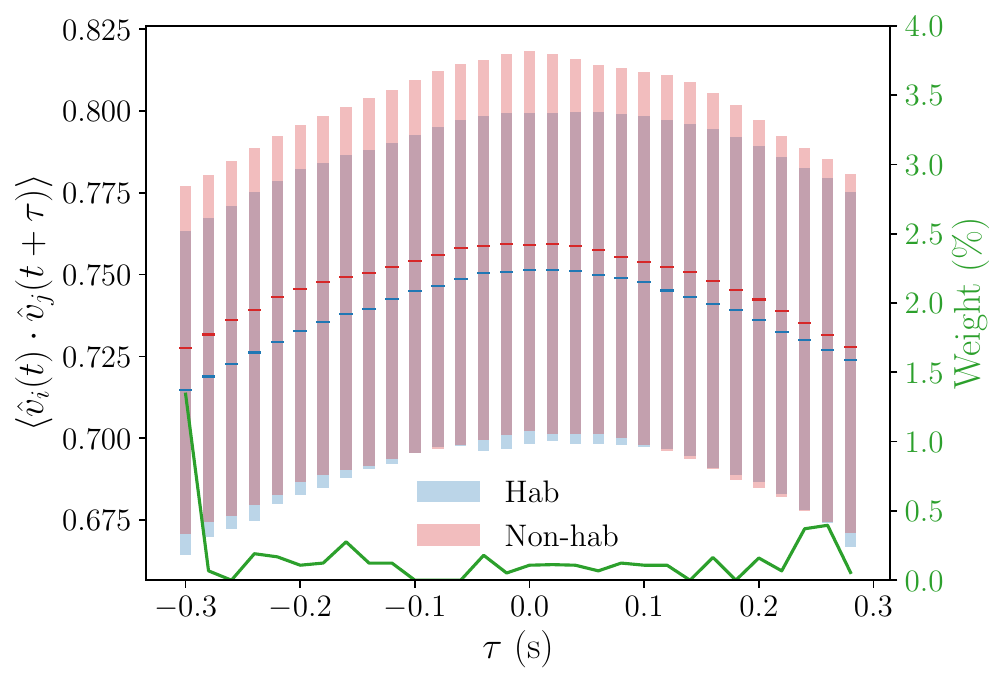}%
}
\subfloat[]{%
  \includegraphics[width=0.3\textwidth]{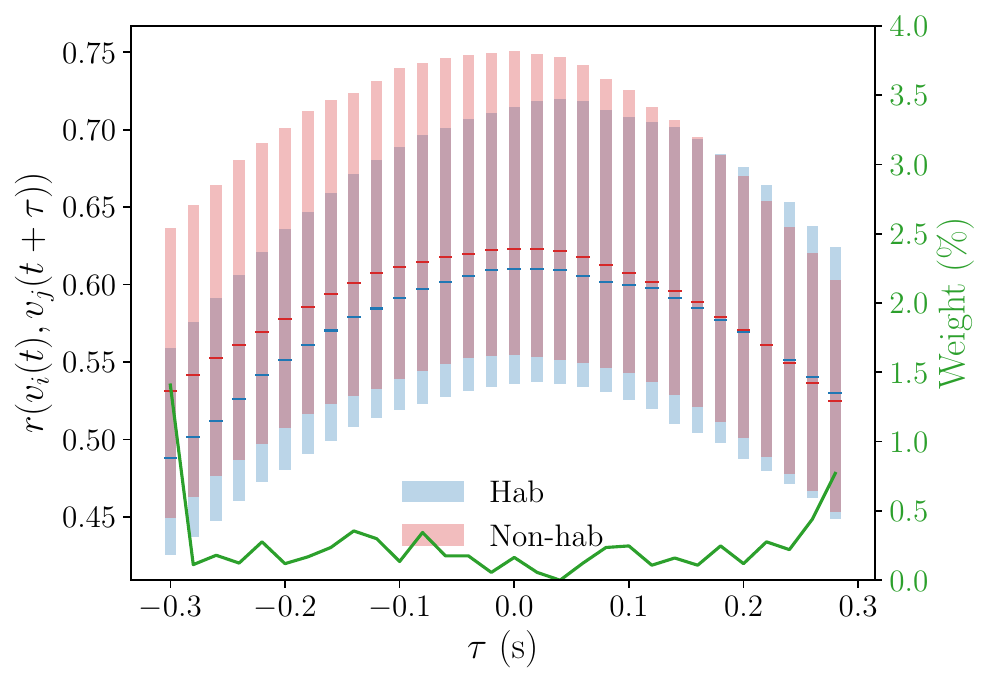}%
}

\subfloat[]{%
  \includegraphics[width=0.3\textwidth]{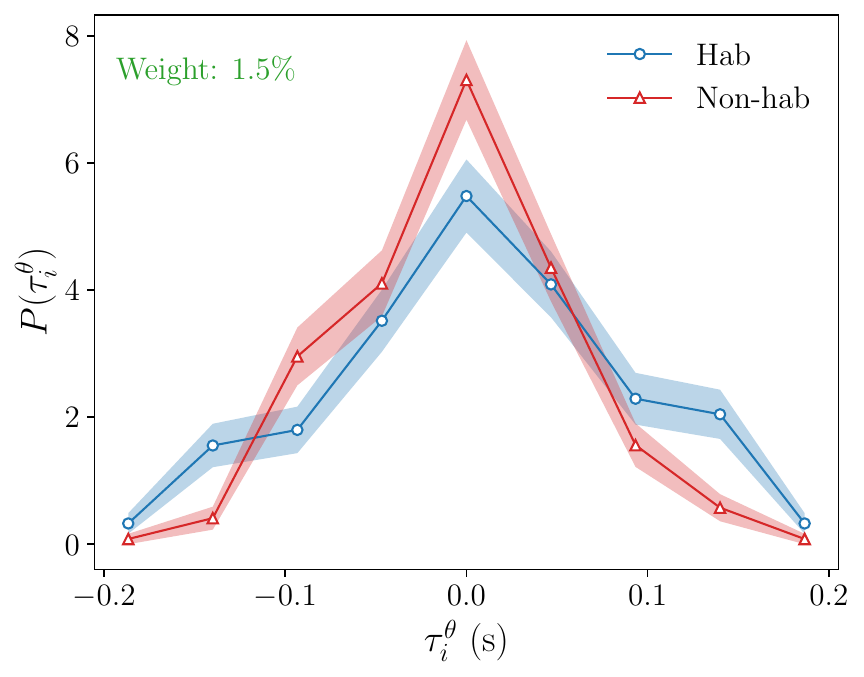}%
}
\subfloat[]{%
  \includegraphics[width=0.3\textwidth]{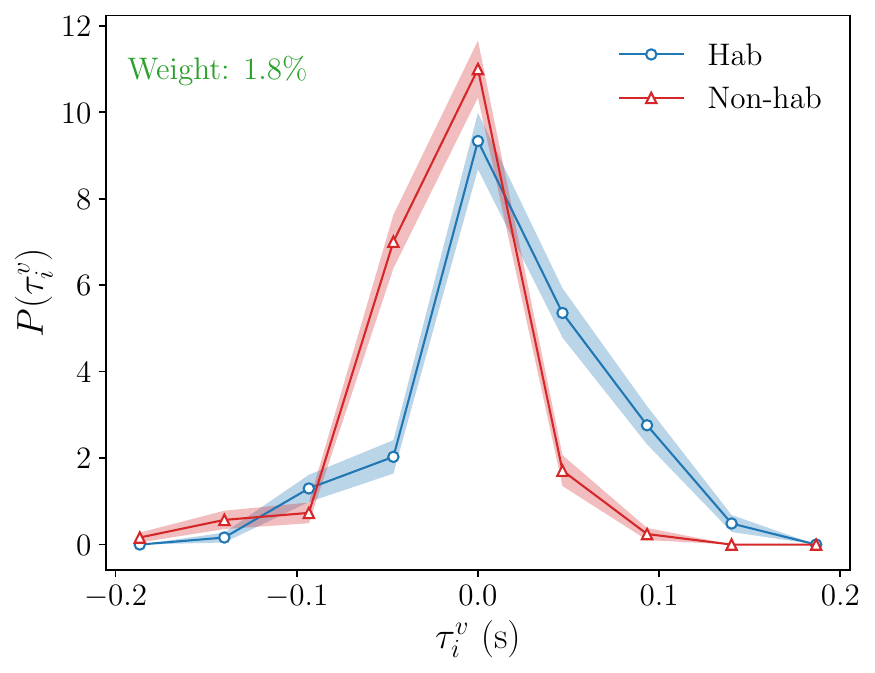}%
}

\subfloat[]{%
  \includegraphics[width=0.3\textwidth]{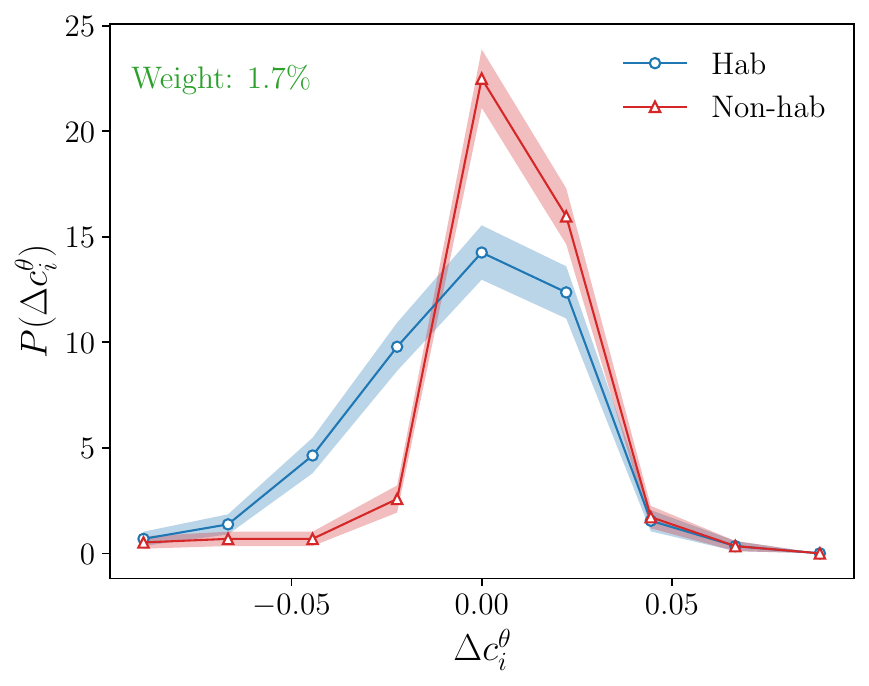}%
}
\subfloat[]{%
\includegraphics[width=0.3\textwidth]{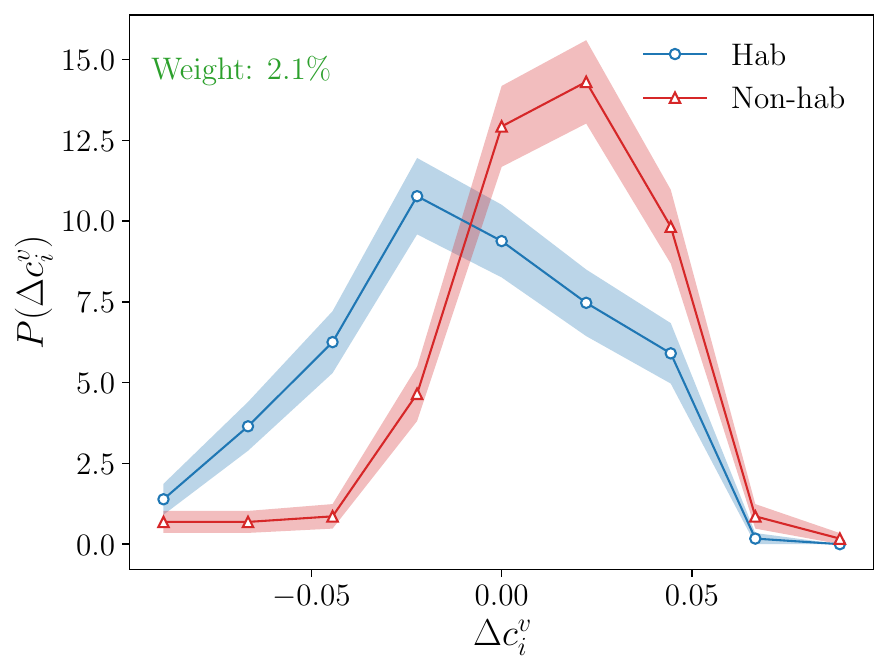}%
}
	\caption{Features of the temporal window machine learning tool. (a) Histogram of the normalized distance to the center of  mass $\left| \vec{r}_i - \vec{r}_{CM} \right|/R_g$, (b) histogram of the average normalized distance to other individuals $\left< \left| \vec{r}_i - \vec{r}_{j}\right|/R_g \right>$, (c) histogram of the speed $v_i$, (d)-(e) leader-follower correlations averaged across other individuals $j$ (d) in the orientation $\left< \hat{v}_i (t) \cdot \hat{v}_{j} (t+\tau) \right>$ and (e) in the speed $r(v_i(t), v_{j}(t+\tau))$, (f)-(g) PDF of the delay of the maximum correlations in (f) the orientation $\tau^\theta_i$ and (g) the speed $\tau^v_i$ and (h)-(i) relative difference of the maximum correlation value with respect to the group in (h) the orientation $\Delta c^\theta_i$ and (i) the speed $\Delta c^v_i$. In (a)-(e), box plots indicate the first, second, and third quartiles of the function for individuals within the subgroups. In green we show the feature weights of the machine learning algorithm, averaged across leave-one-series-out cross-validation folds.}\label{supp:fig:machineLearning_plots_tWindow}
\end{figure}

\clearpage

\end{document}